\documentclass[epj,nopacs,final]{svjour}
\usepackage{graphics}
\usepackage{latexsym,overpic,rotating,float}
\usepackage{subfigure,color,showkeys}
\usepackage{epsfig,color,rotating,amsmath,delarray,array,multirow}
\usepackage{makeidx,pifont,float,amssymb}
\usepackage{ulem} 
\newcommand{\er}{$\pm$}
\newcommand{\be}{\begin{eqnarray}}
\newcommand{\ee}{\end{eqnarray}}
\newcommand{\bea}{\begin{eqnarray}}
\newcommand{\eea}{\end{eqnarray}}
\newcommand{\bc}{\begin{center}}
\newcommand{\ec}{\end{center}}

\newcommand{\beq}{\begin{equation}}
\newcommand{\eeq}{\end{equation}}
\newcommand{\ba}{\begin{eqnarray}}
\newcommand{\ea}{\nonumber \end{eqnarray}}

\newcommand{\bi}{\begin{enumerate}}
\newcommand{\ei}{\end{enumerate}}

\begin{document}

\title{\boldmath Hyperon II: Properties of excited hyperons}
\titlerunning{Hyperon II: Properties of excited hyperons}

\author{ A.V.~Sarantsev\inst{1,2}, M.~Matveev\inst{1,2}, V.A. Nikonov\inst{1,2}, A.V. Anisovich\inst{1,2}, U. Thoma\inst{1}, and
E.~Klempt\inst{1} }
\authorrunning{M.~Matveev {\it et al.}}

\institute{\inst{1}Helmholtz--Institut f\"ur Strahlen--
                     und Kernphysik, Universit\"at Bonn, 53115 Bonn, Germany\\
\inst{2}National Research Centre ``Kurchatov Institute'', Petersburg Nuclear Physics Institute,
Gatchina, 188300 Russia}

\date{\today}

\abstract{We report properties of $\Lambda$ and $\Sigma$ hyperon resonances formed in $K^-$ induced
reactions. Special emphasis is laid on the analysis of the three-body final states $2\pi^0\Lambda$ and $2\pi^0\Sigma$
and of the quasi-two-body final states $\pi\Lambda(1520)$,  $\bar K\Delta(1232)$, $\pi\Sigma(1385)$, $\bar K^*N$,
and $\omega\Lambda$. We
give pole positions of $\Lambda$ and $\Sigma$ hyperon resonances and transition residues from the $K^-p$
initial to various final states as well as Breit-Wigner masses and widths and decay branching
ratios. Twenty resonances and ``bumps'' reported in the Review of Particle Physics
are not required in our fits, evidence for five new resonances is reported.
The observed mass spectrum is compared to the spectrum calculated in the Bonn quark model.
Three spin doublets, six $\Lambda$ hyperons, are tentatively assigned to the SU(3) singlet system. }


\maketitle

\section{Introduction}

The nature of hadron resonances is of topical interest, important questions need to be answered. Do
conventional quark models provide a complete picture when they interpret meson resonances as
composed of a quark and an antiquark and baryon resonances as composed of three quarks? Are there
resonances beyond this picture, glueballs, i.e. bound states of glue without constituent quarks;
are there hybrids in which the gluon string between quarks may carry additional excitation? Are
there tetraquarks or pentaquarks? Modern approaches are based on effective field theories and
describe an increasing number of resonances as hadronic molecules bound by strong interactions. The
approach provides a systematic access to the production and to the decay processes of many
resonances. These resonances are called ``dynamically generated''. Well-known examples in the
baryon sector are $\Lambda(1405)1/2^-$ that is generated from $\bar K N-\Sigma\pi$ coupled channel chiral
dynamics
\cite{Dalitz:1959dn,Kaiser:1995eg,Oset:1997it,Oller:2000fj,Jido:2003cb,Ikeda:2012au,Guo:2012vv,Mai:2014xna,Miyahara:2018onh},
$N(1535)1/2^-$ can be interpreted as dynamically generated quasi-bound $\Lambda K -\Sigma K$ state 
\cite{Kaiser:1995cy,Inoue:2001ip,Mai:2012wy}, $\Delta(1700)3/2^-$ from $\Delta(1232)\eta$
\cite{Doring:2010fw}, and $\Xi$(2030) and $\Xi$(2120) are interpreted as $\bar{K}^*\Sigma$
molecular states~\cite{Huang:2018uox}. Further examples can be found in Ref.~\cite{Guo:2017jvc}.
These observations lead to the question which resonances can be generated dynamically from
appropriate decay products and which ones not.

Dynamically generated states are often observed close to or in between two-particle thresholds. It is
hence important to measure all important decay modes of a resonance. High-mass resonances are close
to an opening threshold only for massive decay products. It is hence particularly interesting to study decay
modes of resonances into excited intermediate states like $\rho$, $\omega$,
or $K^*(892)$, or into $\Delta(1232)3/2^+$,
$\Lambda(1520)3/2^-$ or $\Sigma(1385)3/2^+$. From~now~on\-wards, these resonances will be
abbreviated as $K^*$, $\Delta(1232)$, $\Lambda(1520)$ or $\Sigma(1385)$ (and as
$K^*$, $\Delta$, $\Lambda^*$ or $\Sigma^*$ in the Tables). Coupled-channel
techniques involving vector mesons or baryons with higher spin are being developed
\cite{Lutz:2018kaz} with the aim to test the hadrogenesis conjecture. This conjecture expects that it might be possible
to generate the full spectrum of meson and baryon resonances by final-state interactions
of mesons and baryons including their respective excitations. 

In the preceding paper~\cite{Matveev:2019igl} we reported a coupled-channel analysis of data on
$K^-p$ scattering into two-body final states like elastic ($K^-p\to K^-p$) or charge exchange
($K^-p\to \bar K^0n$) scattering, or in inelastic reactions like $K^-p$ $\to \pi^0\Lambda$, $\pi^{\pm
0}\Sigma^{\mp 0}$, $\eta\Lambda$, $\eta\Sigma$, and $K^{0+} \Xi^{0-}$. References to these data
and a detailed description of the analysis method are given in Ref.~\cite{Matveev:2019igl}. In this
paper we extend the report to three-body final states $\pi^0\pi^0\Lambda$~\cite{Prakhov:2004ri},
$\pi^0\pi^0\Sigma$~\cite{Prakhov:2004an}, and the quasi-two-body final states
$\pi^0\Lambda(1520)$~\cite{Cameron:1977jr,Litchfield:1973ap}, $\bar K\Delta$~\cite{Litchfield:1973ey}, 
$\pi\Sigma(1385)$~\cite{Cameron:1978en}, $\bar K^*N$~\cite{Cameron:1978qi}, 
and $\omega\Lambda$ \cite{Brandstetter:1972xp,Nakkasyan:1975yz,Baccari:1976ik}. We emphasize that in both 
papers, all data are included in the partial wave analysis. In
Section~\ref{Data} we show the data in comparison to our fit. The results, decay modes of $\Lambda$
and $\Sigma$ resonances into various quasi-two-body final states, are presented in
Section~\ref{Results}. In Section~\ref{sec:class}, the spectrum of hyperon resonances is compared
to the Bonn quark model~\cite{Loring:2001ky}. The paper ends with a short Summary (Section~\ref{Summary}).

\section{\label{Data}\boldmath Data on $K^-p\to$ three-body finals states}
\subsection{\label{BNL}Reactions $K^-p\to \pi^0\pi^0\Lambda$ and $K^-p\to\pi^0\pi^0\Sigma$}

\begin{figure*}[pt]
\begin{center}
\begin{tabular}{ccc}
\includegraphics[width=0.32\textwidth,height=0.25\textwidth]{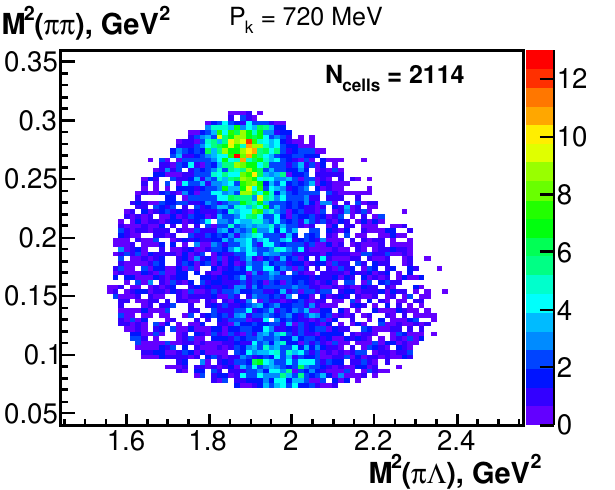}&
\hspace{-2mm}\includegraphics[width=0.32\textwidth,height=0.25\textwidth]{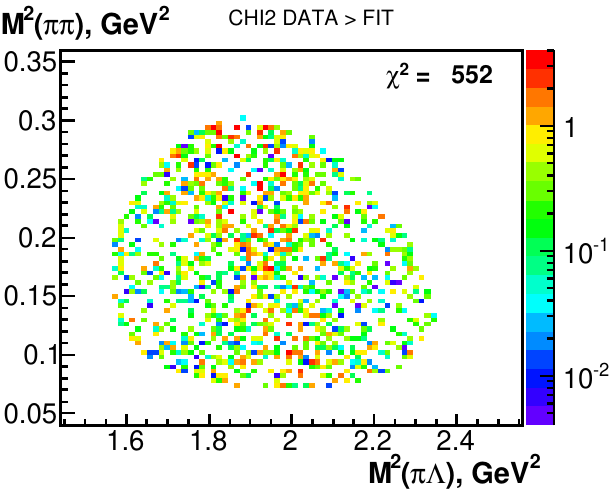}&
\hspace{-2mm}\includegraphics[width=0.32\textwidth,height=0.25\textwidth]{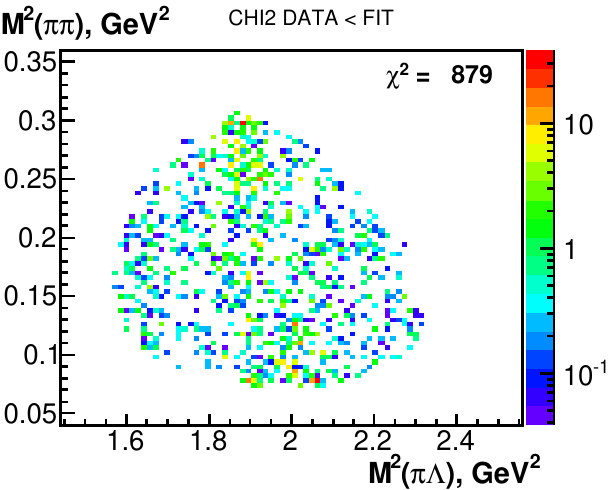}\\
\includegraphics[width=0.32\textwidth,height=0.25\textwidth]{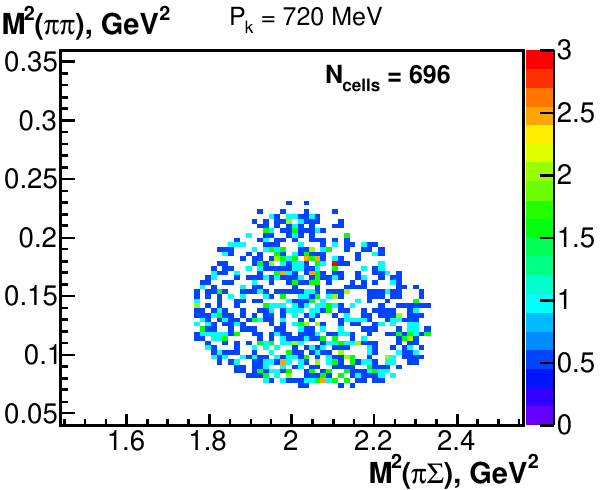}&
\hspace{-2mm}\includegraphics[width=0.32\textwidth,height=0.25\textwidth]{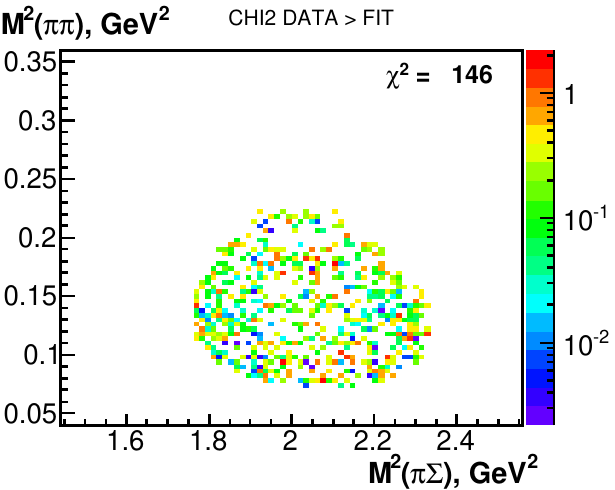}&
\hspace{-2mm}\includegraphics[width=0.32\textwidth,height=0.25\textwidth]{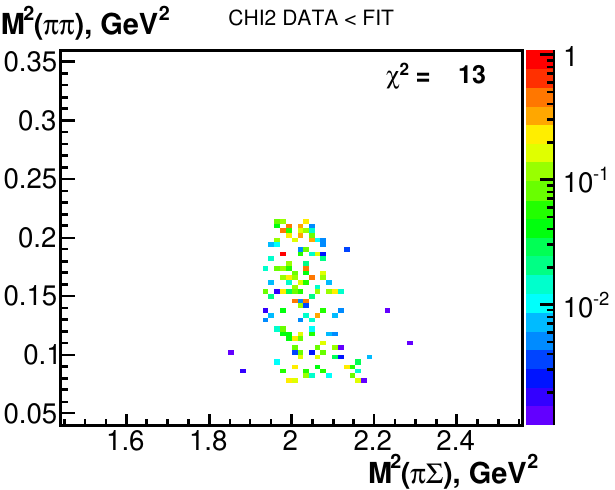}\\
\end{tabular}
 \vspace{-3mm}
\end{center}
\caption{\label{diff} The $\pi\pi\Lambda$ (top) and $\pi\pi\Sigma$ (bottom) Dalitz plots for the reactions
$K^-p\to \pi^0\pi^0\Lambda$ (2114 events) and $K^-p\to\pi^0\pi^0\Sigma$ (696 events) at 720\,MeV kaon
momentum~\cite{Prakhov:2004an}. 
Left subfigures: reconstructed data without 
acceptance correction, center/right: $\chi^2$-distributions for the case where the 
data exceeds the fit (center row) and where the fit exceeds the data (lower row). 
}
\end{figure*}
\begin{figure}[pt]
\includegraphics[width=0.48\textwidth,height=0.43\textheight]{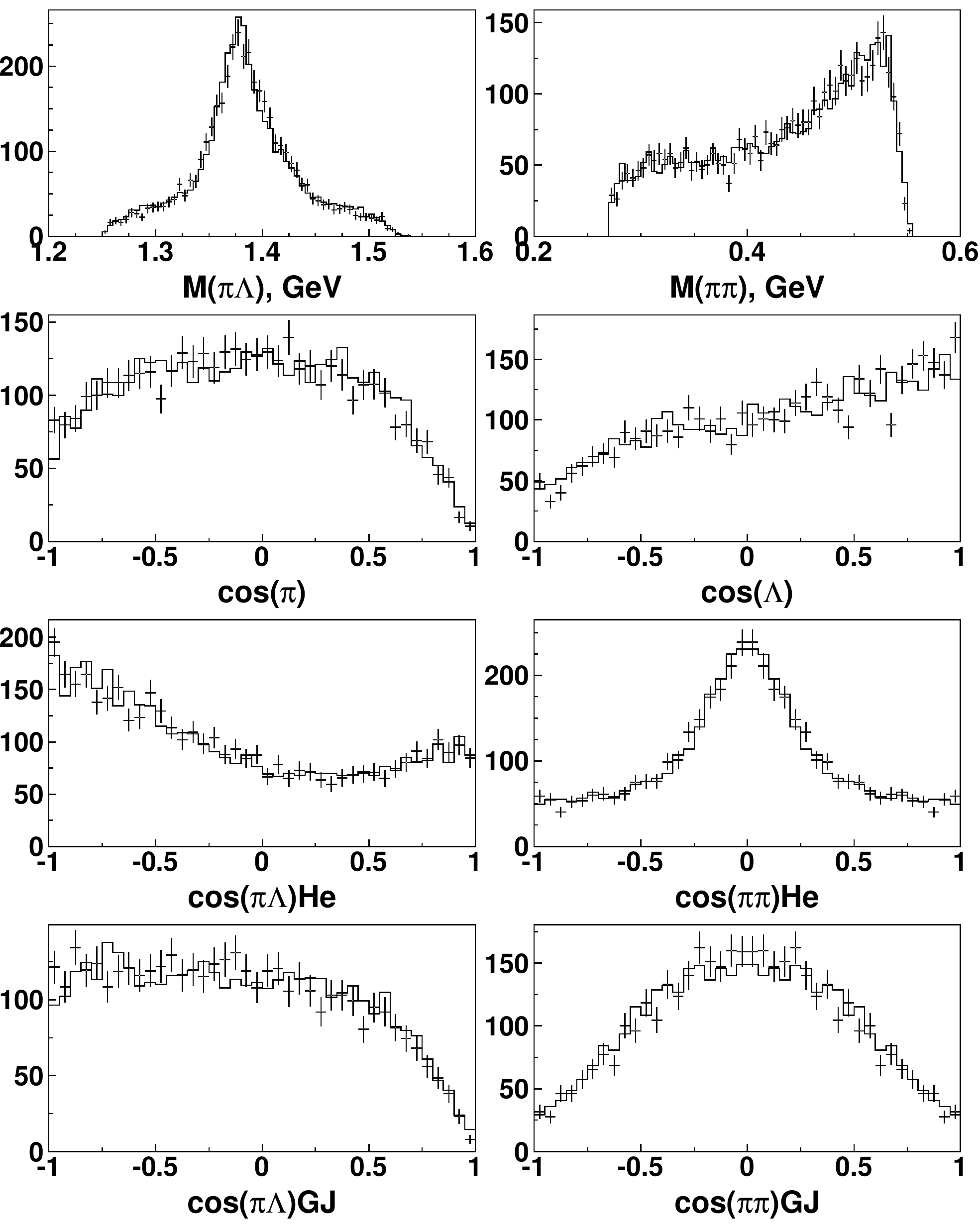}\vspace{1mm}
\caption{\label{BNL1}The $\Lambda\pi$ invariant mass and some angular distributions for the reactions
$K^-p\to \pi^0\pi^0\Lambda$ and $K^-p\to\pi^0\pi^0\Sigma$ at 720\,MeV kaon
momentum~\cite{Prakhov:2004an}. The fit is represented by the solid line.
The third row shows $\cos{\theta}$ in the helicity frame, the last row in the Gottfried-Jackson frame. }
\end{figure}
\begin{figure}[pt]
\includegraphics[width=0.48\textwidth,height=0.43\textheight]{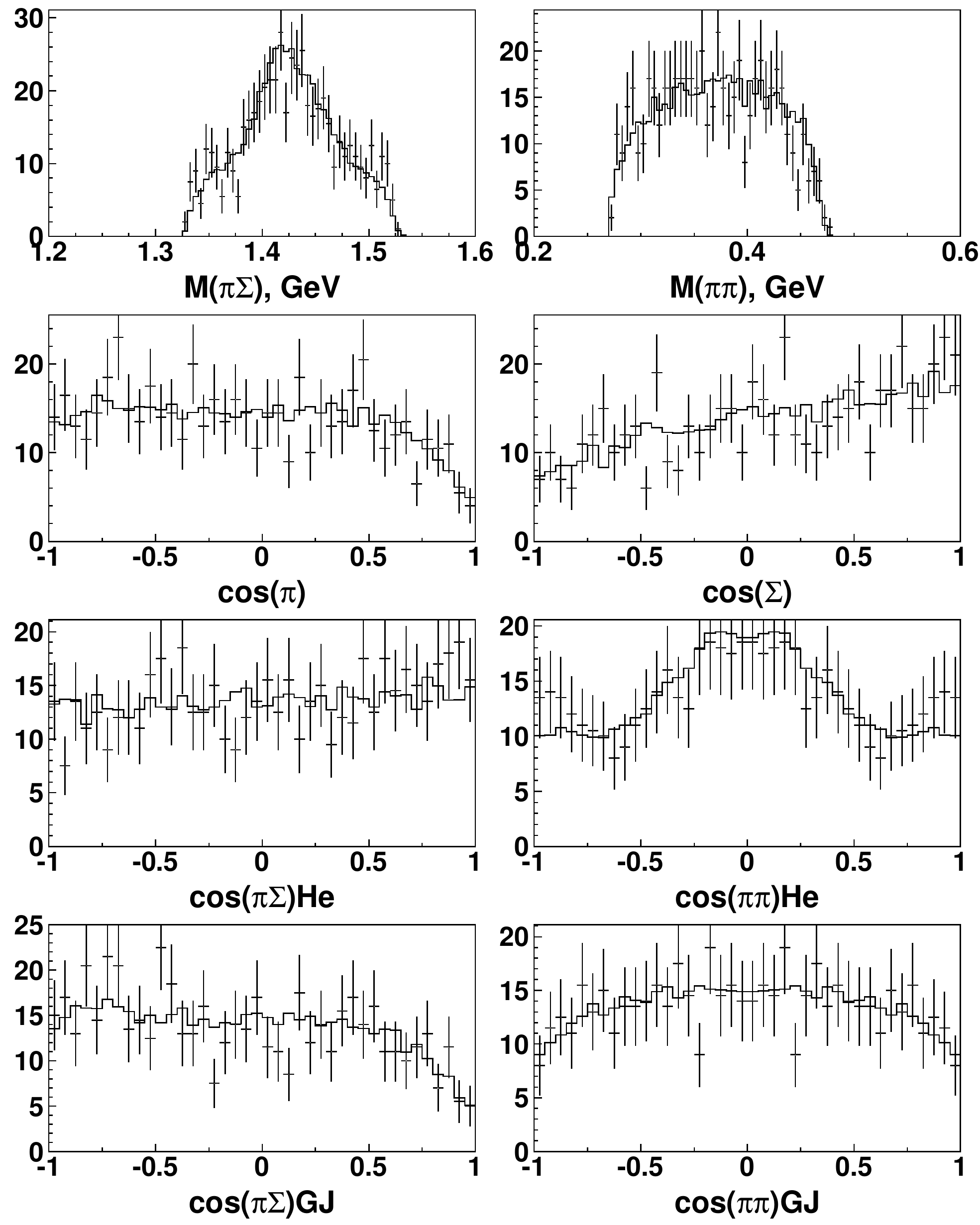}\vspace{1mm}
\caption{\label{BNL2}The $\Sigma\pi$ invariant mass and some angular distributions for the reactions
$K^-p\to \pi^0\pi^0\Lambda$ and $K^-p\to\pi^0\pi^0\Sigma$ at 720\,MeV kaon
momentum~\cite{Prakhov:2004an}. The fit is represented by the solid line. 
The third row shows $\cos{\theta}$ in the helicity frame, the last row in the Gottfried-Jackson frame.} 
\end{figure}
The reactions $K^-p\to \pi^0\pi^0\Lambda$~\cite{Prakhov:2004ri} and
$K^-p\to\pi^0\pi^0\Sigma$~\cite{Prakhov:2004an} were studied at BNL at eight incident $K^-$ momenta
between 514 and 750\,MeV/c using the Crystal Ball multiphoton spectrometer. Figure~\ref{diff} shows the Dalitz
plots for the two reactions. The data were made
available to us on an event-by-event basis. This allowed us to include the data in an event-based
likelihood fit which takes into account all correlations between the kinematical variables describing
the reaction. The $\chi^2$ differences between data and fit for cells in which the data exceed the fit 
or the fit exceeds the data are shown in separate Dalitz plots. No unexplained structures can be seen.
The $\Lambda\pi$ and $\Sigma\pi$ invariant mass distributions
for the highest incident Kaon momentum are shown in Figs.~\ref{BNL1} and \ref{BNL2}. 
The former reaction is dominated by formation of the
$\Sigma(1385)$ resonance while the latter one has a large $\sigma(500)\Sigma$ contribution
where the $\sigma(500)$ stands for the $\pi\pi$ $S$-wave interactions; in
addition, some $\Lambda(1405)$ can be seen.

\subsection{$K^-p\to$ quasi-two-body final states}
The data on quasi-two-body final states were taken  in the 60ties and 70ties of last century in
bubble chambers at CERN and Rutherford. In $K^-p$ scattering, $\Lambda$ and $\Sigma$
resonances can be formed. If they have a large mass, they may have a large number of different
decay modes.

Bubble chamber events are classified according to their topology. A fraction of the events with
two tracks emerging from the interaction point can be assigned to
\be
\nonumber
K^-p\to K^-p \pi^0
\ee
by a measurement of the bubble density (for particle identification) and using kinematic
constraints to construct the missing $\pi^0$. In the invariant mass distribution of the final-state $K^-p$ pair, $\Lambda(1520)$ with spin-parity $J^P=3/2^-$
and $\Lambda(1820)$ with $J^P=5/2^+$ were observed. Studying the $p\pi^0$ and $K^-\pi^0$ mass
distributions, $K^-\Delta^+(1232)$ or $K^{*-}\,p$ can be seen.

In events with a topology with a primary interaction point, from that two
tracks emerge, and a secondary vertex with two tracks, the reaction
\be
\nonumber
K^-p\to K^-p \pi^+\pi^-
\ee
can be identified. The two secondary particles may form a $K^0_s$ -- these events can be discarded
-- or may stem from a excited $\Lambda$ which decayed into $K^-p$. The $\Lambda\pi^\pm$ invariant mass
peaks at the $\Sigma^\pm$. In this way, the reaction $K^-p\to \pi^\pm\Sigma^\mp$ can be studied as
well. The competing reactions $K^-p\to \bar K^0p\pi^-$, $\bar K^0\to \pi^+\pi^-$ and $K^-p\to
\Sigma^0\pi^+\pi^-$, $\Sigma^0\to\Lambda\gamma$ can be separated safely.

Summarizing, the following reactions were studied:
\begin{subequations}
\label{reactions}
\begin{align}
\label{a}
K^-p\to\ & \pi^0\Lambda(1520),\\
\label{b}
K^-p\to\ & K^{-}\Delta^+(1232),\\
\label{c}
K^-p\to\ & \pi^\mp\Sigma^\pm(1385),\\
\label{d}
K^-p\to\ & \bar K^{*}\,N. \\  
\label{e}\hspace{-18mm}
{\rm The\ reaction}\qquad K^-p\to\ & \omega\Lambda
\end{align}
\end{subequations}
is extracted from events with four charged tracks in
the final state where two tracks from a secondary vertex form a $\Lambda$. The missing $\pi^0$ is
identified in a kinematical fit to the $K^-p\to \pi^+\pi^-\pi^0\Lambda$ hypothesis. The three-pion
invariant mass shows a very clear $\omega$ meson above a small background. The number of $\omega$ 
mesons is determined for each data bin. 

The data for reactions~(\ref{a}-\ref{e}) stem from
Refs.~\cite{Cameron:1977jr,Litchfield:1973ap,Litchfield:1973ey,Cameron:1978en,Cameron:1978qi,Brandstetter:1972xp,Nakkasyan:1975yz,Baccari:1976ik}.
The data on $K^-p\to \pi^0\Lambda(1520)$, $\pi\Sigma(1385)$, and $\bar K\Delta(1232)$ cover the 
mass range from the respective threshold to 2170\,MeV. 
Data on $K^-p\to\  K^{*-}p$ are given up to 1955 MeV. In some cases, the papers present 
an inclusive analysis of all data available at that time.

The intermediate resonances in the reactions~(\ref{reactions}) carry spin alignment which reflects
itself in the spin density matrix elements. In the case of an unpolarized target, three density
matrix elements can be measured from the decay angular distributions. The probability distribution
for these reaction is given by
\be\nonumber
W(\cos \Theta^*, \theta, \phi, s) = \frac{3}{4\pi\sigma} A(s) \frac{d\sigma}{d\Omega} \times\\
\nonumber
\left\{\frac12 (\frac13 + \cos^2\theta) + 2(\frac13 - \cos^2\theta)\rho_{\frac32 \frac32}(\cos\Theta^*)\right. \\
\nonumber
-\ 2\sqrt{\frac13} \sin2\theta\cos\phi\ \Re e\rho_{\frac32 \frac12}(\cos\Theta^*) \\
\nonumber \left.-\ 2\sqrt{\frac13} \sin^2\theta\cos2\phi\ \Re e\rho_{\frac32-\frac12}
(\cos\Theta^*) \right\}\,.
\ee
In this equation, $\Theta^*$ is the production angle in the c.m.s. of the intermediate resonance,
$\theta$ and $\phi$ are the decay angles in the helicity frame of the intermediate resonance,
$s=M^2$ the squared invariant mass of the two resonating particles and $A(s)$ the dynamical
amplitude described in Ref.~\cite{Matveev:2019igl}. The expression in curly brackets represents the
decay angular distribution of the intermediate  resonance in terms of density matrix elements
$\rho_{\frac32 \frac32}$, $\Re e\rho_{\frac32 \frac12}$, and $\Re e\rho_{\frac32 \frac12}$. For
reaction $K^-p \to \bar K\,\Delta(1232)$, $\rho_{\frac32 \frac32}$ was substituted by $\rho_{\frac12
\frac12}$  = $\frac12 - \rho_{\frac32 \frac32}$, a substitution which reduces the correlation
between the parameters. Note that the probability $W$ to find, e.g., a $\Lambda(1520)$ in a particular
bin was determined from the fit to the $K^-p$ invariant mass distribution. Interference with 
other amplitudes like the $K^{*-}_0(700)$ -- the $(K^-\pi^0)_{S\rm-wave}$ -- was neglected.
The same method was applied to extract all the reactions~\ref{reactions}. 

\begin{figure*}[ph]
\begin{tabular}{cc}
\hspace{-1mm}\includegraphics[width=0.48\textwidth]{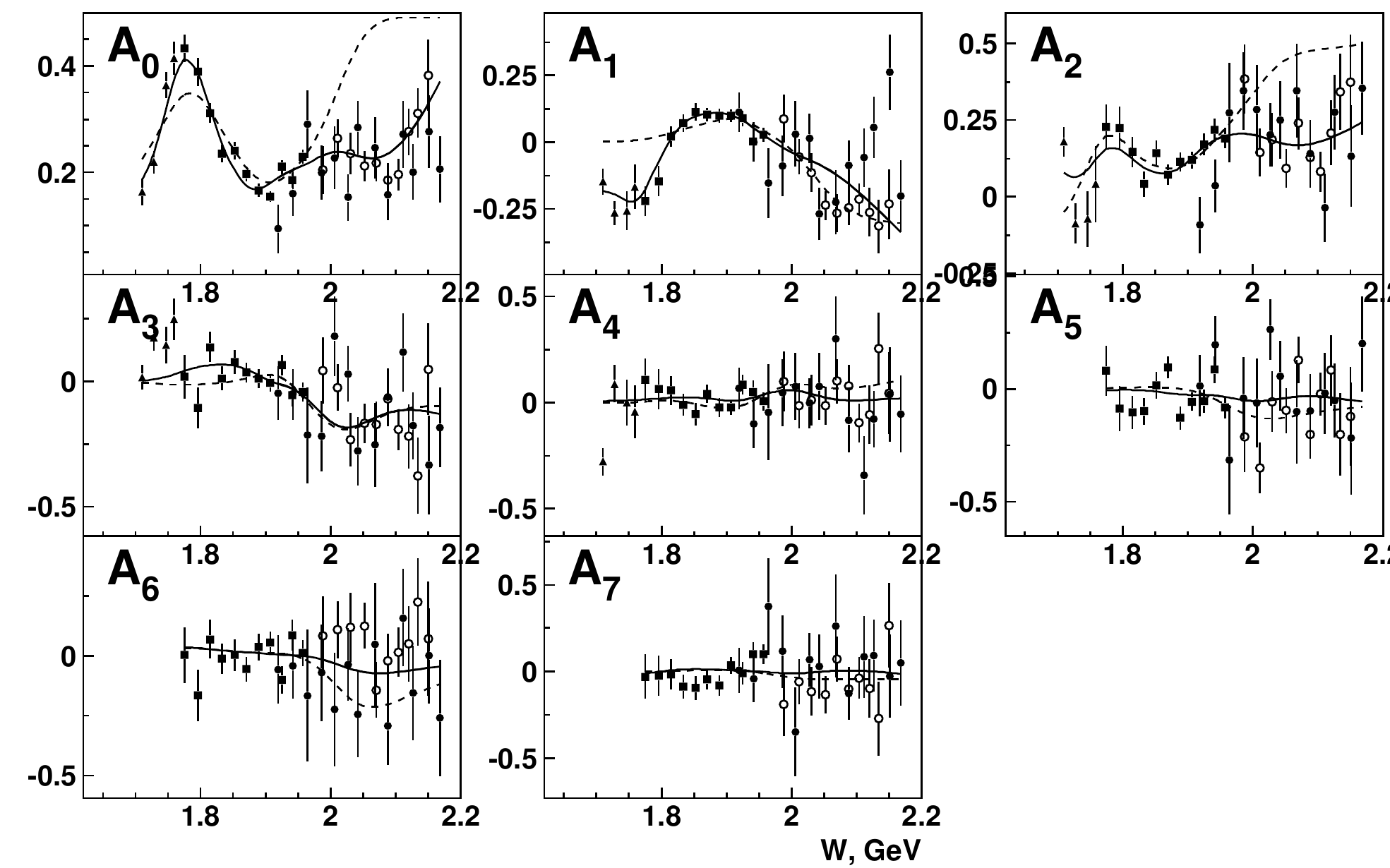}&
\hspace{-1mm}\includegraphics[width=0.48\textwidth]{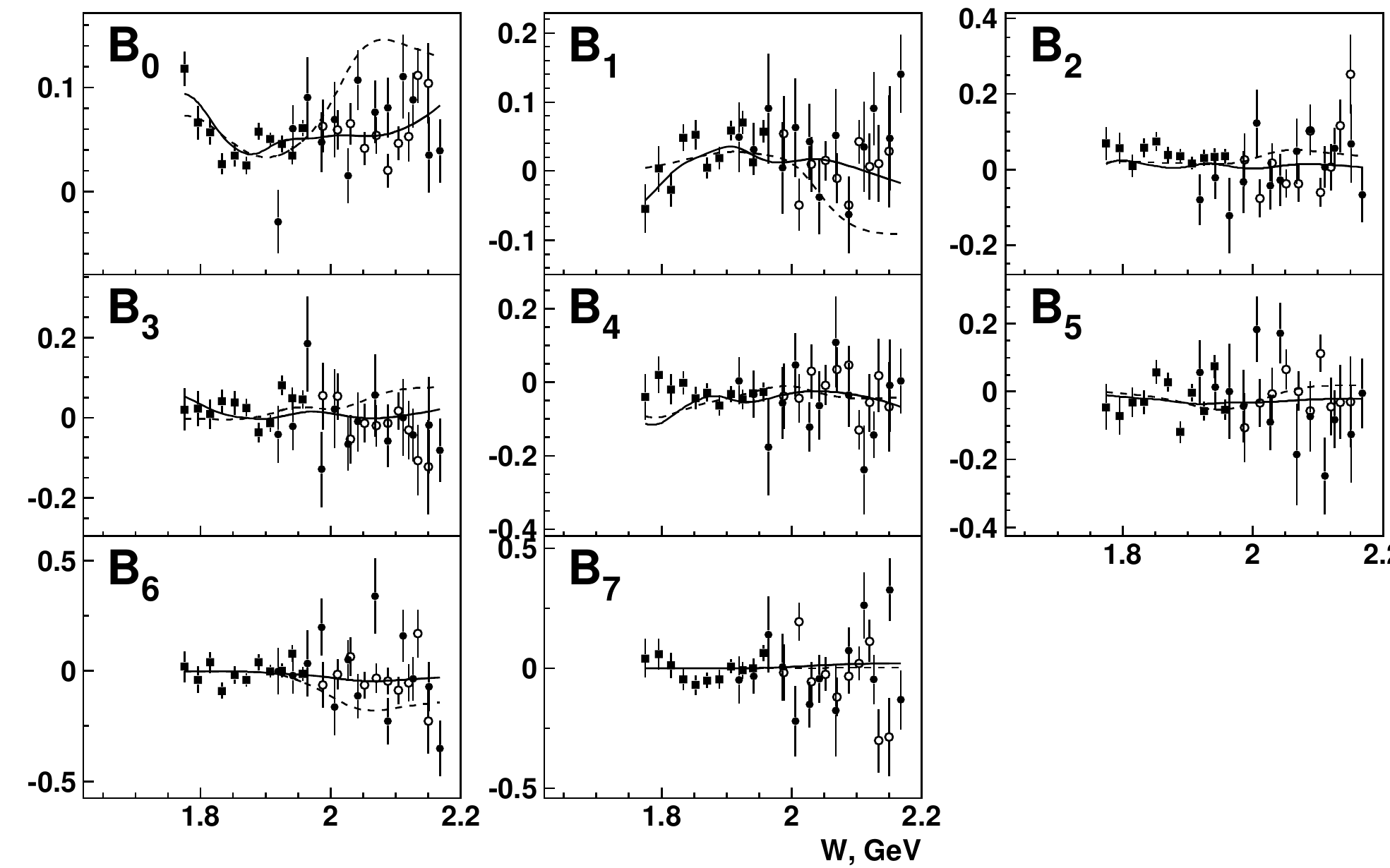}\\
\hspace{-1mm}\includegraphics[width=0.48\textwidth]{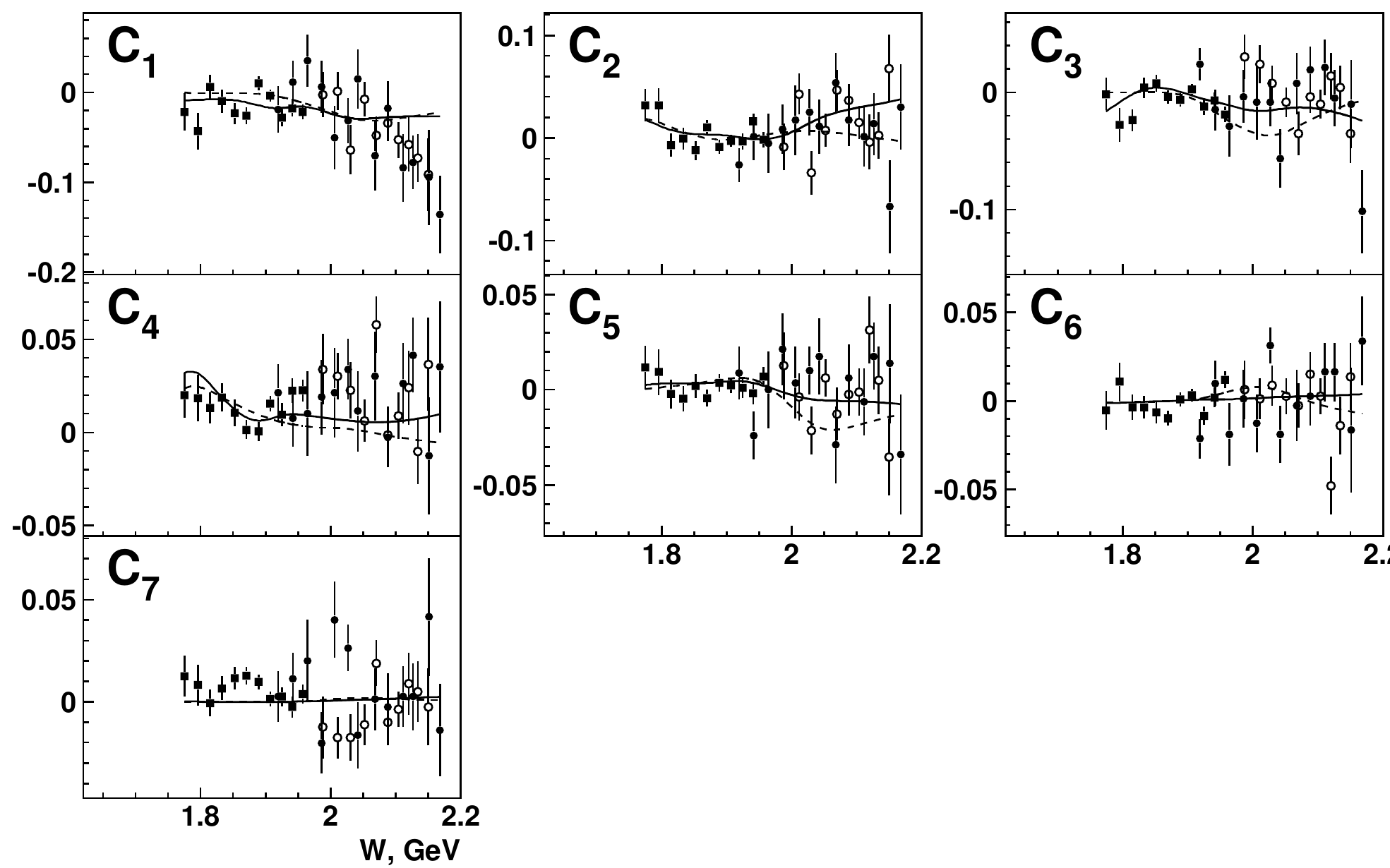}&
\hspace{-1mm}\includegraphics[width=0.48\textwidth]{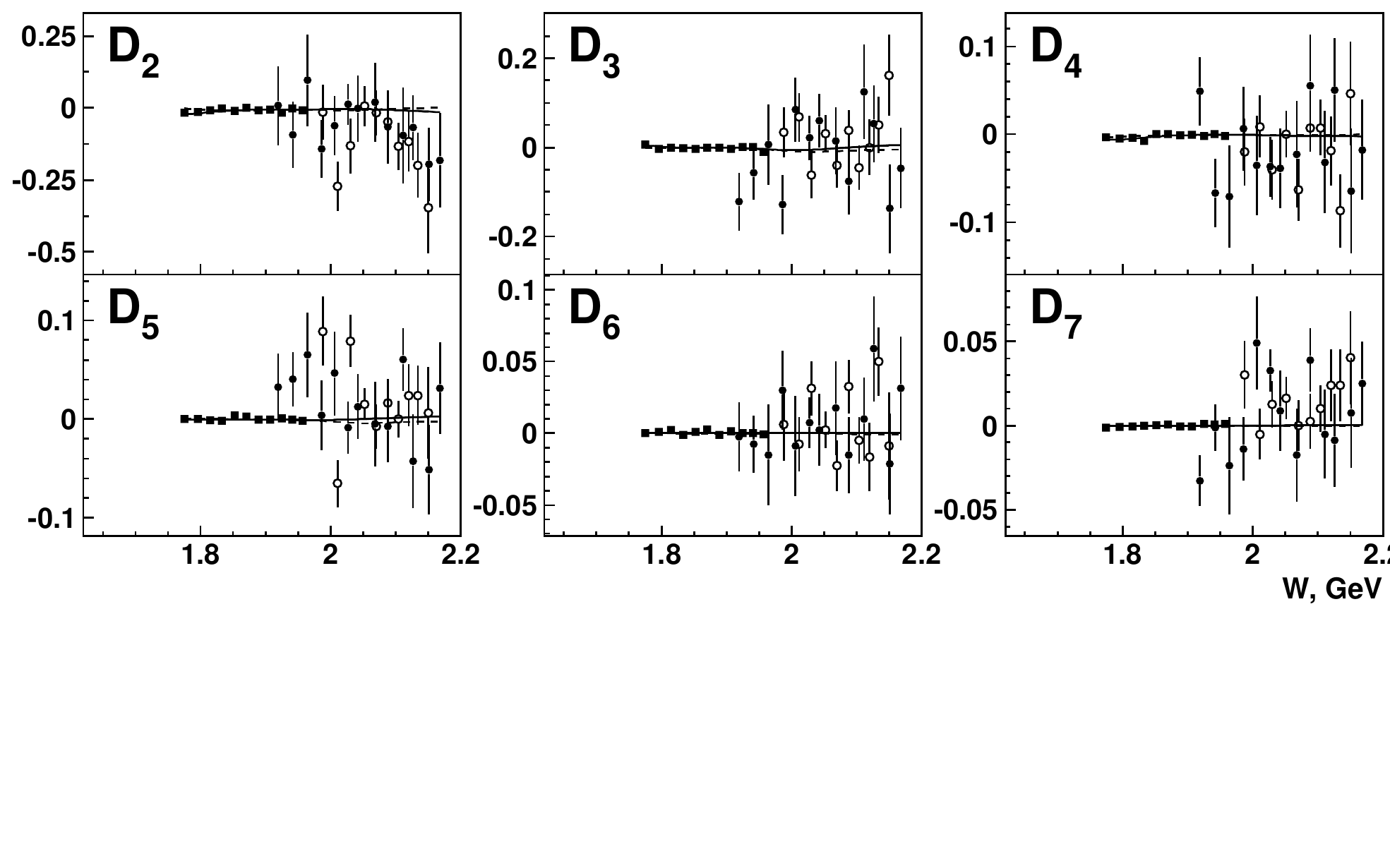}\\
\end{tabular}
\caption{\label{Lambda1520}$K^-p\to \Lambda(1520)\pi^0$~\cite{Cameron:1977jr,Litchfield:1973ap}: 
The associated Legendre coefficients for the differential cross sections and the density matrix elements 
$\rho_{\frac32 \frac32}$, $\Re e\rho_{\frac32 \frac12}$, and $\Re e\rho_{\frac32 \frac12}$. 
Data: $\bigcirc$~\cite{Cameron:1977jr}; {{\LARGE $\bullet$}}~\cite{Cameron:1977jr}; 
$\bigtriangleup$~\cite{Litchfield:1973ap}; $\blacksquare$~\cite{Litchfield:1973ap}.}
\begin{tabular}{cc}
\hspace{-1mm}\includegraphics[width=0.48\textwidth]{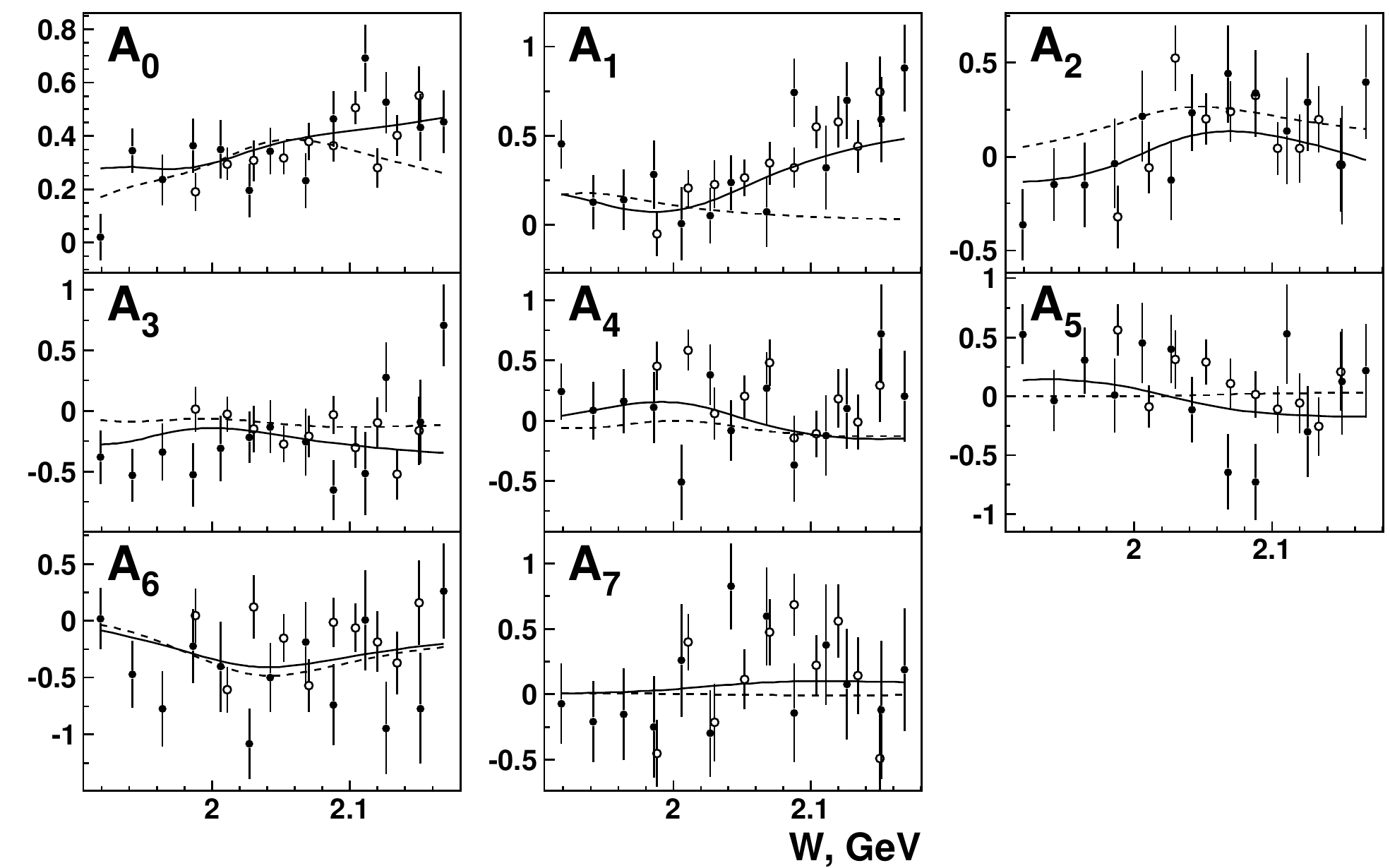}&
\hspace{-1mm}\includegraphics[width=0.48\textwidth]{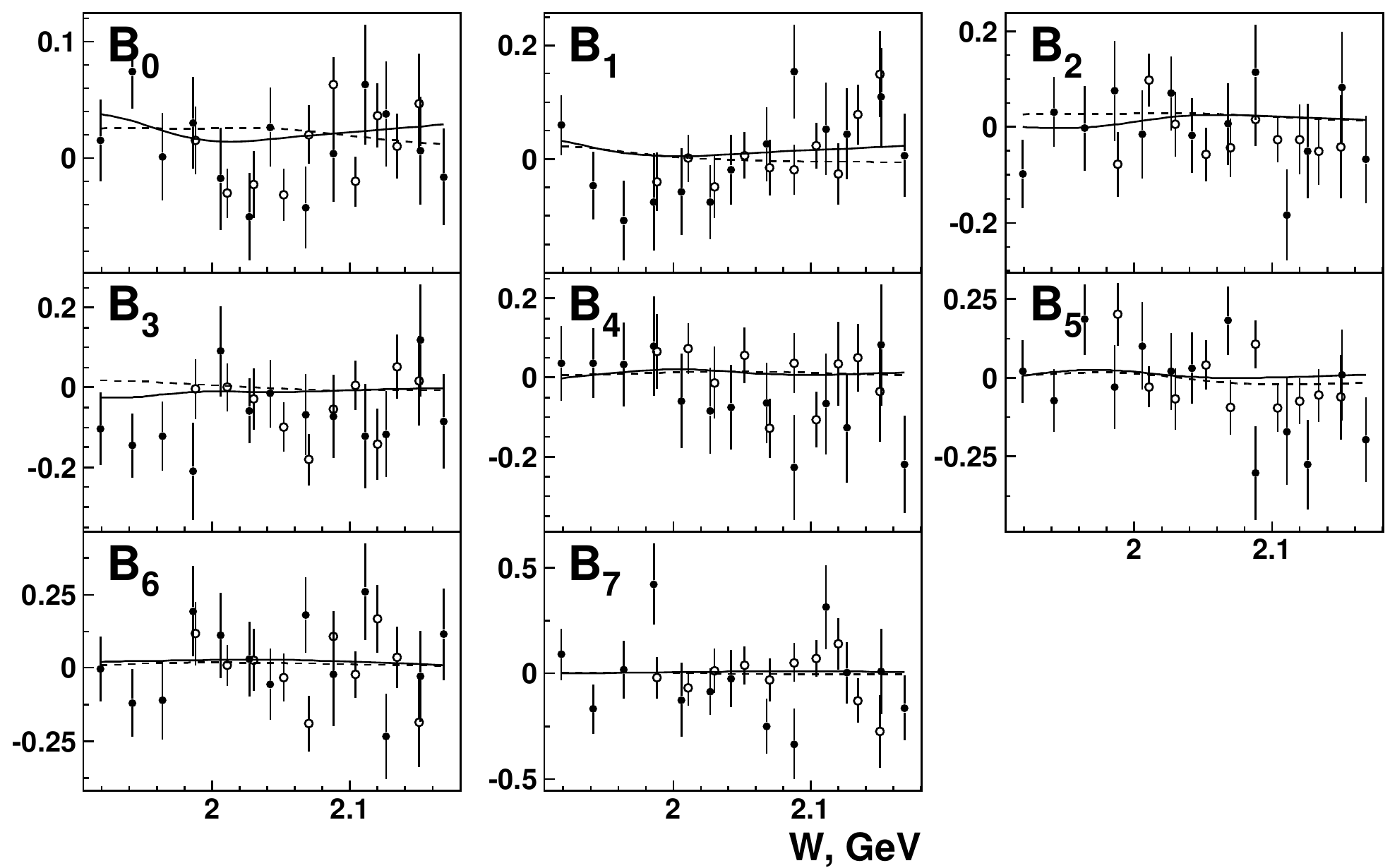}\\
\hspace{-1mm}\includegraphics[width=0.48\textwidth]{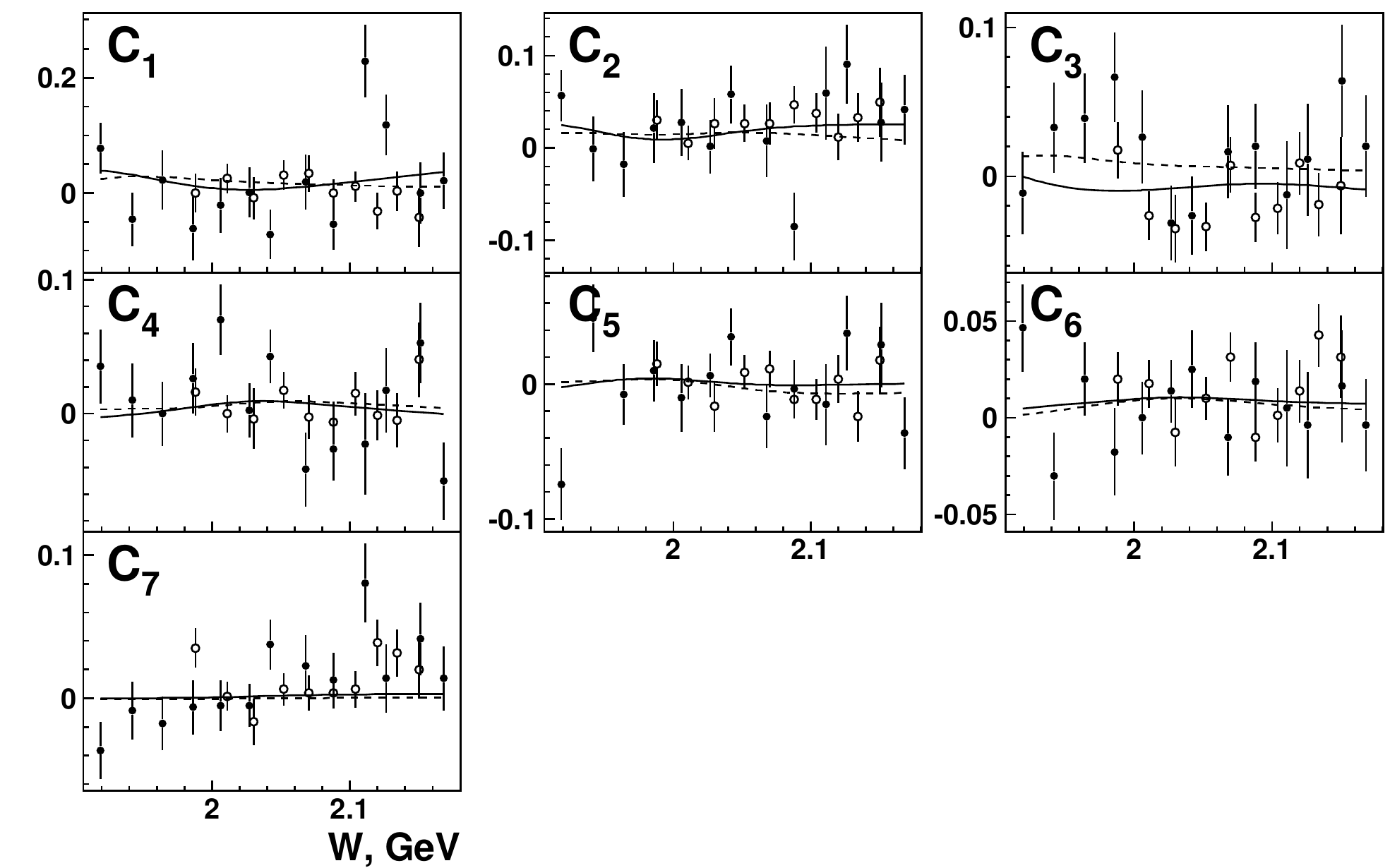}&
\hspace{-1mm}\includegraphics[width=0.48\textwidth]{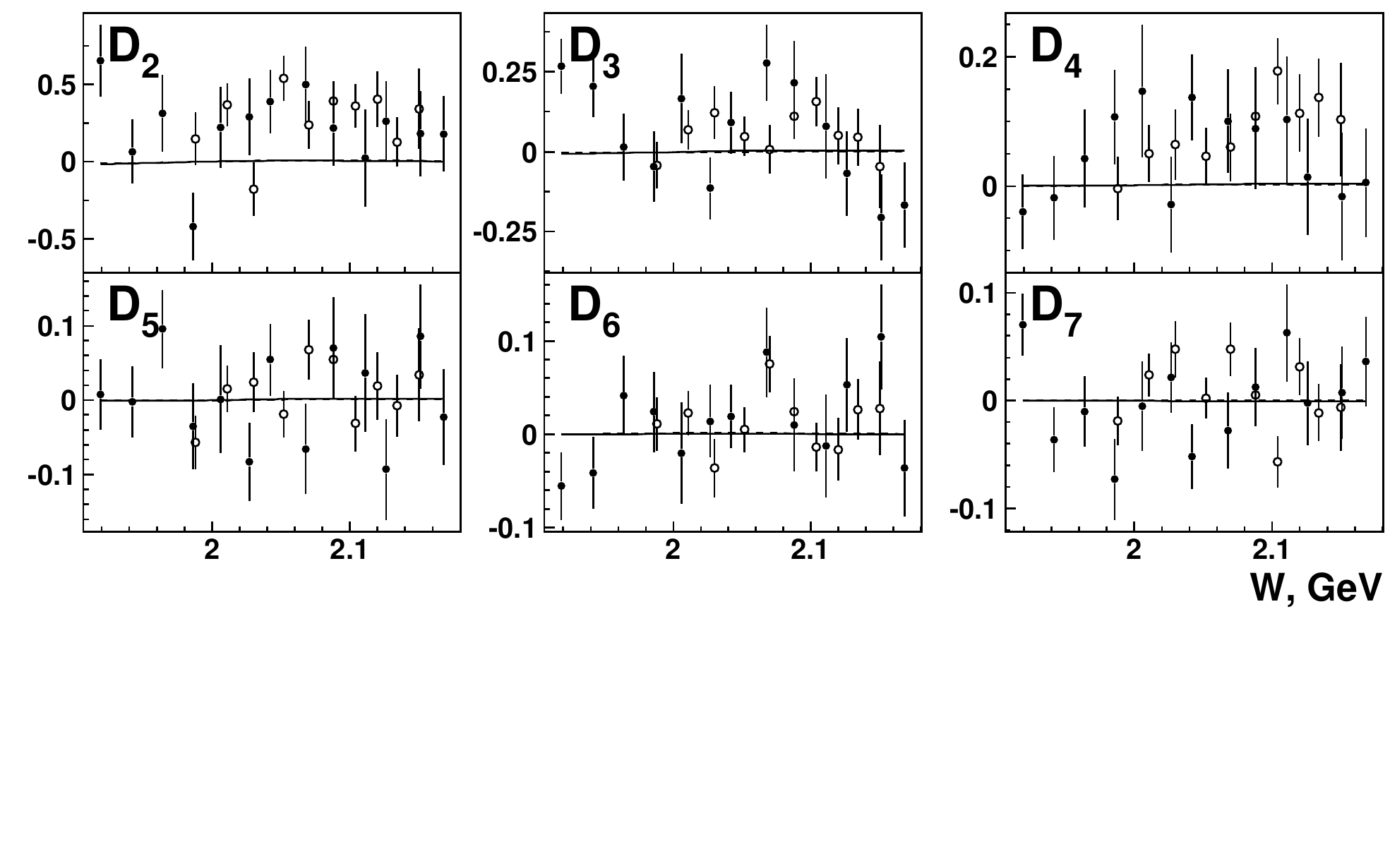}\\
\end{tabular}
\caption{\label{Delta1232}$K^-p\to \Delta^+(1232)K^-$~\cite{Litchfield:1973ey}: 
The associated Legendre coefficients for the differential cross sections and the density matrix elements 
$\rho_{\frac32 \frac32}$, $\Re e\rho_{\frac32 \frac12}$, and $\Re e\rho_{\frac32 \frac12}$. }
\end{figure*}

\begin{figure*}[ph]
\begin{tabular}{cc}
\hspace{-1mm}\includegraphics[width=0.48\textwidth]{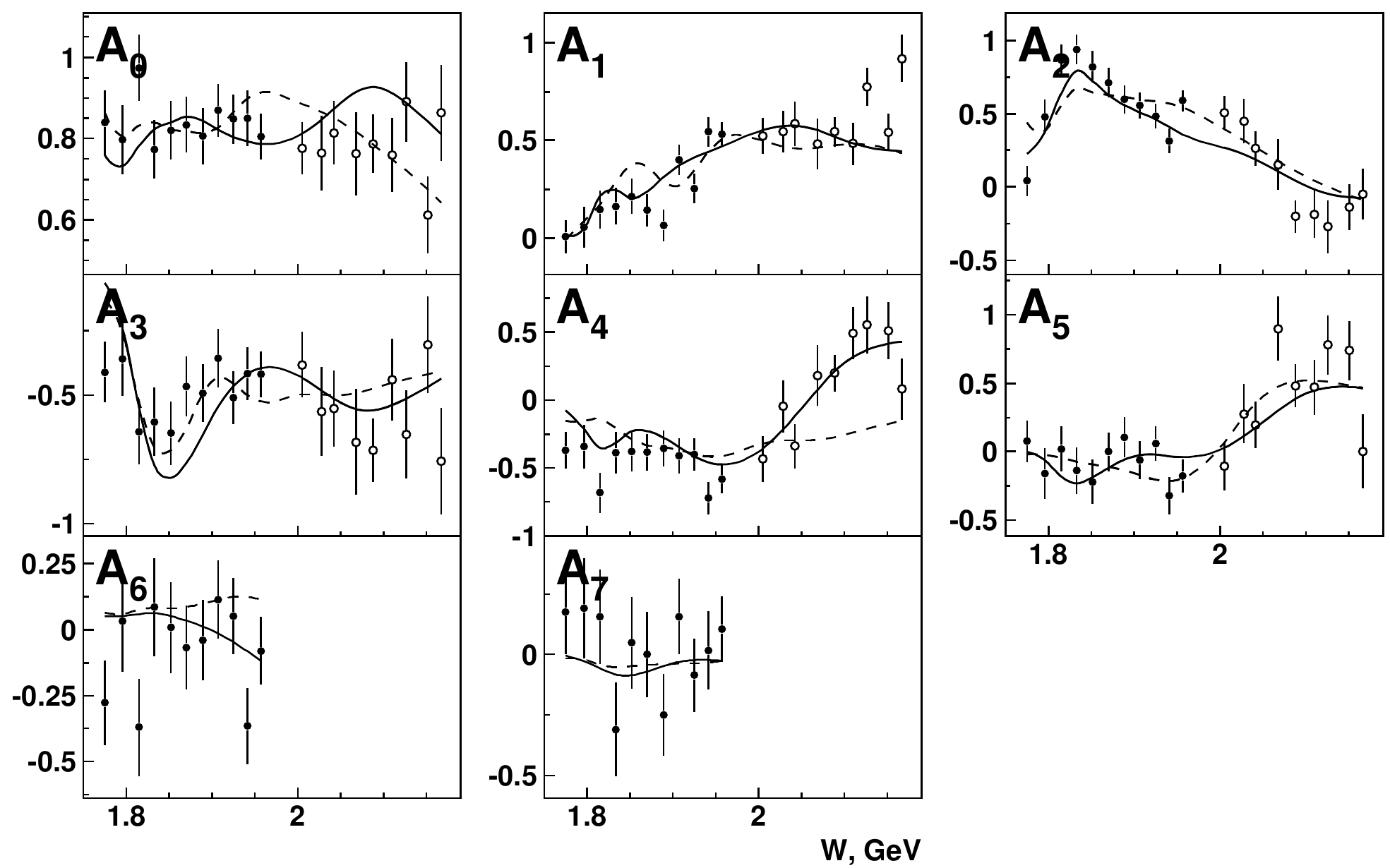}&
\hspace{-1mm}\includegraphics[width=0.48\textwidth]{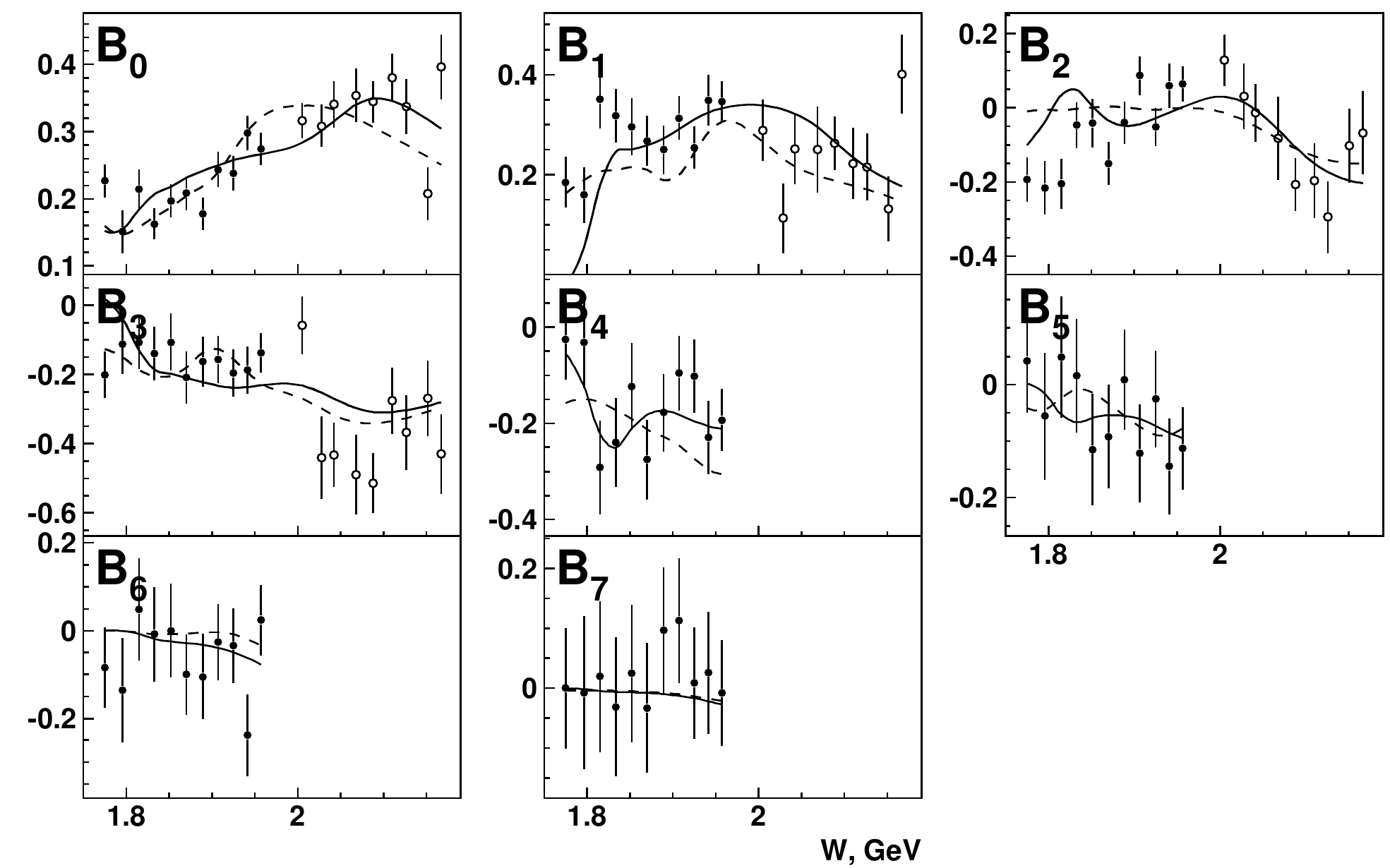}\\
\hspace{-1mm}\includegraphics[width=0.48\textwidth]{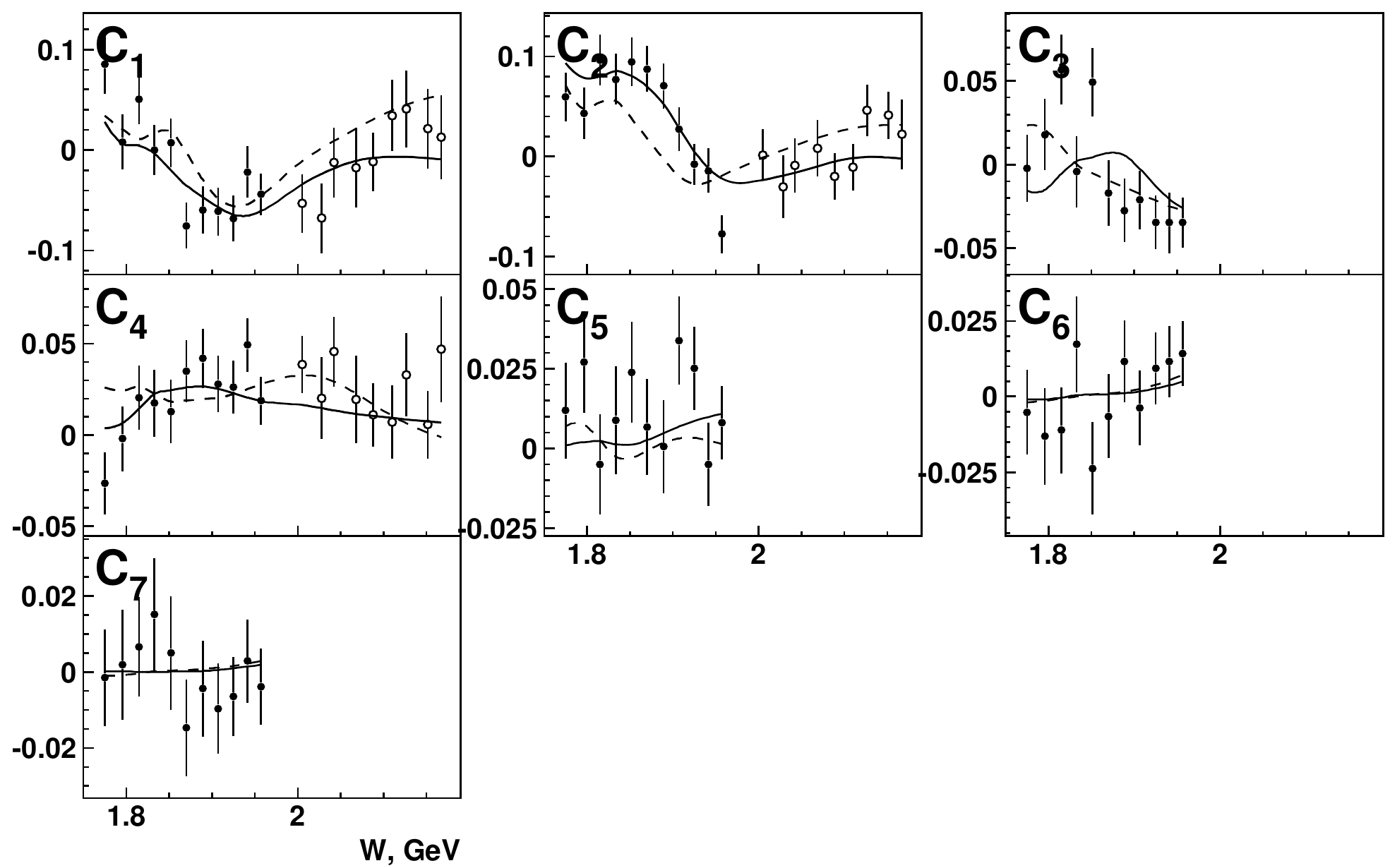}&
\hspace{-1mm}\includegraphics[width=0.48\textwidth]{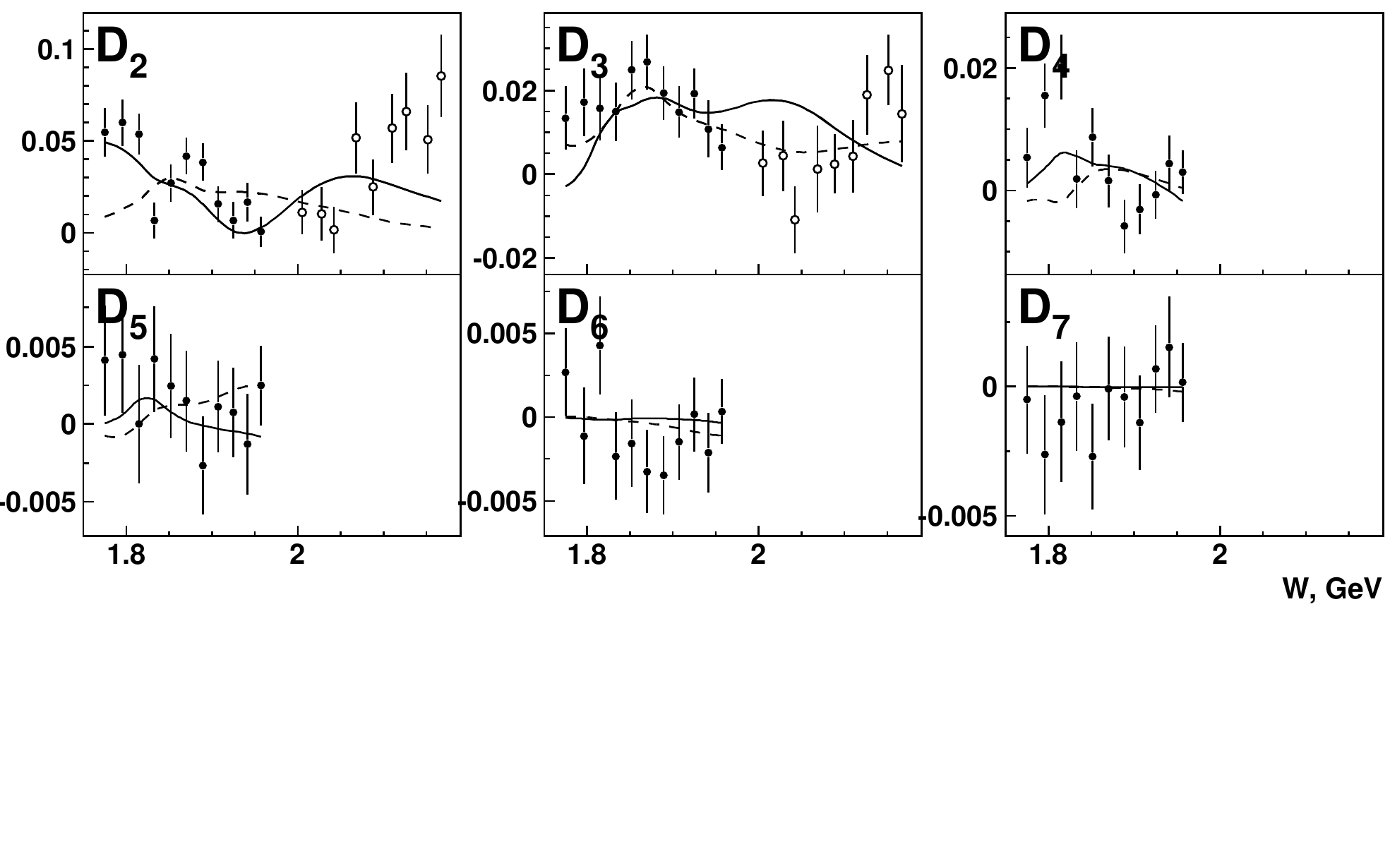}\\
\end{tabular}
\caption{\label{Sigmam}$K^-p\to \Sigma^{+}(1385)\pi^-$~\cite{Cameron:1978en}: 
The associated Legendre coefficients for the differential cross sections and the density matrix elements 
$\rho_{\frac32 \frac32}$, $\Re e\rho_{\frac32 \frac12}$, and $\Re e\rho_{\frac32 -\frac12}$. }
\begin{tabular}{cc}
\hspace{-1mm}\includegraphics[width=0.48\textwidth]{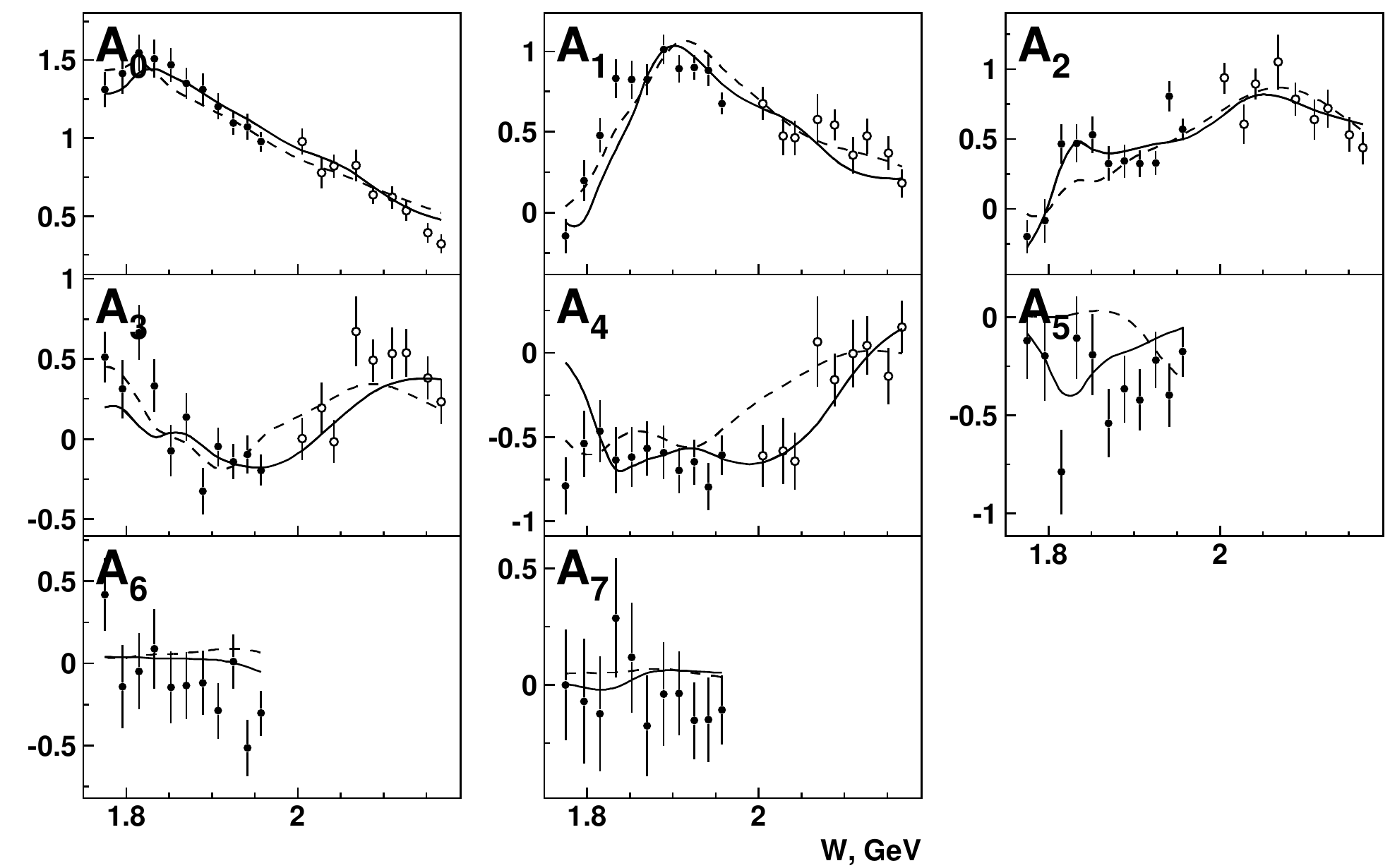}&
\hspace{-1mm}\includegraphics[width=0.48\textwidth]{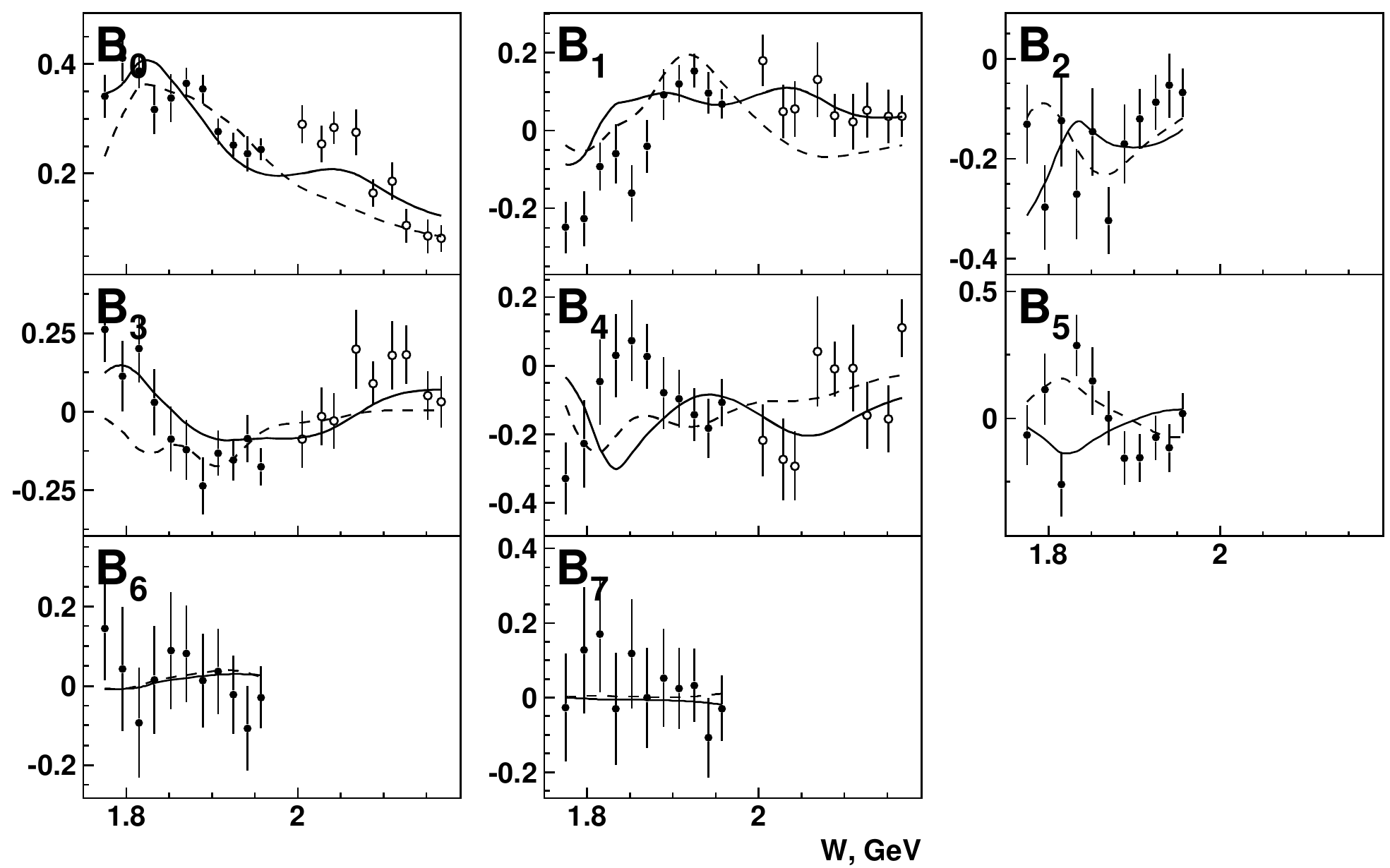}\\
\hspace{-1mm}\includegraphics[width=0.48\textwidth]{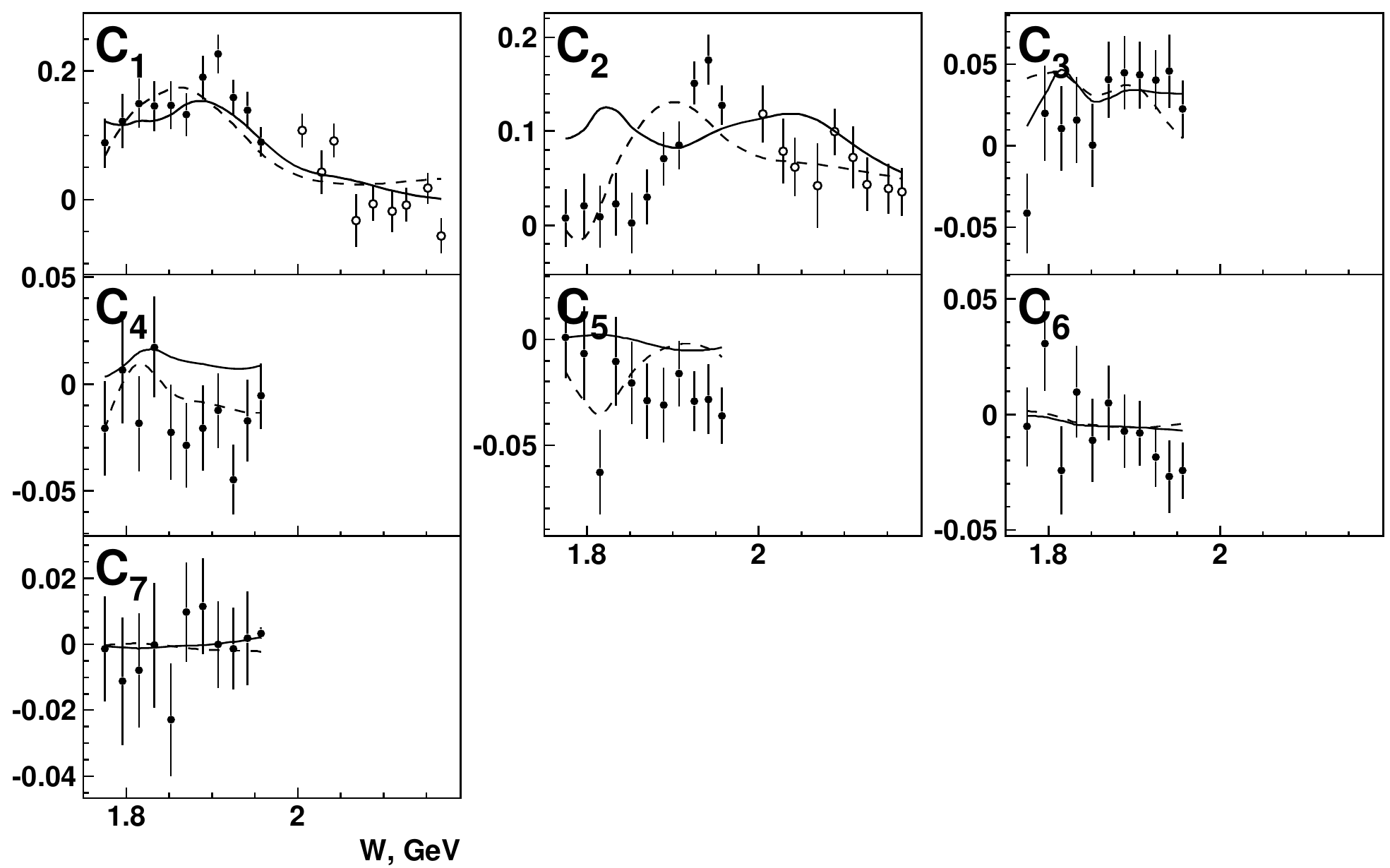}&
\hspace{-1mm}\includegraphics[width=0.48\textwidth]{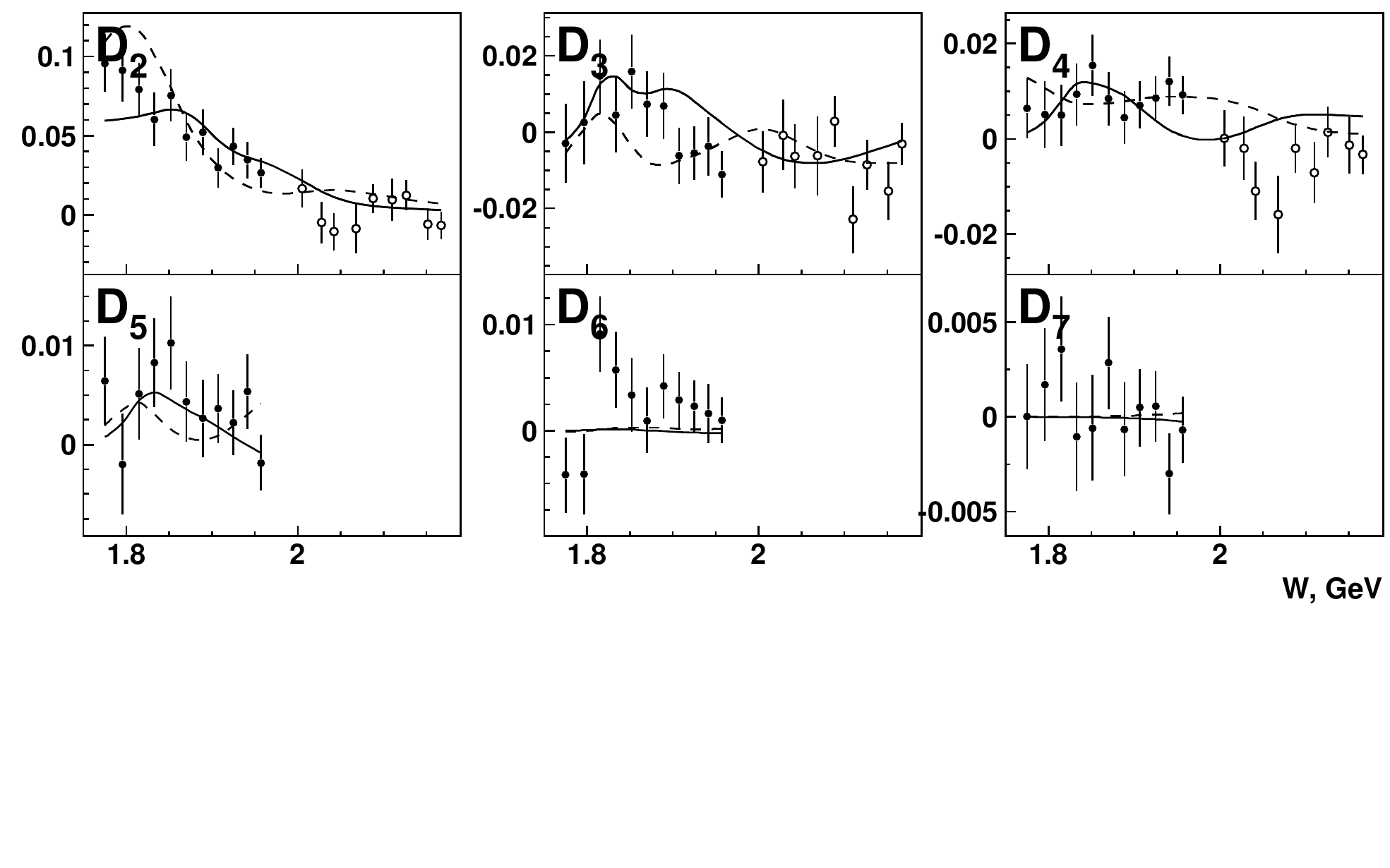}\\
\end{tabular}
\caption{\label{Sigmap}$K^-p\to \Sigma^{-}(1385)\pi^+$~\cite{Cameron:1978en}: 
The associated Legendre coefficients for the differential cross sections and the density matrix elements 
$\rho_{\frac32 \frac32}$, $\Re e\rho_{\frac32 \frac12}$, and $\Re e\rho_{\frac32 -\frac12}$. }
\end{figure*}
\begin{figure*}[ph]
\begin{tabular}{cc}
\hspace{-1mm}\includegraphics[width=0.48\textwidth]{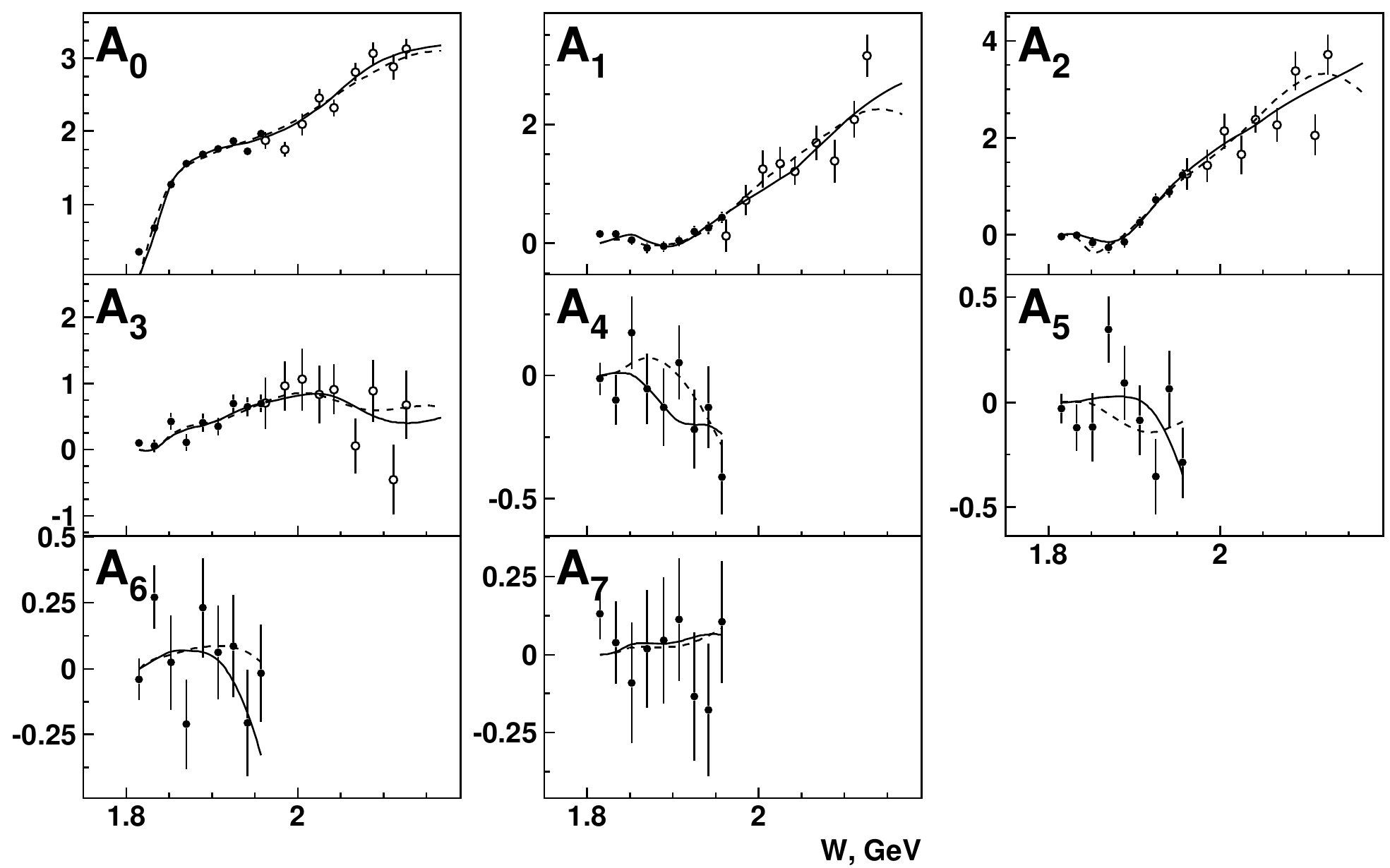}&
\hspace{-1mm}\includegraphics[width=0.48\textwidth]{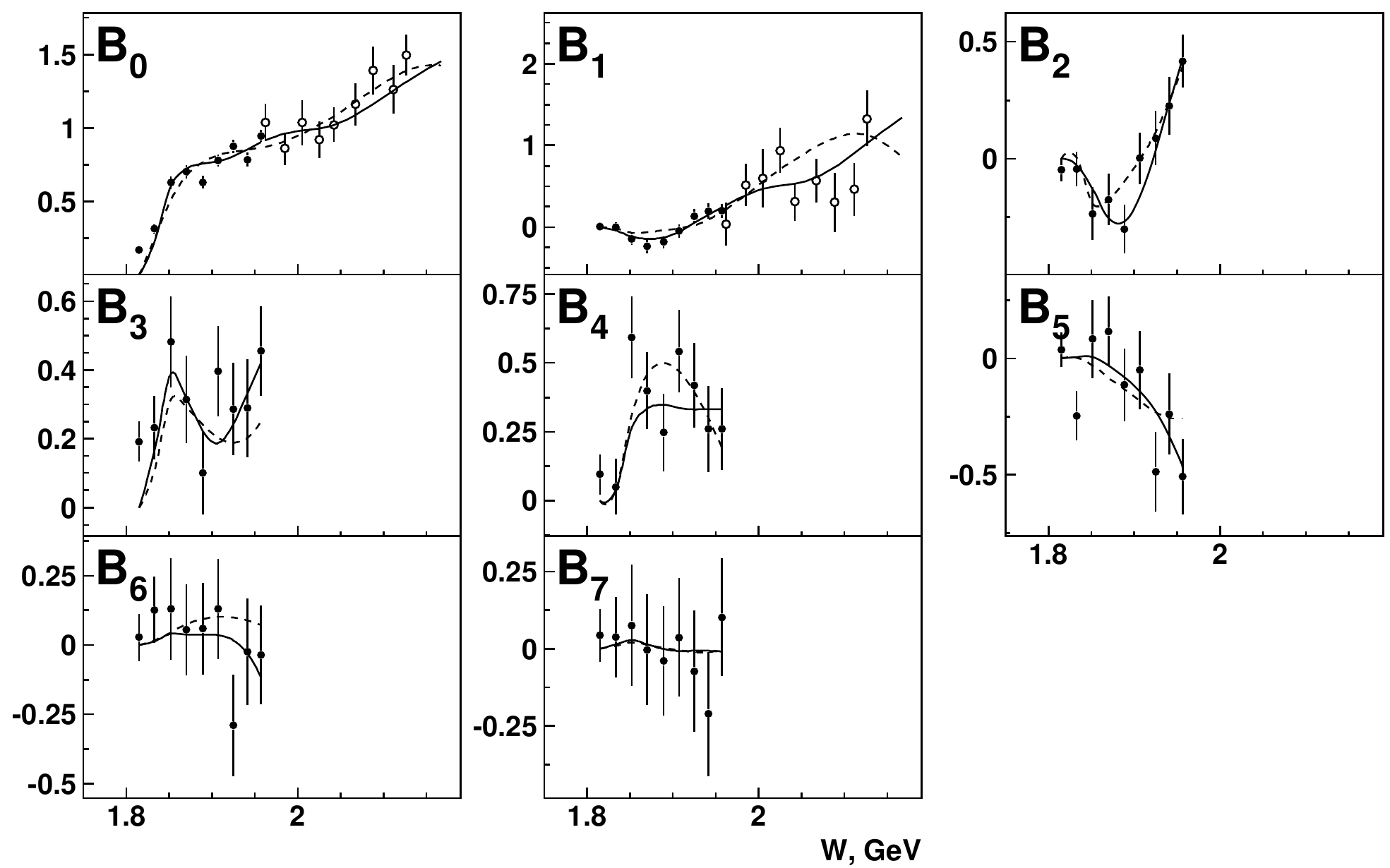}\\
\hspace{-1mm}\includegraphics[width=0.48\textwidth]{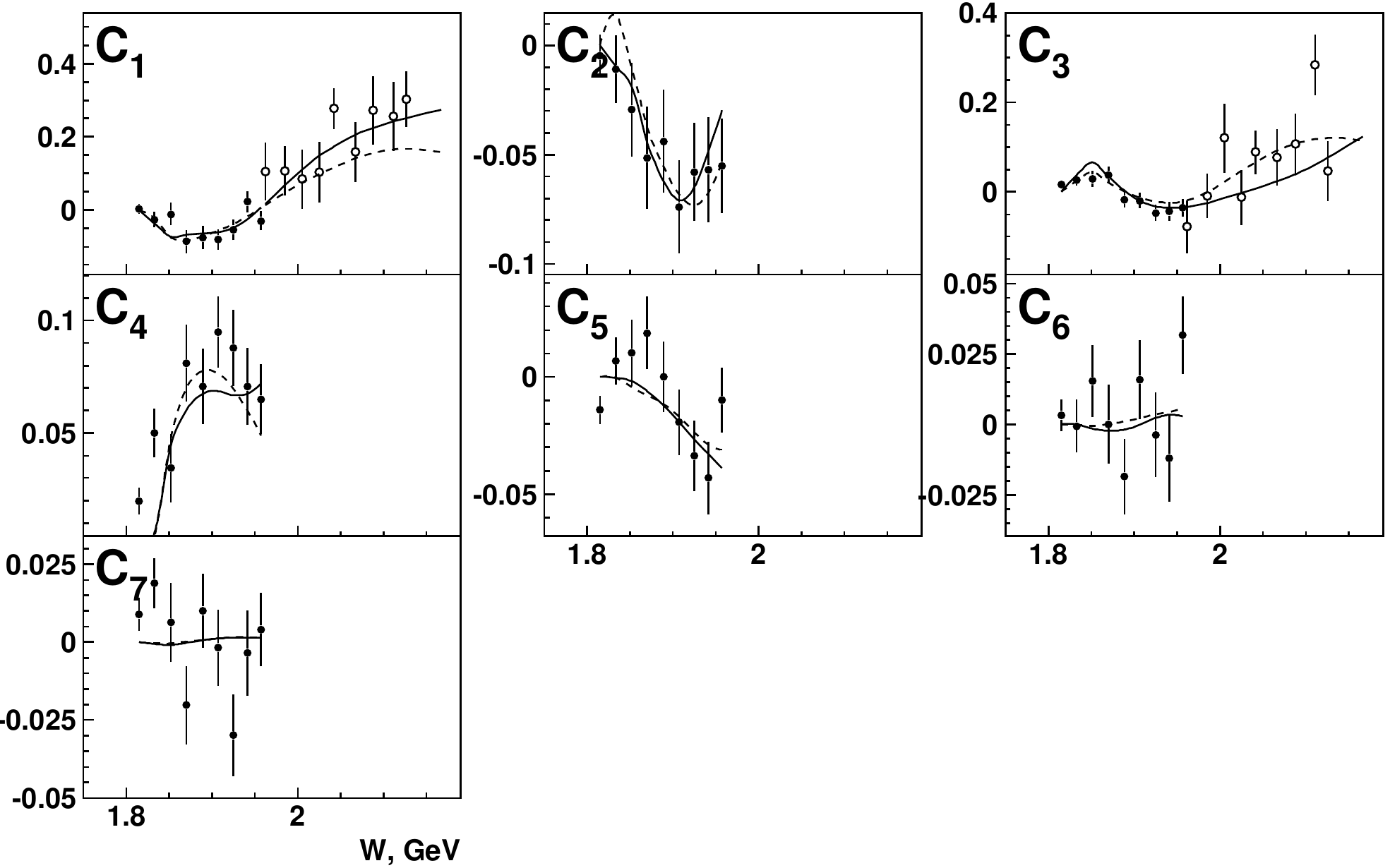}&
\hspace{-1mm}\includegraphics[width=0.48\textwidth]{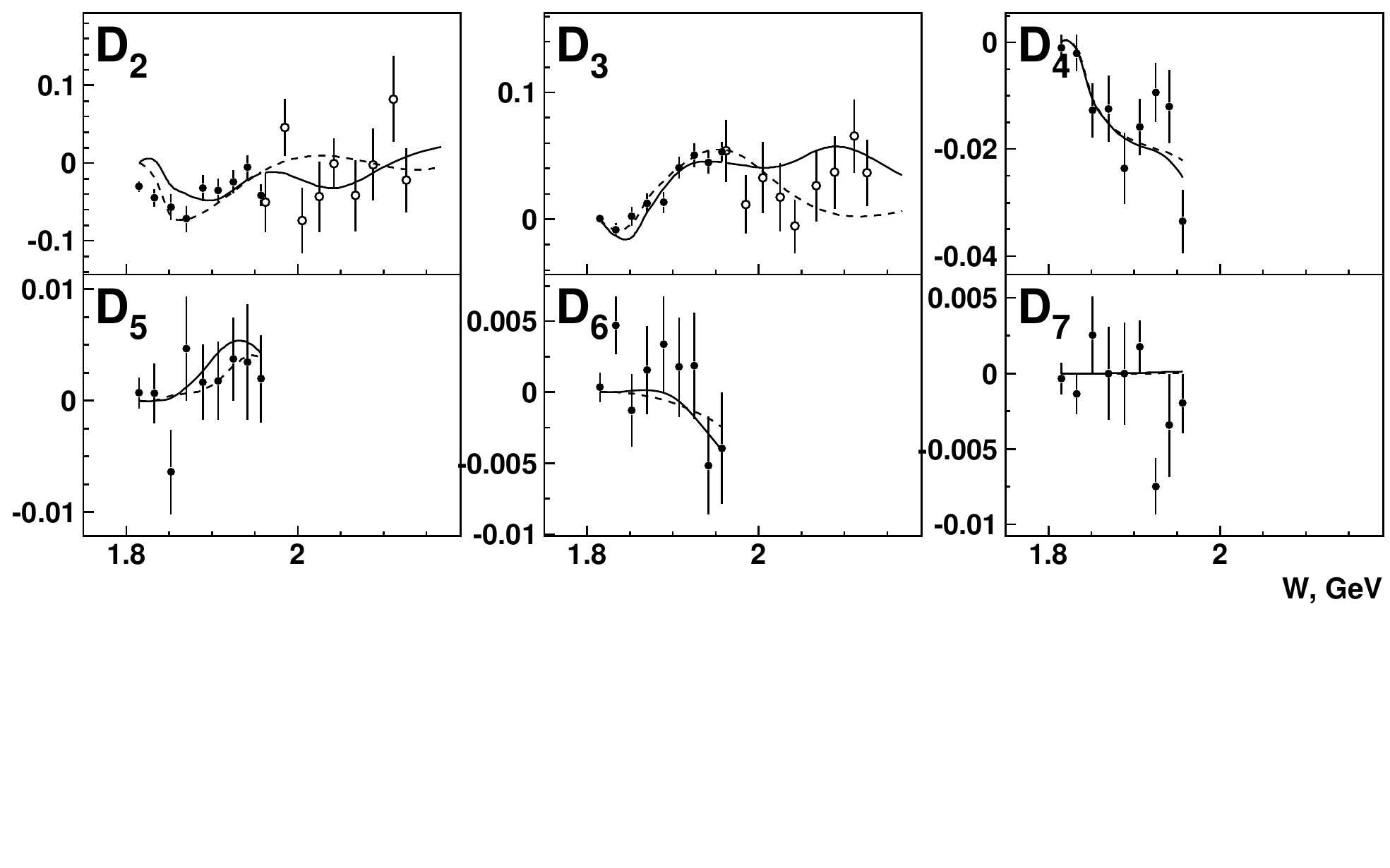}\\
\end{tabular}
\caption{\label{Kstar0}$K^-p\to\bar K^{*0}n$~\cite{Cameron:1978qi}: 
The associated Legendre coefficients for the differential cross sections and the density matrix elements 
$\rho_{\frac32 \frac32}$, $\Re e\rho_{\frac32 \frac12}$, and $\Re e\rho_{\frac32 -\frac12}$. }
\begin{tabular}{cc}
\hspace{-1mm}\includegraphics[width=0.48\textwidth]{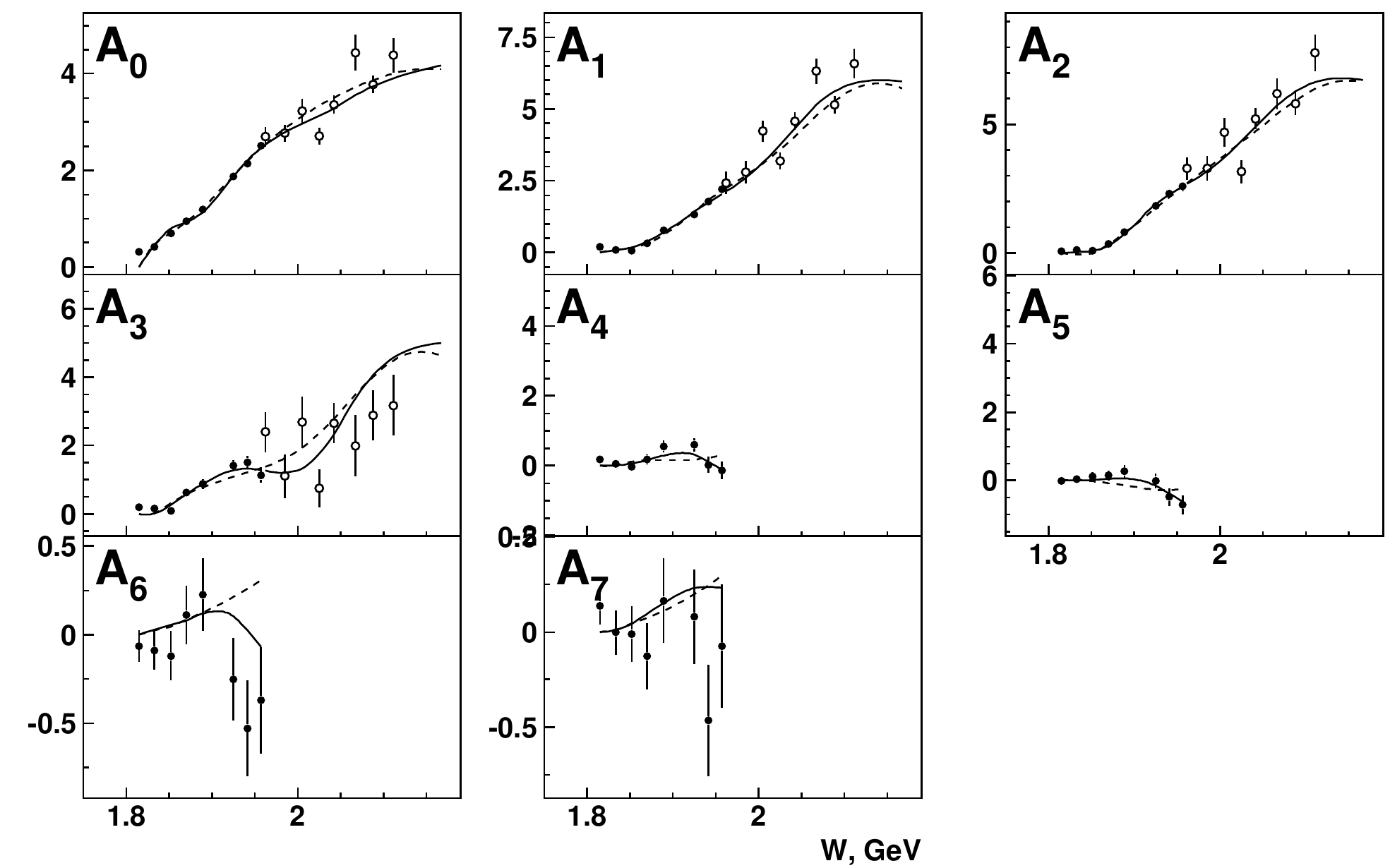}&
\hspace{-1mm}\includegraphics[width=0.48\textwidth]{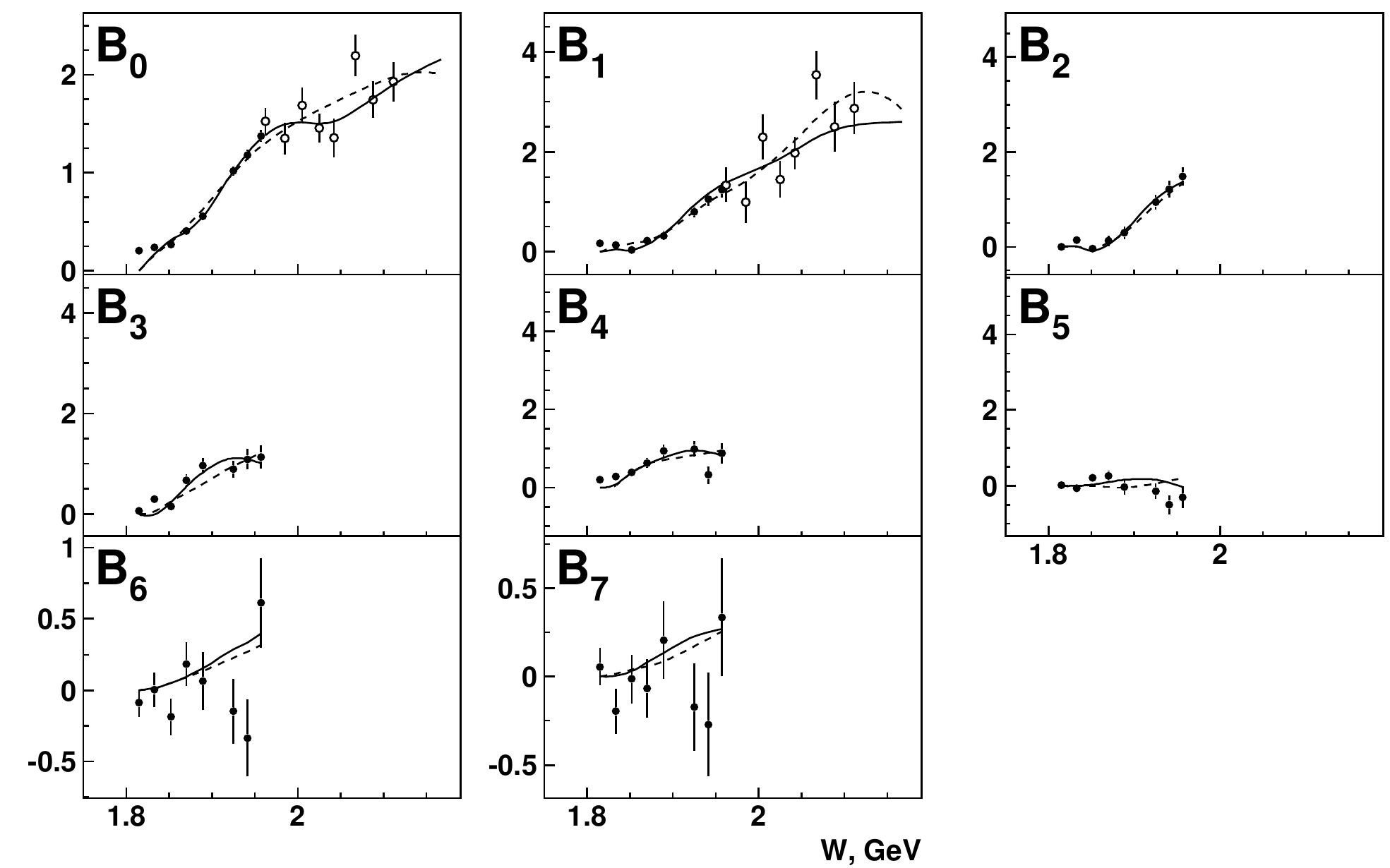}\\
\hspace{-1mm}\includegraphics[width=0.48\textwidth]{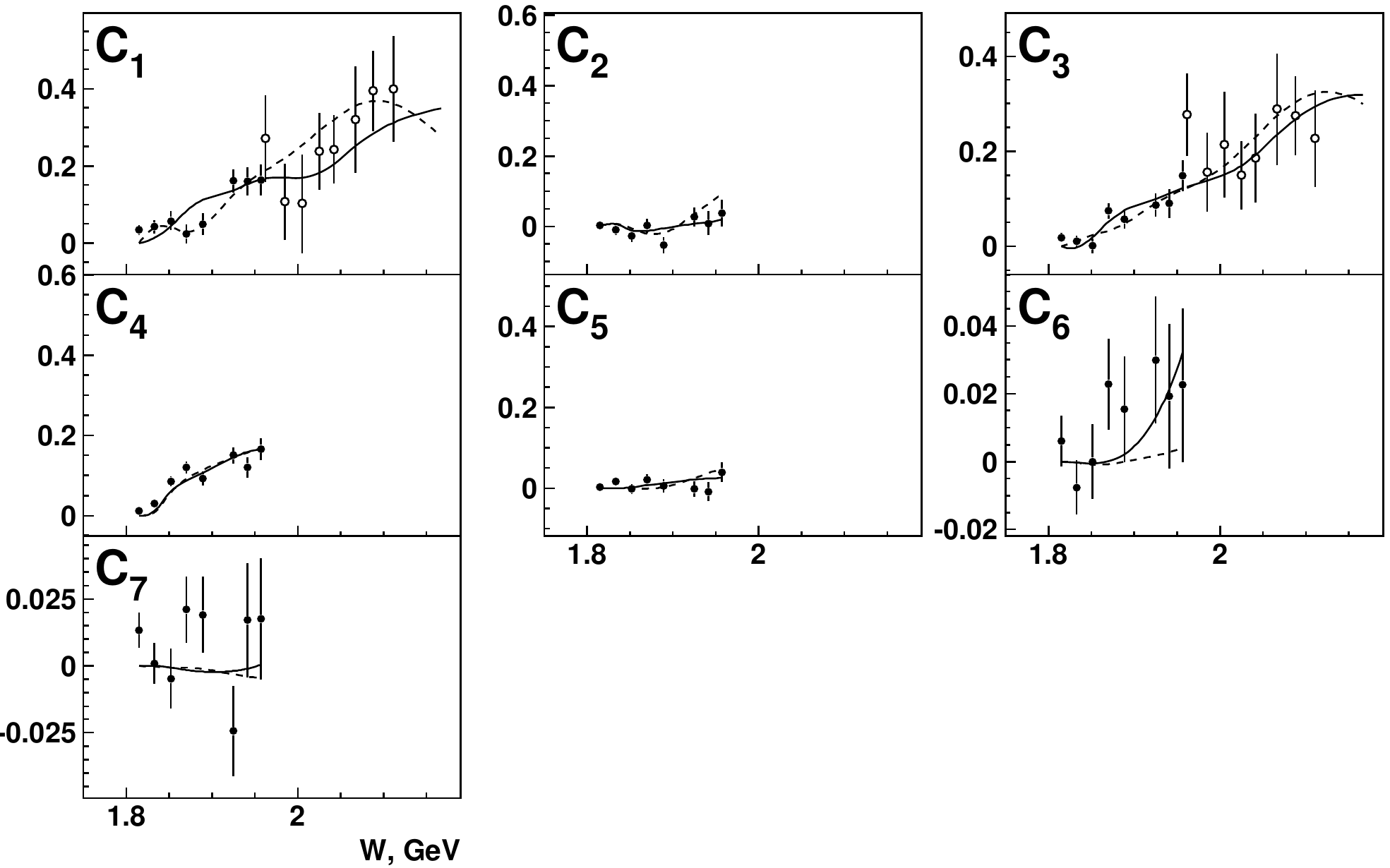}&
\hspace{-1mm}\includegraphics[width=0.48\textwidth]{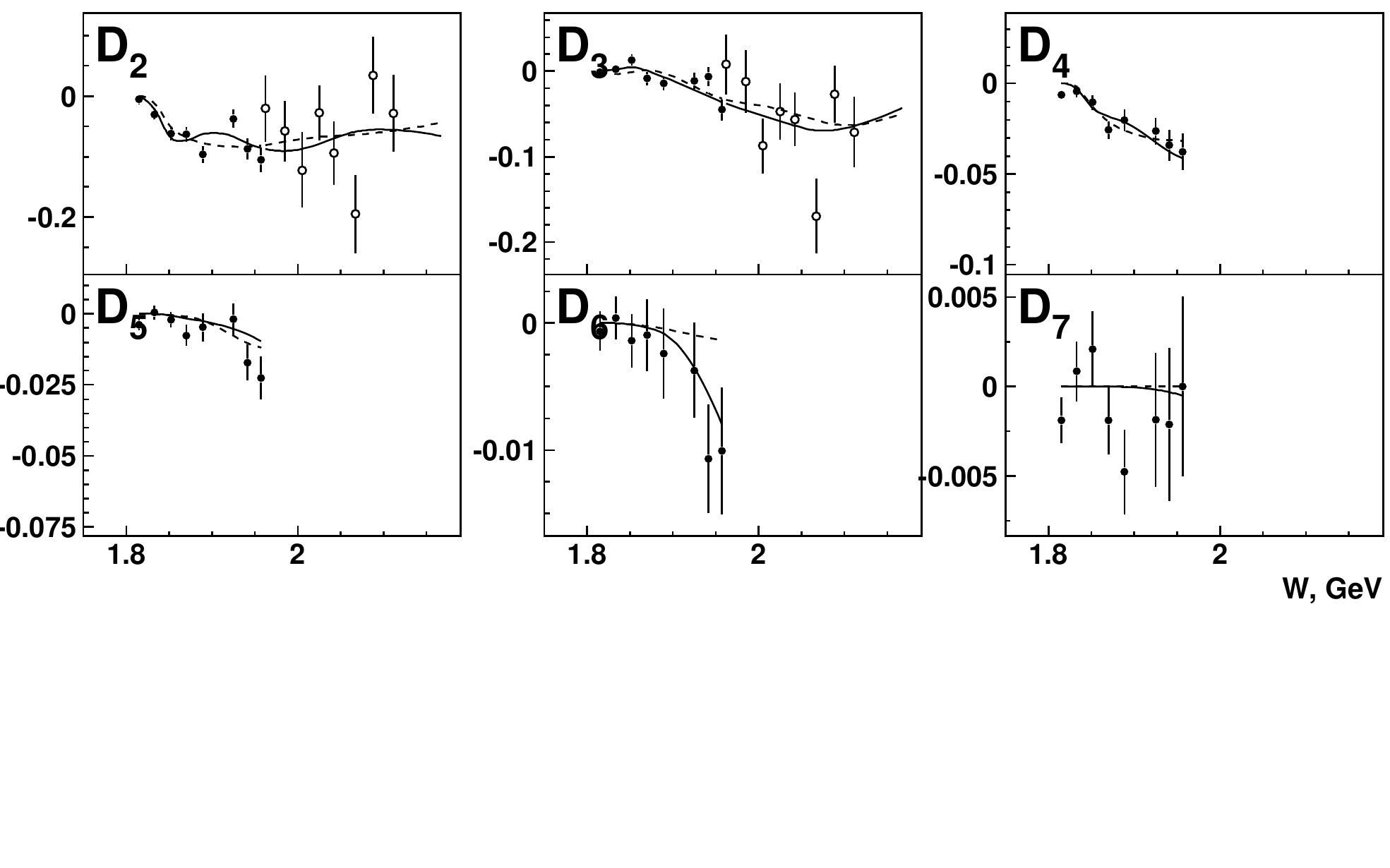}\\
\end{tabular}
\caption{\label{Kstarp}$K^-p\to  K^{*-}p$~\cite{Cameron:1978qi}: 
The associated Legendre coefficients for the differential cross sections and the density matrix elements 
$\rho_{\frac32 \frac32}$, $\Re e\rho_{\frac32 \frac12}$, and $\Re e\rho_{\frac32 -\frac12}$.}
\end{figure*}
\begin{figure*}[pt]
\begin{tabular}{cc}
\hspace{-1mm}\includegraphics[width=0.48\textwidth]{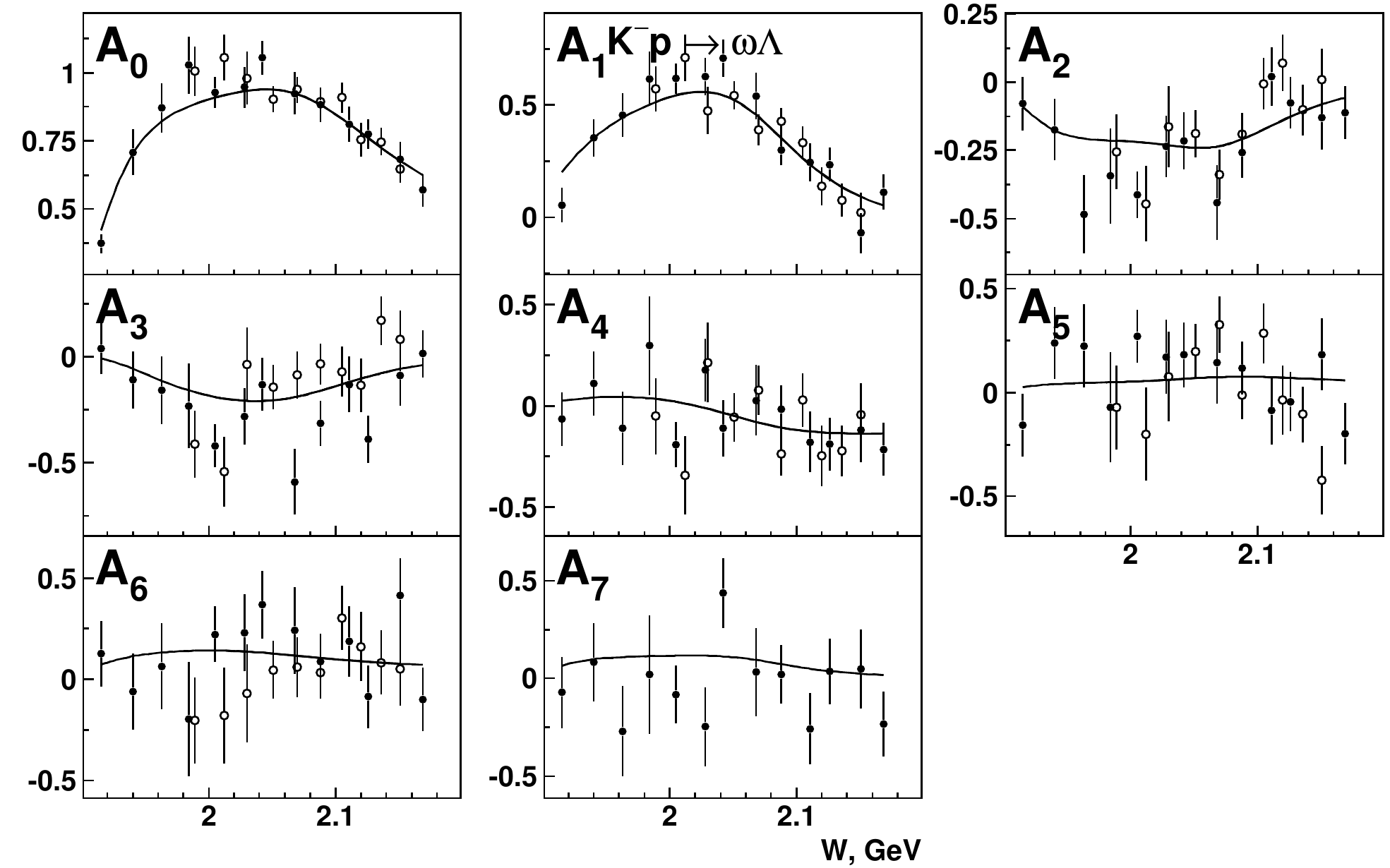}&
\hspace{-1mm}\includegraphics[width=0.48\textwidth]{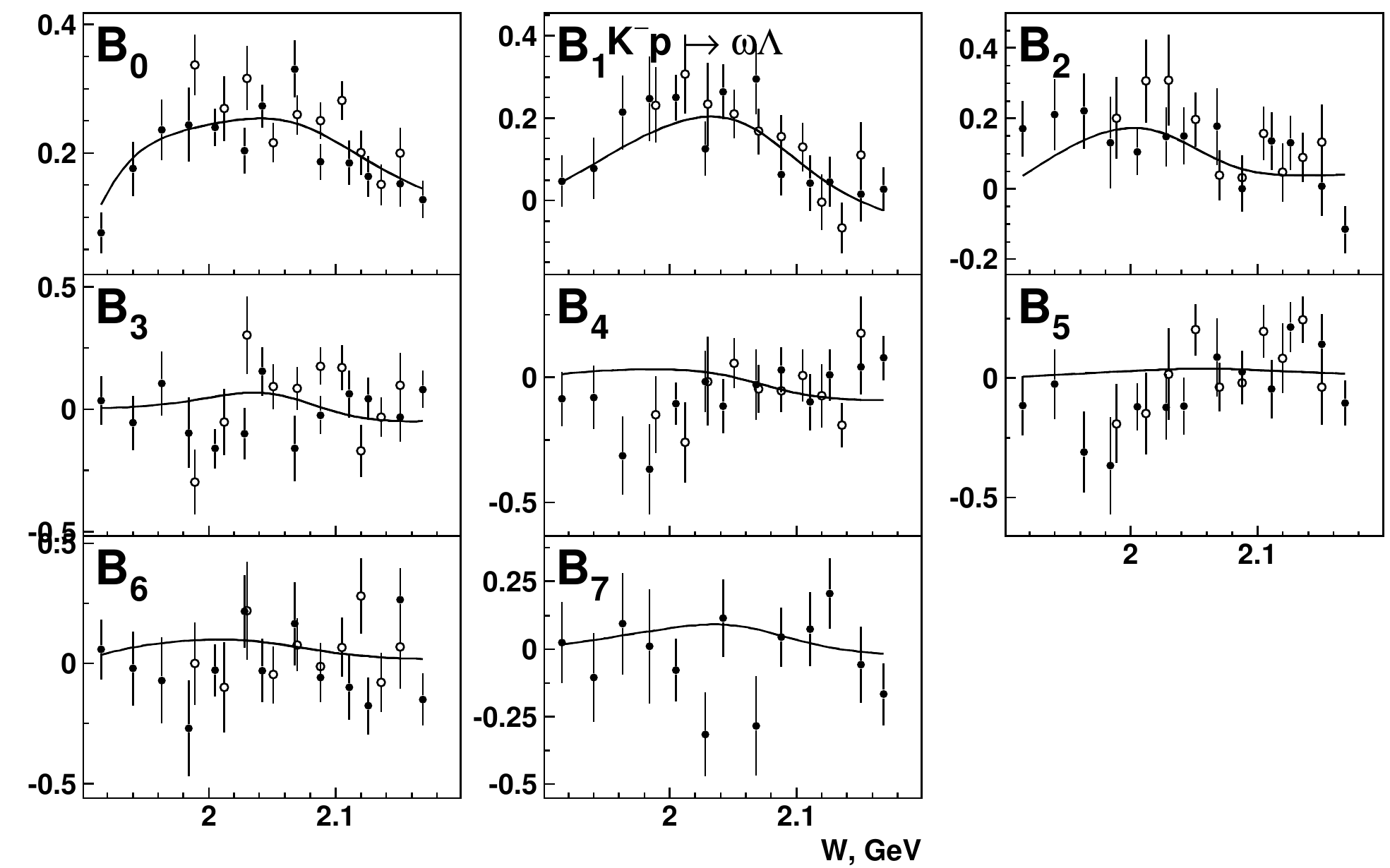}\\
\hspace{-1mm}\includegraphics[width=0.48\textwidth]{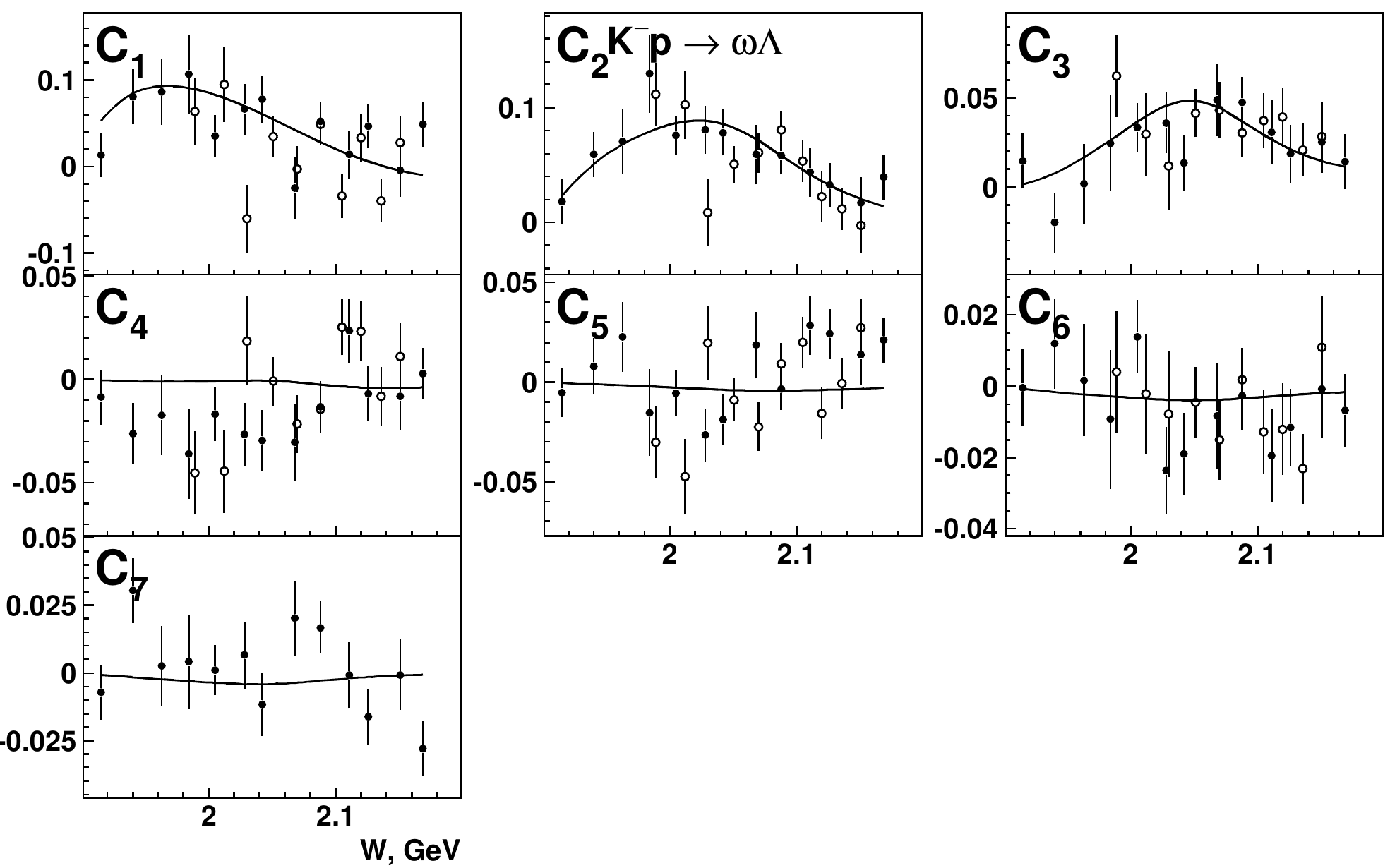}&
\hspace{-1mm}\includegraphics[width=0.48\textwidth]{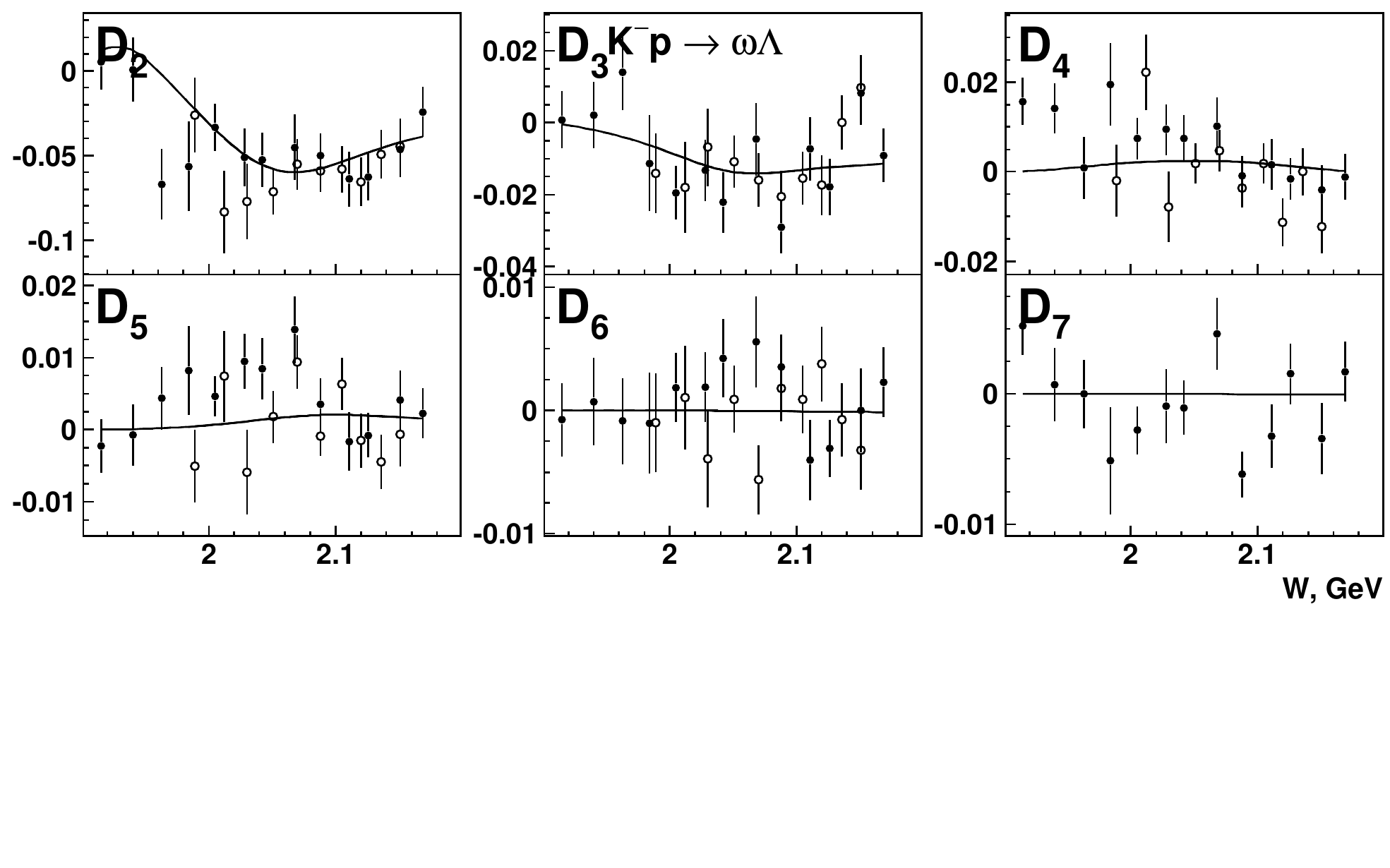}\\
\end{tabular}
\caption{\label{omegaL}$K^-p\to \omega\Lambda$~\cite{Brandstetter:1972xp,Nakkasyan:1975yz,Baccari:1976ik}: 
The associated Legendre coefficients for the differential cross sections and the density matrix elements 
$\rho_{\frac32 \frac32}$, $\Re e\rho_{\frac32 \frac12}$, and $\Re e\rho_{\frac31 -\frac12}$. }
\end{figure*}

The differential cross sections and the density matrix elements were expanded into
associated Legendre polynomials:
\begin{subequations}
\begin{align}
\label{Legendre}
 &\frac{d\sigma}{d\Omega}                &=\qquad&  \sum_l A_l P^0_l(\cos\theta)\\
\rho_{\frac32\frac32}&\frac{d\sigma}{d\Omega}  &=\qquad&  \sum_l B_l P^0_l(\cos\theta)\\
\rho_{\frac32\frac12}&\frac{d\sigma}{d\Omega}  &=\qquad&  \sum_l C_l P^1_l(\cos\theta)\\
\rho_{\frac32-\frac12}&\frac{d\sigma}{d\Omega} &=\qquad&  \sum_l D_l P^2_l(\cos\theta)
\end{align}
\end{subequations}

The expansions were limited to $l=0, 1, \cdots 7$.
The results of the analyses were given in the form of the coefficients
$A_l; \cdots D_l$.

The results on the Legendre coefficients for the fits to differential cross sections and to the
$\rho$ density matrix elements for the various reactions are shown in Figs.~\ref{Sigmam}~to~\ref{omegaL}
and compared to our final fit. The experimental uncertainties in the Legendre
coefficients are comparably large, the fit reproduces the data with a $\chi^2=5812$ for 4611
 data points. 

The authors of Ref.~\cite{Litchfield:1973ap} analzyed the data from two experiments, of the CERN-Heidelberg and
of the Collège de France-Rutherford-Saclay-Strasbourg collaboration which seem not be published. 

\section{\label{Results}Results}

The number and positions of poles of the resonances used in the fits stem mostly from the fits to the 
two-body reactions described in Ref.~\cite{Matveev:2019igl}. 
The fit distributes the intensities, observed in the reaction $K^-p\to \pi^0\Lambda(1520)$, $K^-\Delta^+(1232)$, 
$\pi^\pm\Sigma^\mp(1385)$,  $\bar K^{*}N$ and
$\omega\Lambda$, between the con\-tributing resonances. The $\pi\Lambda^*$ final
state is reached by a number of hyperon resonances, none very significant. Most intensity in the 
$\bar K \Delta(1232)$  final state
stems from four resonances which have a significant $\bar K\Delta(1232)$ branching ratio.
$\Sigma(1915)5/2^+$: \ (21\er6)\%, $\Sigma(2000)$ $3/2^-$: \ (12\er5)\%, $\Sigma(2030)7/2^+$:\ (16\er5)\%, and $\Sigma(2230)3/2^+$: \ (22\er5)\%. 

The branching ratio for the $\Lambda(1810)1/2^+\to \pi\Sigma(1385)$ decay: (40\er15)\% is large. 
This is a remarkable confirmation 
of the 60\%  branching ratio for this decay from Ref.~\cite{Kamano:2014zba,Kamano:2015hxa,Kamano:2016djv}.
It supports the existence of this resonance for which the scan gave only marginal evidence (see~\cite{Matveev:2019igl}). For the following resonances we find a branching ratio of at least 15\% within uncertainties 
into this final state: $\Lambda(1690)3/2^-$: (12\er5)\%, $\Lambda(1830)5/2^-$: (22\er8)\%,  \\
$\Lambda(2070)3/2^+$: (12\er5)\%,  and $\Lambda(2080)5/2^-$: (18\er5)\%. 

The $\bar K^*N$ final state is produced with a high yield via pion exchange in
the $t$-channel. A significant structure is observed in at least $A_0$ -- $A_2$ and $B_0$ -- $ B_3$ at about 1.9\,GeV,
in some coefficients with opposite signs for $\bar K^{*0}n$ and $K^{*-}p$. The structure is assigned to 
$\Lambda(1820)5/2^+$. Its branching ratio (BR) for $\bar K^*N$ decays vanishes by definition since the sum of nucleon
and $K^*$ masses of 1830\,MeV exceeds the $\Lambda(1820)5/2^+$ mass. 
Further significant $\bar K^*N$ branching ratios, reaching within uncertainties 25\%, are observed 
for the following hyperon resonances: \\
$\Lambda(2070)3/2^+$: \ (66\er12)\%,  $\Lambda(2080)5/2^-$: \ (42\er18)\%, \\
$\Sigma(1915)5/2^+$: \ (42\er18)\%, $\Sigma(2000)3/2^-$: \ (53\er11)\%, and $\Sigma(2230)3/2^+$: \ (38\er6)\%. \\
The $\omega\Lambda$
intensity is distributed among several hyperons.

The first point which needs to be made is that we find no evidence for a large
number of resonances reported in the Review of Particle Physics (RPP)~\cite{Tanabashi:2018oca}.   
It needs to be underlined that we fit nearly all published data on
medium-energy $K^-p$ elastic, charge exchange and inelastic scattering.
In the mass range below 2200\,MeV, we find neither evidence for the $\Lambda$ 
resonances $\Lambda(1710)1/2^+$,
$\Lambda(2000)$ (unknown spin-parity),  $\Lambda(2020)7/2^+$, $\Lambda(2050)3/2^-$ 
nor for the $\Sigma^*$'s  $\Sigma(1580)3/2^-$, $\Sigma(1620)1/2^-$, $\Sigma(1730)$ $3/2^+$,  
$\Sigma(1770)$ $1/2^+$,  $\Sigma(1840)3/2^+$,
$\Sigma(1940)3/2^+$, $\Sigma(2000)1/2^-$, 
$\Sigma(2070)5/2^+$, 
Also, we do not observe 
$\Sigma(1480)$, $\Sigma(1560)$, $\Sigma(1620)$, $\Sigma(1670)$, and $\Sigma(1690)$, 
the so-called ``bumps''   seen in 
production experiments. There is
no evidence for the 2* resonances, $\Sigma(1880)1/2^+$ and $\Sigma(2080)3/2^+$.
Also the 3*-resonance $\Lambda(1810)1/2^+$ is not required in our analysis.  
It is seen, however, in several
analyses and ranked as 3* resonance in the RPP. Most of its properties reported in the RPP are confirmed here
when we introduce it in our fits. Hence
we keep it  in our fits and give it (1*). There is also the 1* candidate $\Lambda(1620)1/2^-$ in the RPP. 
If we include it in the fit, the $\chi^2$ gain is just below the limit above which we would consider 
it a 1* resonance (see\cite{Matveev:2019igl}), we therefore also keep this state in our fits and give 
it (1*). Overall, this is an important ``cleaning'' of the resonance spectrum.
Table~\ref{starrating} shows a comparison of the RPP star rating and our rating.


\begin{table}[pt]
\caption{\label{starrating}Star rating of hyperons resonances below 2200\,MeV
of the RPP \cite{Tanabashi:2018oca} and from this work, see~\cite{Matveev:2019igl}. 
'(*)' indicates states which we keep in the fit even though we find no clear evidence for 
their existence (see text). }
\renewcommand{\arraystretch}{1.4}
\bc
\begin{tabular}{lcclcc}
\hline\hline
                                  & \cite{Tanabashi:2018oca}&\hspace{-2mm}\cite{Matveev:2019igl}&& \cite{Tanabashi:2018oca}&\hspace{-2mm}\cite{Matveev:2019igl}\\\hline
$\Lambda(1405)1/2^-$&****&\hspace{-2mm}****&$\Sigma(1580)3/2^-$&* &\hspace{-2mm}  -    \\ 
$\Lambda(1520)3/2^-$&****&\hspace{-2mm}****&$\Sigma(1620)1/2^-$&*&\hspace{-2mm}  (*)   \\ 
$\Lambda(1600)1/2^+$&*** &\hspace{-2mm}****&$\Sigma(1660)1/2^+$&***&\hspace{-2mm}***\\           
$\Lambda(1670)1/2^-$&****&\hspace{-2mm}****&$\Sigma(1670)3/2^-$&****&\hspace{-2mm}****\\           
$\Lambda(1690)3/2^-$&****&\hspace{-2mm}****&$\Sigma(1730)3/2^+$&*&\hspace{-2mm}  -   \\           
$\Lambda(1710)1/2^+$&  *  &\hspace{-2mm}  -     &$\Sigma(1750)1/2^-$&***&\hspace{-2mm}****\\  
$\Lambda(1800)1/2^-$&***  &\hspace{-2mm}*** &$\Sigma(1770)1/2^+$&*&\hspace{-2mm}  -   \\         
$\Lambda(1810)1/2^+$&*** &\hspace{-2mm} (*)    &$\Sigma(1775)5/2^-$&****&\hspace{-2mm}****\\                
$\Lambda(1820)5/2^+$&****&\hspace{-2mm}****&$\Sigma(1840)3/2^+$&\hspace{-2mm}*&\hspace{-2mm}  -   \\              
$\Lambda(1830)5/2^-$&**** &\hspace{-2mm}***&$\Sigma(1880)1/2^+$&\hspace{-2mm}**&\hspace{-2mm}  -   \\             
$\Lambda(1890)3/2^+$&****&\hspace{-2mm}****&$\Sigma(1900)1/2^-$&*&\hspace{-2mm}**\\              
$\Lambda(2000)$     &  *    & \hspace{-2mm}  -          &$\Sigma(1915)5/2^+$&****&\hspace{-2mm}****\\  
$\Lambda(2020)7/2^+$&  *    &\hspace{-2mm}  -         &$\Sigma(1940)3/2^+$&*&\hspace{-2mm}  -   \\       
$\Lambda(2050)3/2^-$&  *    &\hspace{-2mm}  -          &$\Sigma(1940)3/2^-$&***&\hspace{-2mm}***\\  
$\Lambda(2070)3/2^+$&\hspace{-2mm}  - &\hspace{-2mm} * &$\Sigma(2000)1/2^-$&*&\hspace{-2mm}  -    \\
$\Lambda(2080)5/2^-$& \hspace{-2mm}  -  &\hspace{-2mm} * &$\Sigma(2000)3/2^-$&-&\hspace{-2mm}  *    \\
$\Lambda(2100)7/2^-$&****&\hspace{-2mm}**** &$\Sigma(2030)7/2^+$&****&\hspace{-2mm}****\\
$\Lambda(2110)5/2^+$&***& \hspace{-2mm} **   &$\Sigma(2070)5/2^+$&*&\hspace{-2mm}  -   \\  
                                   &     &          &$\Sigma(2080)3/2^+$&**&\hspace{-2mm}  -   \\  
                                   &     &          &$\Sigma(2100)7/2^-$&\hspace{-2mm}  *   &\hspace{-2mm}* \\
                                   &     &          &$\Sigma(2160)1/2^-$&\hspace{-2mm}  -   &\hspace{-2mm}* \\\hline\hline
\multicolumn{6}{c}{The $\Sigma$ ``bumps''  at 1480 (*), 1560 (**), 1690 (**)\,MeV }\\
\multicolumn{6}{c}{and the claims at 1620 and 1670\,MeV from production}\\ 
\multicolumn{6}{c}{experiments are also not seen.}\\ 
\hline\hline
\end{tabular}
\ec
\renewcommand{\arraystretch}{1.}
\end{table}

Table~\ref{Results-1} summarizes our results obtained from the fit to the data listed in
\cite{Matveev:2019igl}. 
For most established hyperon resonances (with three or four stars in the RPP), our results on
masses, widths and on the branching ratios for decays into $N\bar K$, $\Sigma\pi$ and -- for $\Sigma^*$ resonances -- into $\Lambda\pi$ agree well with earlier results. 
In some cases, the pattern (hierarchy) of decay modes is reproduced 
even though there is no quantitative agreement. 
In a few cases, there are significant discrepancies. 

For hyperon masses, widths, and branching ratios, the RPP gives mostly a range which covers most observations.
Our uncertainties give the spread of results from different solutions where single resonances of minor significance are taken into account additionally. When significant resonances are omitted, the fit results often 
change drastically. We do not 
include these fits in the evaluation of uncertainties. Hence our uncertainties may be underestimated.
Therefore we increase the uncertainties in the branching ratios to a minimum of 20\% (except for the 
highly constrained $\Lambda(1520)$).  


Significant decay modes are compared to the RPP listings. In our discussion 
below, ``compatible'' or ``agree''
means 1\,$\sigma$ compatible, ``not inconsistent'' 2\,$\sigma$ compatible. The properties of hyperons 
at the pole position are mostly given by the Kent~\cite{Zhang:2013cua,Zhang:2013sva}
and Osaka-Argonne~\cite{Kamano:2015hxa} group only, often with no uncertainty or
statistical uncertainties only, and the RPP gives no ranges. Here we comment discrepancies only when 
the difference in the modulus exceeds 3$\sigma$. The phases depend critically on
the background model and are very often discrepant. Hence we do not comment on the phases.

Below, we give the sum of all measured branching ratios. The uncertainties in the BR sum are determined from the sum of the
squared individual uncertainties, even though the uncertainties are correlated: their sum must not exceed unity. 

\subsection{The \boldmath$\Lambda$ hyperons}

\paragraph{\boldmath$\Lambda(1520)3/2^-$:} 
Our mass, width and branching ratios (BRs) of the well-known $\Lambda(1520)3/2^-$ are compatible with the RPP range.
Its decays into $N\bar K$ and $\Sigma\pi$ add up to 88\er2\% (BnGa),  the RPP
reports $\Lambda\pi\pi$ with BRs of 10\er1\%, $\Sigma\pi\pi$ with 0.9\er 0.1\% 
and $\Sigma^0\gamma$ with 0.85\er0.15\% as further decay modes. The $\Sigma(1385)\pi$
decay is reported to be {\it seen} in RPP; it signals an SU(3) octet component in the $\Lambda(1520)$ wave function. 
The BR for decays into $\Sigma(1385)\pi$ vanishes in our definition since the sum $M_{\Sigma(1385)}+M_\pi$ exceeds
$M_{\Lambda(1520)}$. Our pole properties agree very well with those from 
the Kent~\cite{Zhang:2013sva} and the Osaka-Argonne 
group~\cite{Kamano:2015hxa}.\vspace{-6mm}

\paragraph{\boldmath$\Lambda(1600)1/2^+$:} 
Our properties of $\Lambda(1600)1/2^+$ fall into the range of values reported in the RPP. 
The sum of the decay fractions is found to be  83-100\%, 
thus nearly no intensity is missed. The RPP Breit-Wigner 
width ranges from 50 to 250\,MeV; we find a width just below the upper value. 
Osaka-Argonne~\cite{Kamano:2015hxa} find a pole width which is a factor two smaller than our value;  
our normalized residues are also smaller than those reported in Ref.~\cite{Kamano:2015hxa}.  
Their squared ratio of the normalized residues
in  $\bar K N\to \pi\Sigma$ over $\bar K N\to \bar K N$ is nearly 4.9, the ratio for the corresponding BR's is 13.3.
Apart from the phase space difference, these two numbers should be the same. 
Our values for these ratios are 1.17 and 1.28, respectively.\vspace{-6mm}

\paragraph{\boldmath$\Lambda(1670)1/2^-$:} 
Our $\Lambda(1670)1/2^-$ properties are mostly fully compatible 
with RPP values except for the $\Sigma\pi$
decay fraction where we find (12\er3)\%, outside of the 25 to 55\% RPP range. The strong $\Lambda\eta$ decay
mode of (20\er8)\% reminds of the strong coupling of $N(1535)1/2^-\to N\eta$. The decay fractions sum up to 
77-100\%. Our normalized transition residues are not inconsistent with those 
from Ref.~\cite{Kamano:2015hxa}.\vspace{-6mm}

\paragraph{\boldmath$\Lambda(1690)3/2^-$:\ } 
Mass, width and pole position of the $\Lambda(1690)3/2^-$ agree well with the values reported in the RPP.
The sum of all BR's is 78-100\%. The BR for $N\bar K$ is consistent with RPP,
the one for $\Sigma\pi$ exceeds the RPP range slightly. We find a (5\er2)\% BR
for $\Lambda\sigma$ decays, the RPP reports a 25\% BR for decays into $\Lambda\pi\pi$. This
number is just 2$\sigma$ compatible with our sum of the contributions from
$\Lambda\sigma$  and $\Sigma(1385)\pi\to\Lambda\pi\pi$. 
Our normalized transition residues agree well with those from Ref.~\cite{Kamano:2015hxa}.\vspace{-6mm}

\begin{table*}
\caption{\label{Results-1}Resonance parameters of  $\Lambda$ and $\Sigma$ hyperons: Pole positions
and normalized transition amplitudes (in \%),
Breit-Wigner masses and widths, and decay branching ratios (in \%). The pole positions and Breit-Wigner masses and widths
are given in MeV; the transition amplitudes are normalized to $\Gamma_{\rm pole}/2$. In the Tables
$\Delta(1232)3/2^+$ is abbreviated as $\Delta$, $\Sigma(1385)3/2^+$ as $\Sigma^*$,
$\Lambda(1520)3/2^-$ as $\Lambda^*$, and the $\pi\pi$ $S$-wave or $f_0(500)$ as $\sigma$.
A subscript $S, P, \cdots$ denotes the orbital angular momentum between the outgoing baryon and meson,
a subscript 1/2 or 3/2 the sum of the spins of baryon and meson.
}
\renewcommand{\arraystretch}{1.3}
\bc
\begin{scriptsize}
\begin{tabular}{|l|l|}
\begin{tabular}{lrr|lrr}
\multicolumn{5}{l}{\boldmath\fbox{\fbox{$\Lambda(1670){1/2^-}$}}
\unboldmath
\hfill****}\\[4ex]
\hline\hline\\[-2.7ex]
\multicolumn{3}{c|}{Pole parameters}&\multicolumn{3}{c}{Breit-Wigner parameters}\\[0.3ex]
\hline\\[-2.7ex]
\multicolumn{3}{l|}{M=1676\er2\qquad\qquad$\Gamma$=33\er4} &\multicolumn{3}{l}{M=1677\er 2\qquad\quad $\Gamma$=33\er4}\\
\multicolumn{3}{l|}{norm. tran. res. (\%) for $\bar K N\to$ } & \multicolumn{3}{l}{Branchings (\%)}\\[0.2ex]
$N\bar K$                              &(30\er 6)&-(145\er 11)$^\circ$&Br($N\bar K$)                      &\qquad\ 33\er 7\\
$\Sigma\pi$                     &(19\er 6)& (145\er 14)$^\circ$&Br($\Sigma\pi$)              &12\er 3\\
$\Lambda\eta$                 &(26\er 9)& (104\er 14)$^\circ$& Br($\Lambda\eta$)         &20\er 8 \\
$\Xi K$                             &  (2\er2) & (100\er25)$^\circ$ & Br($\Xi K$)                     & 0 \\
$\Lambda\omega_{S1/2}$& (9\er4)  &-(60\er35)$^\circ$   & Br($\Lambda\omega_S$)& 0 \\
$\Lambda\omega_{D3/2}$& (5\er4)  &    & Br($\Lambda\omega_D$)& 0 \\
$\Lambda\sigma$             & (25\er 8)&(160\er15)$^\circ$ & Br($\Lambda\sigma)$      & 20\er 8 \\
$\Sigma^*\pi_D$             & (13\er 6)& (110\er12)$^\circ$  & Br($\Sigma^*\pi_D$)     & 6\er2 \\
$N\bar K^*_S$             & (31\er 14)&(100\er45)$^\circ$  & Br($N\bar K^*_S$)     & 0\\
$N\bar K^*_D$             & (6\er 3)&-(85\er40)$^\circ$ & Br($N\bar K^*_D$) & 0
\\
\hline\hline\\[-1.5ex]
\end{tabular}
&
\begin{tabular}{lrr|lrr}
\multicolumn{5}{l}{\boldmath\fbox{\fbox{$\Lambda(1800){1/2^-}$}}
\unboldmath
\hfill***}\\[4ex]
\hline\hline\\[-2.7ex]
\multicolumn{3}{c|}{Pole parameters}&\multicolumn{3}{c}{Breit-Wigner parameters}\\[0.3ex]
\hline\\[-2.7ex]
\multicolumn{3}{l|}{M=1809\er9\qquad\qquad$\Gamma$=205\er 16} &\multicolumn{3}{l}{M=1811\er 10\qquad\ \ $\Gamma$=209\er18}\\
\multicolumn{3}{l|}{norm. tran. res. (\%) for $\bar K N\to$ } & \multicolumn{3}{l}{Branchings (\%)}\\[0.2ex]
$N\bar K$                    &(34\er 7)&(103\er 8)$^\circ$ & Br($N\bar K$)              &\qquad\quad 35\er 7 \\
$\Sigma\pi$              &(30\er6)&-(123\er8)$^\circ$ & Br($\Sigma\pi$)              & 27\er6  \\
$\Lambda\eta$            & (6\er3)&(75\er10)$^\circ$ & Br($\Lambda\eta$)            & 1.0\er0.5  \\
$\Xi K$                   &(1.5\er 1)&(40\er40)$^\circ$ & Br($\Xi K$)                   & 0  \\
$\Lambda\omega_{S1/2}$   &(12\er4)&-(114\er30)$^\circ$ & Br($\Lambda\omega_{S1/2}$)   & 0  \\
$\Lambda\omega_{D1/2}$   & (8\er3)&-(90\er17)$^\circ$ & Br($\Lambda\omega_{D1/2}$)   & 0  \\
$\Lambda \sigma$         &(24\er5)&(25\er10)$^\circ$ & Br($\Lambda \sigma$)      & 15\er4  \\
$\Sigma^*\pi_D$          &(16\er6)&-(140\er35)$^\circ$ & Br($\Sigma^*\pi_D$)      & 9\er4  \\
$N\bar K^*_S$           &(18\er 6)&(65\er40)$^\circ$  & Br($N\bar K^*_S$)     & 0 \\
$N\bar K^*_D$           & (9\er 7)& & Br($N\bar K^*_D$)     & 0\\
\hline\hline\\[-1.5ex]
\end{tabular} \\
\begin{tabular}{lrr|lrr}
\vspace{-2.2cm}\\
\multicolumn{5}{l}{\boldmath\fbox{\fbox{$\Lambda(1520){3/2^-}$}}
\unboldmath
\hfill****}\\[4ex]
\hline\hline\\[-2.7ex]
\multicolumn{3}{c|}{Pole parameters}&\multicolumn{3}{c}{Breit-Wigner parameters}\\[0.3ex]
\hline\\[-2.7ex]
\multicolumn{3}{l|}{M=1517.5\er0.4\qquad$\Gamma$=15.3\er0.9} &\multicolumn{3}{l}{M=1518.5\er 0.5\,\,$\Gamma$=15.7\er1.0}\\
\multicolumn{3}{l|}{norm. tran. res. (\%) for $\bar K N\to$ } & \multicolumn{3}{l}{Branchings (\%)}\\[0.2ex]
$N\bar K$                    &\qquad\  (45\er1)&\quad -(10\er3)$^\circ$ & Br($N\bar K$)            & \qquad\quad\ \ 45\er1\\
$\Sigma\pi$              & (44\er1)&-(15\er3)$^\circ$ & Br($\Sigma\pi$)     & 43\er1\\
$\Lambda\eta$            & (1.3\er0.3)&(116\er3)$^\circ$ & Br($\Lambda\eta$)  & 0  \\
\hline\hline\\[-1.5ex]
\end{tabular}
&
\begin{tabular}{lrr|lrr}
\multicolumn{5}{l}{\boldmath\fbox{\fbox{$\Lambda(1690){3/2^-}$}}
\unboldmath
\hfill****}\\[4ex]
\hline\hline\\[-2.7ex]
\multicolumn{3}{c}{Pole parameters}&\multicolumn{3}{c}{Breit-Wigner parameters}\\[0.3ex]
\hline\\[-2.7ex]
\multicolumn{3}{l|}{M=1683\er3\qquad\qquad$\Gamma$=72\er 5} &\multicolumn{3}{l}{M=1689\er 3\qquad\qquad $\Gamma$=75\er5}\\
\multicolumn{3}{l|}{norm. tran. res. (\%) for $\bar K N\to$ } & \multicolumn{3}{l}{Branchings (\%)}\\[0.2ex]
$N\bar K$                    &\quad (24\er5)&-(28\er5)$^\circ$ & Br($N\bar K$)                 &\qquad\qquad 23\er5\\
$\Sigma\pi$              & (35\er7)&(175\er6)$^\circ$ & Br($\Sigma\pi$)         & 50\er10 \\
$\Lambda\eta$            & (5\er2)&(88\er8)$^\circ$ & Br($\Lambda\eta$)         & $\sim$1 \\
$\Lambda \sigma$      & (8\er2)&-(10\er6)$^\circ$ & Br($\Lambda \sigma$) & 5\er2 \\
$\Sigma^*\pi_S$      & (11\er6)& (170\er70)$^\circ$  & Br($\Sigma^*\pi_S$) & 9\er5 \\
$\Sigma^*\pi_D$      & (6\er4)&(164\er15)$^\circ$ & Br($\Sigma^*\pi_D$) & 3\er2 \\
$N\bar K^*_S$           & (5\er 4)&              & Br($N\bar K^*_S$)     & 0 \\
$N\bar K^*_D$           & (18\er 5)&-(110\er45)$^\circ$ & Br($N\bar K^*_D$)     & 0\\
\hline\hline\\[-1.5ex]
\end{tabular}\\
\begin{tabular}{lrr|lrr}
\vspace{-4.2cm}\\
\multicolumn{5}{l}{\boldmath\fbox{\fbox{$\Lambda(1830){5/2^-}$}}
\unboldmath
\hfill***}\\[4ex]
\hline\hline\\[-2.7ex]
\multicolumn{3}{c|}{Pole parameters}&\multicolumn{3}{c}{Breit-Wigner parameters}\\[0.3ex]
\hline\\[-2.7ex]
\multicolumn{3}{l|}{M=1819.5\er3\qquad\qquad$\Gamma$=62\er 5} &\multicolumn{3}{l}{M=1821\er 3\qquad\ \ \ $\Gamma$=64\er 7}\\
\multicolumn{3}{l|}{norm. tran. res. (\%) for $\bar K N\to$ } & \multicolumn{3}{l}{Branchings (\%)}\\[0.2ex]
$N\bar K$                    & (5.5\er1)&(20\er14)$^\circ$ & Br($N\bar K$)                  & \qquad 5.5\er1 \\
$\Sigma\pi$              & (15\er3)&(180\er10)$^\circ$ & Br($\Sigma\pi$)              & 42\er 8 \\
$\Xi K$                   & (1.0\er0.5)&(65\er20)$^\circ$ & Br($\Xi K$)                   & 0  \\
$\Sigma^*\pi_D$      & (10\er4)&(10\er25)$^\circ$ & Br($\Sigma^*\pi_D$) & 20\er8 \\
$\Sigma^*\pi_G$      & (3\er2)&                  & Br($\Sigma^*\pi_G$) & 2\er1.5 \\
$\Lambda\omega_{1/2,D}$      & (4\er3)& & Br($\Lambda\omega_{1/2D}$) & 0 \\
$\Lambda\omega_{3/2,D}$      & (5\er3)&-(110\er35)$^\circ$ &  Br($\Lambda\omega_{3/2D}$) & 0 \\
\hline\hline\\[-1.5ex]
\end{tabular}
&
\begin{tabular}{lrr|lrr}
\multicolumn{5}{l}{\boldmath\fbox{\fbox{$\Lambda(2080){5/2^-}$}}
\unboldmath
\hfill * (new)}\\[4ex]
\hline\hline\\[-2.7ex]
\multicolumn{3}{c|}{Pole parameters}&\multicolumn{3}{c}{Breit-Wigner parameters}\\[0.3ex]
\hline\\[-2.7ex]
\multicolumn{3}{l|}{M=2070\er15\qquad\quad$\Gamma$=172\er28 } &\multicolumn{3}{l}{M=2082\er 13\qquad $\Gamma$=181\er29}\\
\multicolumn{3}{l|}{norm. tran. res. (\%) for $\bar K N\to$ } & \multicolumn{3}{l}{Branchings (\%)}\\[0.2ex]
$N\bar K$                    & (12\er3)&-(35\er22)$^\circ$ & Br($N\bar K$)                    &  \qquad 11\er 3 \\
$\Sigma\pi$              & (7\er3)&(11\er16)$^\circ$ & Br($\Sigma\pi$)              & 5\er 2 \\
$\Xi K$                   & (6\er2)&(115\er20)$^\circ$ & Br($\Xi K$)                   & 4\er1  \\
$\Lambda\omega_{D1/2}$   & (6\er3)&(115\er25)$^\circ$ & Br($\Lambda\omega_{D1/2}$)   & 4\er2  \\
$\Lambda\omega_{D3/2}$   & (9\er3)&-(10\er35)$^\circ$ & Br($\Lambda\omega_{D3/2}$)   & 8\er3  \\
$\Sigma^*\pi_D$      & (14\er4)&(155\er45)$^\circ$ & Br($\Sigma^*\pi_D$) & 15\er5 \\
$\Sigma^*\pi_G$      & (5\er3)& (30\er45)$^\circ$ & Br($\Sigma^*\pi_G$) & 3\er2 \\
$N\bar K^*_{D1/2}$        & (16\er8)&-(120\er50)$^\circ$& Br($N\bar K^*_{D1/2}$)     & 17\er9 \\
$N\bar K^*_{D3/2}$        & (20\er14)&(60\er50)$^\circ$& Br($N\bar K^*_{D3/2}$)     & 25\er16 \\
\hline\hline\\[-1.5ex]
\end{tabular}\\
\begin{tabular}{lrr|lrr}
\vspace{-2.5cm}\\
\multicolumn{5}{l}{\boldmath\fbox{\fbox{$\Lambda(2100){7/2^-}$}}
\unboldmath
\hfill****}\\[4ex]
\hline\hline\\[-2.7ex]
\multicolumn{3}{c|}{Pole parameters}&\multicolumn{3}{c}{Breit-Wigner parameters}\\[0.3ex]
\hline\\[-2.7ex]
\multicolumn{3}{l|}{M=2040\er14\qquad\quad$\Gamma$=215\er 29} &\multicolumn{3}{l}{M=2090\er 15\quad\ \ $\Gamma$=290\er 30}\\
\multicolumn{3}{l|}{norm. tran. res. (\%) for $\bar K N\to$ } & \multicolumn{3}{l}{Branchings (\%)}\\[0.2ex]
$N\bar K$                    & (28\er6)&\ -(40\er10)$^\circ$ & Br($N\bar K$)                    &\qquad\ 24\er 5\\
$\Sigma\pi$              & (9\er2)&-(35\er15)$^\circ$ & Br($\Sigma\pi$)              & 3\er1.5\\
$\Sigma^*\pi_D$      & (4\er3)& & Br($\Sigma^*\pi_D$) & $<1$ \\
$\Sigma^*\pi_G$      & (6\er3)&-(45\er15)$^\circ$                  & Br($\Sigma^*\pi_G$) & 1\er1 \\
$N\bar K^*_{D3/2}$        & (11\er6)&-(30\er30)$^\circ$& Br($N\bar K^*_{D3/2}$)     & 4\er2 \\
\hline\hline\\[-1.5ex]
\end{tabular}
\end{tabular}
\end{scriptsize}
\ec
\end{table*}


\begin{table*}[pt]
Table~\ref{Results-1} continued.\\[2ex]
\renewcommand{\arraystretch}{1.2}\begin{scriptsize}
\begin{tabular}{l|l|}
\begin{tabular}{lrr|lrr}
\multicolumn{5}{l}{\boldmath\fbox{\fbox{$\Lambda(1600){1/2^+}$}}
\unboldmath
\hfill****}\\[4ex]
\hline\hline\\[-2.7ex]
\multicolumn{3}{c|}{Pole parameters}&\multicolumn{3}{c}{Breit-Wigner parameters}\\[0.3ex]
\hline\\[-2.7ex]
\multicolumn{3}{l|}{M=1562\er 8\qquad\qquad$\Gamma$=232\er15} &\multicolumn{3}{l}{M=1605\er 8\qquad $\Gamma$=245\er15}\\
\multicolumn{3}{l|}{norm. tran. res. (\%) for $\bar K N\to$ } & \multicolumn{3}{l}{Branchings (\%)}\\[0.2ex]
$N\bar K$                    & (36\er7)&-(63\er10)$^\circ$ & Br($N\bar K$)                   & \quad\ \ 29\er 6 \\
$\Sigma\pi$              & (39\er8)&(148\er10)$^\circ$ & Br($\Sigma\pi$)              & 37\er 7 \\
$\Lambda\eta$            & (22\er13)&(180\er20)$^\circ$ & Br($\Lambda\eta$)            & 0 \\
$\Lambda \sigma$      & (30\er6)&-(70\er10)$^\circ$ & Br($\Lambda \sigma$)      & 19\er 4 \\
$\Sigma^*\pi_P$      & (37\er7)&(103\er12)$^\circ$ & Br($\Sigma^*\pi_P$)      &9\er 4 \\
$N\bar K^*_{P1/2}$        & (2\er 1)& (126\er45)$^\circ$ & Br($N\bar K^*_{P1/2}$)     & 0 \\
$N\bar K^*_{P3/2}$        & (2\er 1)& -(135\er45)$^\circ$& Br($N\bar K^*_{P3/2}$)     & 0 \\
\hline\hline\\[-1.5ex]
\end{tabular}
&
\begin{tabular}{lrr|lrr}
\multicolumn{5}{l}{\boldmath\fbox{\fbox{$\Lambda(1810){1/2^+}$}}
\unboldmath
\hfill (*)}\\[4ex]
\hline\hline\\[-2.7ex]
\multicolumn{3}{c|}{Pole parameters}&\multicolumn{3}{c}{Breit-Wigner parameters}\\[0.3ex]
\hline\\[-2.7ex]
\multicolumn{3}{l|}{M=1773\er7\qquad\qquad$\Gamma$=38\er 14} &\multicolumn{3}{l}{M=1773\er 7\qquad $\Gamma$=39\er15}\\
\multicolumn{3}{l|}{norm. tran. res. (\%) for $\bar K N\to$ } & \multicolumn{3}{l}{Branchings (\%)}\\[0.2ex]
$N\bar K$                    &(1.8\er0.8)&(65\er 26)$^\circ$ & Br($N\bar K$)                 & 2.5\er1.3 \\
$\Sigma\pi$              & (4.5\er2.0)&-(143\er24)$^\circ$ & Br($\Sigma\pi$)              & 16\er5 \\
$\Lambda \sigma$      & (5.5\er2.0)&(30\er16)$^\circ$ & Br($\Lambda \sigma$)      & 25\er9 \\
$\Sigma^*\pi_P$      & (8\er3)&-(50\er30)$^\circ$ & Br($\Sigma^*\pi_P$)      &\quad 40\er15 \\
$N\bar K^*_{P1/2}$        & (3\er 3)&                 & Br($N\bar K^*_{P1/2}$)     & 0 \\
$N\bar K^*_{P3/2}$        & (5\er 4)&                 & Br($N\bar K^*_{P3/2}$)     & 0 \\
                               &             &                &                                   &\\     
\hline\hline\\[-1.5ex]
\end{tabular}\\
\begin{tabular}{lrr|lrr}
\multicolumn{5}{l}{\boldmath\fbox{\fbox{$\Lambda(1890){3/2^+}$}}
\unboldmath
\hfill****}\\[4ex]
\hline\hline\\[-2.7ex]
\multicolumn{3}{c|}{Pole parameters}&\multicolumn{3}{c}{Breit-Wigner parameters}\\[0.3ex]
\hline\\[-2.7ex]
\multicolumn{3}{l|}{M=1872\er 5\qquad\qquad$\Gamma$=101\er 10} &\multicolumn{3}{l}{M=1873\er 5\qquad $\Gamma$=103\er10}\\
\multicolumn{3}{l|}{norm. tran. res. (\%) for $\bar K N\to$ } & \multicolumn{3}{l}{Branchings (\%)}\\[0.2ex]
$N\bar K$                    & (30\er 6)&(0\er10)$^\circ$ & Br($N\bar K$)                   & 30\er 6 \\
$\Sigma\pi$              & (14\er5)&(148\er12 )$^\circ$ & Br($\Sigma\pi$)              &\qquad\ 6\er 2 \\
$\Xi K$                   & (6.5\er2)&(160\er30)$^\circ$ & Br($\Xi K$)                   & $\sim$1  \\
$\Lambda\omega_{P1/2}$   & (24\er6)&(15\er20)$^\circ$ & Br($\Lambda\omega_{P1/2}$)   & 0  \\
$\Lambda\omega_{P3/2}$   & (15\er8)&-(165\er20)$^\circ$ & Br($\Lambda\omega_{P3/2}$)   & 0  \\
$\Lambda\omega_{F3/2}$   & $\sim 0$& & Br($\Lambda\omega_{F3/2}$)   &  0  \\
$\Sigma^*\pi_P$      & (11\er5)&-(160\er45)$^\circ$ & Br($\Sigma^*\pi_P$)      & 6\er 3 \\
$\Sigma^*\pi_F$      & (10\er4)&  (10\er50)$^\circ$& Br($\Sigma^*\pi_F$)      & 4\er2 \\
$N\bar K^*_{P1/2}$        & (3\er 3)&                   & Br($N\bar K^*_{P1/2}$)     & $<1$ \\
$N\bar K^*_{P3/2}$        & (5\er 3)&  (180\er40)$^\circ$ & Br($N\bar K^*_{P3/2}$)     & $\sim 1$ \\
                               &             &                              &                                  &\\       
\hline\hline\\[-1.5ex]
\end{tabular}
&
\begin{tabular}{lrr|lrr}
\multicolumn{5}{l}{\boldmath\fbox{\fbox{$\Lambda(2070){3/2^+}$}}
\unboldmath
\hfill * (new)}\\[4ex]
\hline\hline\\[-2.7ex]
\multicolumn{3}{c|}{Pole parameters}&\multicolumn{3}{c}{Breit-Wigner parameters}\\[0.3ex]
\hline\\[-2.7ex]
\multicolumn{3}{l|}{M=2044\er20\qquad\quad\ $\Gamma$=360\er45} &\multicolumn{3}{l}{M=2070\er24\qquad $\Gamma$=370\er50}\\
\multicolumn{3}{l|}{norm. tran. res. (\%) for $\bar K N\to$ } & \multicolumn{3}{l}{Branchings (\%)}\\[0.2ex]
$N\bar K$                    & (15\er5)&-(37\er10)$^\circ$ & Br($N\bar K$)                    &\qquad 12\er5  \\
$\Sigma\pi$              & (10\er3)&-(47\er8)$^\circ$ & Br($\Sigma\pi$)              & 7\er3  \\
$\Xi K$                   & (11\er3)& (0\er25)$^\circ$ & Br($\Xi K$)                   & 7\er3  \\
$\Lambda\omega_{P1/2}$   & (10\er4)&(150\er17)$^\circ$ & Br($\Lambda\omega_{P1/2}$)   & 7\er4  \\
$\Lambda\omega_{P3/2}$   & (8\er4)& (20\er30)$^\circ$ & Br($\Lambda\omega_{P3/2}$)   & 3\er2  \\
$\Lambda\omega_{F3/2}$   & (4\er2)&-(175\er35)$^\circ$ & Br($\Lambda\omega_{F3/2}$)   & 1\er1  \\
$\Sigma^*\pi_P$      & (12\er7)&-(160\er55)$^\circ$ & Br($\Sigma^*\pi_P$)      & 10\er5 \\
$\Sigma^*\pi_F$      & (7\er4)&-(145\er50)$^\circ$ & Br($\Sigma^*\pi_F$)      & 2\er2 \\
$N\bar K^*_{P1/2}$        & (36\er7)&-(45\er30)$^\circ$ & Br($N\bar K^*_{P1/2}$)     & 42\er 8 \\
$N\bar K^*_{P3/2}$        & (16\er5)&(150\er35)$^\circ$ & Br($N\bar K^*_{P3/2}$)     & 14\er 6\\
$N\bar K^*_{F3/2}$        & (14\er8)&-(50\er30)$^\circ$ & Br($N\bar K^*_{F3/2}$)     & 10\er 6\\
\hline\hline\\[-1.5ex]
\end{tabular}\\
\begin{tabular}{lrr|lrr}
\multicolumn{5}{l}{\boldmath\fbox{\fbox{$\Lambda(1820){5/2^+}$}}
\unboldmath
\hfill****}\\[4ex]
\hline\hline\\[-2.7ex]
\multicolumn{3}{c|}{Pole parameters}&\multicolumn{3}{c}{Breit-Wigner parameters}\\[0.3ex]
\hline\\[-2.7ex]
\hline
\multicolumn{3}{l|}{M=1813\er 3\qquad\qquad$\Gamma$=78\er 7} &\multicolumn{3}{l}{M=1822\er 4\qquad\quad $\Gamma$=80\er8}\\
\multicolumn{3}{l|}{norm. tran. res. (\%) for $\bar K N\to$ } & \multicolumn{3}{l}{Branchings (\%)}\\[0.2ex]
$N\bar K$                    & (60\er12)&-(22\er5)$^\circ$ & Br($N\bar K$)                   & 58\er 12 \\
$\Sigma\pi$              & (34\er7)&(174\er5)$^\circ$ & Br($\Sigma\pi$)              & 19\er 4 \\
$\Xi K$                   & $\sim 0$  && Br($\Xi K$)                   & 0  \\
$\Lambda\omega_{P3/2}$   & (4\er4)& & Br($\Lambda\omega_{P3/2}$)   & 0  \\
$\Sigma^*\pi_P$      & (7\er2)&-(60\er50)$^\circ$ & Br($\Sigma^*\pi_P$)      & $\sim1$ \\
$\Sigma^*\pi_F$      & (11\er4)&(5\er45)$^\circ$                  & Br($\Sigma^*\pi_F$)      &\qquad\ 2\er1 \\
$N\bar K^*_{F1/2}$        & (2\er 2)&                 & Br($N\bar K^*_{F1/2}$)     &  0 \\
$N\bar K^*_{P3/2}$        & (35\er 15)&-(30\er45)$^\circ$                 & Br($N\bar K^*_{P3/2}$)     & 0 \\
$N\bar K^*_{F3/2}$        & (2\er 2)&                 & Br($N\bar K^*_{F3/2}$)     & 0 \\
\hline\hline\\[-1.5ex]
\end{tabular}
&
\begin{tabular}{lrr|lrr}
\multicolumn{5}{l}{\boldmath\fbox{\fbox{$\Lambda(2110){5/2^+}$}}
\unboldmath
\hfill **}\\[4ex]
\hline\hline\\[-2.7ex]
\multicolumn{3}{c|}{Pole parameters}&\multicolumn{3}{c}{Breit-Wigner parameters}\\[0.3ex]
\hline\\[-2.7ex]
\multicolumn{3}{l|}{M=2048\er10\qquad\quad\ $\Gamma$=255\er20} &\multicolumn{3}{l}{M=2086\er12 \qquad $\Gamma$=274\er25}\\
\multicolumn{3}{l|}{norm. tran. res. (\%) for $\bar K N\to$ } & \multicolumn{3}{l}{Branchings (\%)}\\[0.2ex]
$N\bar K$                    & (2.0\er 0.5)&(5\er15)$^\circ$ & Br($N\bar K$)                 & 2.0\er 0.5 \\
$\Sigma\pi$              & (13\er3)&(0\er15)$^\circ$ & Br($\Sigma\pi$)              &\qquad 88\er 20 \\
$\Xi K$                   & (0.5\er0.5)& & Br($\Xi K$)                   & $\sim$0  \\
$\Lambda\omega_{P1/2}$   & (1\er1)& & Br($\Lambda\omega_{P3/2}$)   & $<1$ \\
$\Lambda\omega_{P3/2}$   & (3\er1)&-(7\er16)$^\circ$ & Br($\Lambda\omega_{P3/2}$)   & 5\er2 \\
$\Lambda\omega_{F3/2}$   & (1\er1)& & Br($\Lambda\omega_{F3/2}$)   & $<1$  \\ &&&&\\ &&&&\\ &&&&\\
\hline\hline\\[-1.5ex]
\end{tabular}\\
\begin{tabular}{lrr|lrr}
\multicolumn{5}{l}{\boldmath\fbox{\fbox{$\Sigma(1620){1/2^-}$}}
\unboldmath
\hfill (*)}\\[4ex]
\hline\hline\\[-2.7ex]
\multicolumn{3}{c|}{Pole parameters}&\multicolumn{3}{c}{Breit-Wigner parameters}\\[0.3ex]
\hline\\[-2.7ex]
\multicolumn{3}{l|}{M=1680\er 8\qquad\qquad$\Gamma$=39\er 11} &\multicolumn{3}{l}{M=1681\er 6\qquad\quad\ \ $\Gamma$=40\er12}\\
\multicolumn{3}{l|}{norm. tran. res. (\%) for $\bar K N\to$ } & \multicolumn{3}{l}{Branchings (\%)}\\[0.2ex]
$N\bar K$                    & (11\er3)& (43\er20)$^\circ$ & Br($N\bar K$)                  &\qquad\quad\ 11\er 3 \\
$\Sigma\pi$              & (14\er3)&\quad -(90\er25)$^\circ$ & Br($\Sigma\pi$)          & 17\er 5 \\
$\Lambda\pi$             & (10\er3)&(75\er20)$^\circ$ & Br($\Lambda\pi$)         & 9\er 3 \\
$\Xi K$                   & (2\er1)&(120\er20)$^\circ$  & Br($\Xi K$)               &  0 \\
$\Lambda^*\pi$       & (12\er5)&(140\er40)$^\circ$ & Br($\Lambda^*\pi$)   &10\er5  \\
$\Sigma^*\pi$       &\ (1.5\er1)&(155\er40)$^\circ$ & Br($\Sigma^*\pi$)   & $<$1  \\
$N\bar K^*_{S}$        & (5\er 4)& & Br($N\bar K^*_{S}$)     & 0 \\
$N\bar K^*_{D}$\quad           & (1\er 1)& & Br($N\bar K^*_{D}$)     & 0 \\
&&&&\\
\hline\hline\\[-1.5ex]
\end{tabular}
&
\begin{tabular}{lrr|lrr}
\multicolumn{5}{l}{\boldmath\fbox{\fbox{$\Sigma(1750){1/2^-}$}}
\unboldmath
\hfill ****}\\[4ex]
\hline\hline\\[-2.7ex]
\multicolumn{3}{c|}{Pole parameters}&\multicolumn{3}{c}{Breit-Wigner parameters}\\[0.3ex]
\hline\\[-2.7ex]
\multicolumn{3}{l|}{M=1689\er11\qquad\quad\ $\Gamma$=206\er 18} &\multicolumn{3}{l}{M=1692\er 11\qquad\quad $\Gamma$=208\er 18}\\
\multicolumn{3}{l|}{norm. tran. res. (\%) for $\bar K N\to$ } & \multicolumn{3}{l}{Branchings (\%)}\\[0.2ex]
$N\bar K$                    & (46\er 9)&\  -(144\er 15)$^\circ$ & Br($N\bar K$)               &\qquad\qquad 46\er9 \\
$\Sigma\pi$              & (27\er5)&(100\er18)$^\circ$ & Br($\Sigma\pi$)          & 16\er4 \\
$\Sigma\eta$              & (5\er3)& & Br($\Sigma\eta$)          & 0 \\
$\Lambda\pi$             & (26\er6)&(115\er15)$^\circ$ & Br($\Lambda\pi$)         & 14\er5 \\
$\Xi K$                   & (2\er2)& & Br($\Xi K$)               &  0 \\
$\Lambda^*\pi$    & (15\er7)&-(25\er40)$^\circ$ & Br($\Lambda^*\pi$)   & 2\er1  \\
$\Sigma^*\pi$       & (4\er3)& & Br($\Sigma^*\pi$)   & $<$1  \\
$N\bar K^*_{S}$        & (5\er 3)&-(100\er35)$^\circ$ & Br($N\bar K^*_{S}$)     & 0 \\
$N\bar K^*_{D}$ \quad          & (2\er 2)& & Br($N\bar K^*_{D}$)     & 0 \\
\hline\hline\\[-1.5ex]
\end{tabular}\\
\end{tabular}
\end{scriptsize}
\end{table*}


\begin{table*}[pt]
Table~\ref{Results-1} continued.\\[2ex]
\renewcommand{\arraystretch}{1.2}\begin{scriptsize}
\begin{tabular}{l|l|}
\begin{tabular}{lrr|lrr}
\multicolumn{5}{l}{\boldmath\fbox{\fbox{$\Sigma(1900){1/2^-}$}}
\unboldmath
\hfill **}\\[4ex]
\hline\hline\\[-2.7ex]
\multicolumn{3}{c|}{Pole parameters}&\multicolumn{3}{c}{Breit-Wigner parameters}\\[0.3ex]
\hline\\[-2.7ex]
\multicolumn{3}{l|}{M=1936\er10\qquad\quad\ $\Gamma$=150\er 25} &\multicolumn{3}{l}{M=1938\er 12\qquad\quad $\Gamma$=155\er30}\\
\multicolumn{3}{l|}{norm. tran. res. (\%) for $\bar K N\to$ } & \multicolumn{3}{l}{Branchings (\%)}\\[0.2ex]
$N\bar K$                    &\quad (45\er9)&(90\er25)$^\circ$ & Br($N\bar K$)                  &\qquad\qquad 45\er 9 \\
$\Sigma\pi$              & (38\er8)&(95\er20)$^\circ$ & Br($\Sigma\pi$)          & 33\er7\\
$\Sigma\eta$              & (3\er1)&(20\er20) & Br($\Sigma\eta$)          & 1\er1 \\
$\Lambda\pi$             & (14\er5)& \ -(160\er50)$^\circ$ & Br($\Lambda\pi$)         & 6\er2 \\
$\Xi K$                   & (8\er5)&(75\er25)$^\circ$ & Br($\Xi K$)               &  3\er2 \\
$\Lambda^*\pi$       & (4\er2)&-(25\er40)$^\circ$ & Br($\Lambda^*\pi$)   & $<$1  \\
$\Sigma^*\pi$       & (16\er5)& (40\er30)$^\circ$ & Br($\Sigma^*\pi$)   & 7\er3  \\
$\Delta\bar K$         & (11\er4)&(60\er30)$^\circ$ & Br($\Delta\bar K$)      & 2.5\er1  \\
$N\bar K^*_{S}$        & (17\er6)&(50\er50)$^\circ$ & Br($N\bar K^*_{S}$)     & 7\er3 \\
$N\bar K^*_{D}$        & (5\er4)& & Br($N\bar K^*_{D}$)     & $<1$ \\
$\Sigma\eta$              & (3\er1)&(20\er20)$^\circ$ & Br($\Sigma\eta$)          &  $<$1\\
\hline\hline\\[-1.5ex]
\end{tabular}
&
\begin{tabular}{lrr|lrr}
\multicolumn{5}{l}{\boldmath\fbox{\fbox{$\Sigma(2160){1/2^-}$}}
\unboldmath
\hfill * (new)}\\[4ex]
\hline\hline\\[-2.7ex]
\multicolumn{3}{c|}{Pole parameters}&\multicolumn{3}{c}{Breit-Wigner parameters}\\[0.3ex]
\hline\\[-2.7ex]
\multicolumn{3}{l|}{M=2158\er25\qquad\quad\ $\Gamma=300^{+300}_{-60}$} &\multicolumn{3}{l}{M=2165\er 23\qquad $\Gamma=320^{+300}_{-60}$}\\
\multicolumn{3}{l|}{norm. tran. res. (\%) for $\bar K N\to$ } & \multicolumn{3}{l}{Branchings (\%)}\\[0.2ex]
$N\bar K$                    & (29\er8)&-(20\er35)$^\circ$ & Br($N\bar K$)                  &\qquad\qquad 29\er 7 \\
$\Sigma\pi$              & (14\er4)&-(5\er35)$^\circ$ & Br($\Sigma\pi$)          & 7\er 2 \\
$\Lambda\pi$             & (39\er8)&(85\er25)$^\circ$ & Br($\Lambda\pi$)         & 54\er 12 \\
$\Xi K$                   & (5\er2)&-(85\er35)$^\circ$ & Br($\Xi K$)               & $\sim$1  \\
$\Lambda^*\pi$       & (2.5\er1.5)&& Br($\Lambda^*\pi$)   & $<1$  \\
$\Sigma^*\pi$       & (3\er2)&& Br($\Sigma^*\pi$)   & $<1$  \\
$\Delta\bar K$         & (3.5\er2 )&-(30\er40)$^\circ$ & Br($\Delta\bar K_P$)   & $\sim$1  \\
$N\bar K^*_{S}$        & (9\er3)&-(40\er50)$^\circ$ & Br($N\bar K^*_{S}$)     & 3\er1 \\
$N\bar K^*_{D}$        & (4\er3)& & Br($N\bar K^*_{D}$)     & $\sim1$ \\
&&&&\\
&&&&\\
\hline\hline\\[-1.5ex]
\end{tabular}\\
\begin{tabular}{lrr|lrr}
\multicolumn{5}{l}{\boldmath\fbox{\fbox{$\Sigma(1670){3/2^-}$}}
\unboldmath
\hfill ****}\\[4ex]
\hline\hline\\[-2.7ex]
\multicolumn{3}{c|}{Pole parameters}&\multicolumn{3}{c}{Breit-Wigner parameters}\\[0.3ex]
\hline\\[-2.7ex]
\multicolumn{3}{l|}{M=1661\er 3\qquad\qquad$\Gamma$=52\er 6} &\multicolumn{3}{l}{M=1665\er 3\qquad\qquad $\Gamma$=54\er6}\\
\multicolumn{3}{l|}{norm. tran. res. (\%) for $\bar K N\to$ } & \multicolumn{3}{l}{Branchings (\%)}\\[0.2ex]
$N\bar K$                    & (10\er 2)&-(31\er12)$^\circ$ & Br($N\bar K$)                  &\qquad\ 10\er 2 \\
$\Sigma\pi$              & (25\er5)&-(25\er10)$^\circ$ & Br($\Sigma\pi$)              & 70\er 15 \\
$\Lambda\pi$             & (9\er3)&-(52\er12)$^\circ$ & Br($\Lambda\pi$)             & 9\er 2 \\
$\Xi K$                   & (2\er1)&(160\er20)$^\circ$ & Br($\Xi K$)                   &  0  \\
$\Lambda(1405)\pi$       & (3\er2)&(160\er15)$^\circ$ & Br($\Lambda(1405)\pi$)       & 1\er1  \\
$\Sigma \sigma$       & (8\er3)&-(25\er15)$^\circ$ & Br($\Sigma \sigma$)      & 7\er3  \\
$\Lambda^*\pi_P$     & (4\er2)&(120\er20)$^\circ$ & Br($\Lambda^*\pi_P$)     & $\sim$1  \\
$\Lambda^*\pi_F$     & (1\er1)&& Br($\Lambda^*\pi_F$)     & 0  \\
$\Delta\bar K_S$       & (1\er1)&& Br($\Delta\bar K_S$)        & 0  \\
$N\bar K^*_{S}$        & (5\er3)&(50\er60)$^\circ$ & Br($N\bar K^*_{S}$)     & 0 \\
$N\bar K^*_{D1/2}$        & (3\er2)& & Br($N\bar K^*_{D1/2}$)     & 0 \\
$N\bar K^*_{D3/2}$        & (1\er1)& & Br($N\bar K^*_{D3/2}$)     & 0 \\
\hline\hline\\[-1.5ex]
\end{tabular}
&
\begin{tabular}{lrr|lrr}
\multicolumn{5}{l}{\boldmath\fbox{\fbox{$\Sigma(1940){3/2^-}$}}
\unboldmath
\hfill ***}\\[4ex]
\hline\hline\\[-2.7ex]
\multicolumn{3}{c|}{Pole parameters}&\multicolumn{3}{c}{Breit-Wigner parameters}\\[0.3ex]
\hline\\[-2.7ex]
\multicolumn{3}{l|}{M=1856\er10\qquad\quad\ $\Gamma$=220\er 22} &\multicolumn{3}{l}{M=1878\er 12\qquad $\Gamma$=224\er 25}\\
\multicolumn{3}{l|}{norm. tran. res. (\%) for $\bar K N\to$ } & \multicolumn{3}{l}{Branchings (\%)}\\[0.2ex]
$N\bar K$                    & (3\er2)&-(95\er60)$^\circ$ & Br($N\bar K$)                  &\qquad\ \ 3\er 2 \\
$\Sigma\pi$              & (16\er4)&-(160\er15)$^\circ$ & Br($\Sigma\pi$)              & 86\er 21 \\
$\Lambda\pi$             & (4\er3)&(25\er25)$^\circ$ & Br($\Lambda\pi$)             & 6\er4 \\
$\Xi K$                   & (1\er1)& & Br($\Xi K$)                   & $\sim$ 0  \\
$\Lambda^*\pi_P$     & (1\er1)& & Br($\Lambda^*\pi_P$)     & $\sim0$ \\
$\Lambda^*\pi_F$     & $\sim 0$&\ & Br($\Lambda^*\pi_F$)     &  0  \\
$\Delta\bar K_S$       & (3\er1)&(120\er20)$^\circ$ & Br($\Delta\bar K_S$)        & 3\er1  \\
$N\bar K^*_{S}$        & (3\er2)&(20\er35)$^\circ$  & Br($N\bar K^*_{S}$)     & 3\er2 \\
$N\bar K^*_{D1/2}$        & (2\er1)& & Br($N\bar K^*_{D1/2}$)     & 1\er1 \\
$N\bar K^*_{D3/2}$        & (1\er1)& & Br($N\bar K^*_{D3/2}$)     & $\sim 0$ \\ \\ \\
\hline\hline\\[-1.5ex]
\end{tabular}\\
\begin{tabular}{lrr|lrr}
\multicolumn{5}{l}{\boldmath\fbox{\fbox{$\Sigma(2000){3/2^-}$}}
\unboldmath
\hfill * (new)}\\[4ex]
\hline\hline\\[-2.7ex]
\multicolumn{3}{c|}{Pole parameters}&\multicolumn{3}{c}{Breit-Wigner parameters}\\[0.3ex]
\hline\\[-2.7ex]
\multicolumn{3}{l|}{M=1995\er12\quad\qquad\ $\Gamma$=175\er 24} &\multicolumn{3}{l}{M=2005\er 14\qquad\ \ $\Gamma$=178\er23}\\
\multicolumn{3}{l|}{norm. tran. res. (\%) for $\bar K N\to$ } & \multicolumn{3}{l}{Branchings (\%)}\\[0.2ex]
$N\bar K$                    & (7\er 3)&-(115\er25)$^\circ$ & Br($N\bar K$)                  &\qquad\quad\ 7\er 3 \\
$\Sigma\pi$              & (4\er2)&(130\er22)$^\circ$ & Br($\Sigma\pi$)            & 3\er 2 \\
$\Lambda\pi$             & (6\er3)&(170\er25)$^\circ$ & Br($\Lambda\pi$)           & 5\er 2 \\
$\Xi K$                   & (4\er2)&-(120\er45)$^\circ$ & Br($\Xi K$)                 & 3\er2  \\
$\Lambda^*\pi_P$     & (3\er2)&(80\er35)$^\circ$ & Br($\Lambda^*\pi_P$)   & 2\er2  \\
$\Lambda^*\pi_F$     & (8\er5)&(150\er65)$^\circ$ & Br($\Lambda^*\pi_F$)   & 12\er6  \\
$\Sigma^*\pi_P$     & (4\er2)&(25\er45)$^\circ$ & Br($\Sigma^*\pi_P$)   & 3\er2  \\
$\Sigma^*\pi_F$     & (2\er2)&                         & Br($\Sigma^*\pi_F$)   & 2\er2  \\
$\Delta\bar K_S$       & (8\er4)&(0\er30)$^\circ$& Br($\Delta\bar K_S$)      & 11\er5 \\
$\Delta\bar K_D$       & (2\er2)& & Br($\Delta\bar K_D$)      & 1\er1  \\
$N\bar K^*_{S}$        & (12\er3)&-(60\er60)$^\circ$  & Br($\bar NK^*_{S}$)     & 27\er7 \\
$N\bar K^*_{D1/2}$        & (8\er4)&(55\er60)$^\circ$ & Br($N\bar K^*_{D1/2}$)     &13\er6 \\
$N\bar  K^*_{D3/2}$        & (8\er4)&(15\er60)$^\circ$  & Br($N\bar K^*_{D3/2}$)     &13\er6 \\
\hline\hline\\[-1.5ex]
\end{tabular}
&
\begin{tabular}{lrr|lrr}
\multicolumn{5}{l}{\boldmath\fbox{\fbox{$\Sigma(1775){5/2^-}$}}
\unboldmath
 \hfill ****}\\[4ex]
\hline\hline\\[-2.5ex]
\multicolumn{3}{c|}{Pole parameters}&\multicolumn{3}{c}{Breit-Wigner parameters}\\
[0.5ex]
\hline
\multicolumn{3}{l|}{M=1767\er4\qquad\qquad$\Gamma$=122\er 8} &\multicolumn{3}{l}{M=1776\er 4\qquad $\Gamma$=124\er 8}\\
\multicolumn{3}{l|}{norm. tran. res. (\%) for $\bar K N\to$ } & \multicolumn{3}{l}{Branchings (\%)}\\[0.2ex]
$N\bar K$                    & (44\er 9)&-(17\er 10)$^\circ$ & Br($N\bar K$)                  &\quad\ 43\er 9 \\
$\Sigma\pi$              & (13\er3)&(10\er12)$^\circ$ & Br($\Sigma\pi$)              & 3.5\er1.0 \\
$\Lambda\pi$             & (47\er10)&(130\er15)$^\circ$ & Br($\Lambda\pi$)             & 49\er10 \\
$\Xi K$                   & (2\er1)&-(90\er35)$^\circ$ & Br($\Xi K$)                   & 0  \\
$\Lambda^*\pi_P$     & (9\er3)&(10\er30)$^\circ$ & Br($\Lambda^*\pi_P$)     & 2\er1 \\
$\Lambda^*\pi_F$     & (1\er1)& & Br($\Lambda^*\pi_F$)     & $\sim$0 \\
$\Delta\bar K_D$       & (2\er2)& & Br($\Delta\bar K_D$)        & $\sim$0  \\
$N\bar K^*_{D1/2}$        & (4\er2)&-(100\er60)$^\circ$ & Br($N\bar K^*_{D1/2}$)     & 0 \\
$N\bar K^*_{D3/2}$        & (9\er6)& (10\er50)$^\circ$& Br($N\bar K^*_{D3/2}$)     & 0 \\
$N\bar K^*_{G3/2}$        & (4\er2)&-(100\er60)$^\circ$ & Br($N\bar K^*_{D1/2}$)     & 0 \\ \\ \\ \\
\hline\hline\\[-1.5ex]
\end{tabular}\\
\begin{tabular}{lrr|lrr}
\multicolumn{5}{l}{\boldmath\fbox{\fbox{$\Sigma(2100){7/2^-}$}}
\unboldmath
\hfill *}\\[4ex]
\hline\hline\\[-2.7ex]
\multicolumn{3}{c|}{Pole parameters}&\multicolumn{3}{c}{Breit-Wigner parameters}\\[0.3ex]
\hline\\[-2.7ex]
\multicolumn{3}{l|}{M=2093\er16\qquad\quad\ $\Gamma$=210\er35} &\multicolumn{3}{l}{M=2146\er17 \qquad $\Gamma$=260\er40}\\
\multicolumn{3}{l|}{norm. tran. res. (\%) for $\bar K N\to$ } & \multicolumn{3}{l}{Branchings (\%)}\\[0.2ex]
$N\bar K$                    & (9\er2)&-(110\er15)$^\circ$ & Br($N\bar K$)                    &\quad\qquad\qquad 8\er2  \\
$\Sigma\pi$              & (4\er2)&-(50\er20)$^\circ$ & Br($\Sigma\pi$)              & 2\er1  \\
$\Lambda\pi$             & (3\er2)&-(100\er25)$^\circ$ & Br($\Lambda\pi$)             & 1.5\er1  \\
$\Xi K$                   &\quad\ \ (1\er0.5)&-(120\er35)$^\circ$ & Br($\Xi K$)                   & $<$1  \\
\hline\hline\\[-1.5ex]
\end{tabular}&
\begin{tabular}{lrr|lrr}
\\[0.5ex]\multicolumn{5}{l}{\boldmath$\Sigma(2100)7/2^-$\boldmath\bf continued:}\\
\hline\hline\\[-2.7ex]
\multicolumn{3}{c|}{Pole parameters}&\multicolumn{3}{c}{Breit-Wigner parameters}\\[0.3ex]
\hline\\[-2.7ex]
\multicolumn{3}{l|}{norm. tran. res. (\%) for $\bar K N\to$ } & \multicolumn{3}{l}{Branchings (\%)}\\[0.2ex]
$\Lambda^*\pi_F$     & (2\er1)&-(100\er30)$^\circ$ & Br($\Lambda^*\pi_P$)     &\quad\quad 1\er1 \\
$\Lambda^*\pi_H$     & (1\er1)& & Br($\Lambda^*\pi_F$)     & $\sim$0 \\
$\Sigma^*\pi_D$     & (10\er3)&-(60\er30)$^\circ$ & Br($\Lambda^*\pi_P$)     & 12\er6 \\
$\Sigma^*\pi_G$     & (3\er1)&-(50\er30)$^\circ$ & Br($\Lambda^*\pi_F$)     & $\sim$1 \\
$\Delta\bar K_G$       & (4\er2 )&(75\er35)$^\circ$ & Br($\Delta\bar K_D$)        & 1\er1  \\
$N\bar K^*_{D3/2}$        & (8\er4)& (20\er50)$^\circ$& Br($N\bar K^*_{D3/2}$)     & 6\er3 \\
\hline\hline\\[-1.5ex]
\end{tabular}

\end{tabular}
\end{scriptsize}
\end{table*}


\begin{table*}[pt]
Table~\ref{Results-1} continued.\\[2ex]
\renewcommand{\arraystretch}{1.2}\begin{scriptsize}
\begin{tabular}{l|l|}
\begin{tabular}{lrr|lrr}
\multicolumn{5}{l}{\boldmath\fbox{\fbox{$\Sigma(1660){1/2^+}$}}
\unboldmath
\hfill ***}\\[4ex]
\hline\hline\\[-2.7ex]
\multicolumn{3}{c|}{Pole parameters}&\multicolumn{3}{c}{Breit-Wigner parameters}\\[0.3ex]
\hline\\[-2.7ex]
\multicolumn{3}{l|}{M=1585\er 20\qquad\quad\ $\Gamma=290^{+140}_{-\ 40}$} &\multicolumn{3}{l}{M=1665\er20\quad$\Gamma=300^{+140}_{-\ 40}$}\\
\multicolumn{3}{l|}{norm. tran. res. (\%) for $\bar K N\to$ } & \multicolumn{3}{l}{Branchings (\%)}\\[0.2ex]
$N\bar K$                    & (7\er3)&-(165\er35)$^\circ$ & Br($N\bar K$)                &\quad\qquad\ 7\er3 \\
$\Sigma\pi$              & (17\er4)&(150\er20)$^\circ$ & Br($\Sigma\pi$)          & 37\er10 \\
$\Lambda\pi$             & (16\er5)&(0\er25)$^\circ$ & Br($\Lambda\pi$)         & 35\er12 \\
$\Lambda(1405)\pi$       & (6\er3)&-(90\er25)$^\circ$ & Br($\Lambda(1405)\pi$)   & 4\er 2 \\
$\Sigma \sigma$      & (14\er6)&-(150\er30)$^\circ$ & Br($\Sigma \sigma$)  & 20\er8 \\
$\Lambda^*\pi$       & (4\er2 )&\-(5\er20)$^\circ$ & Br($\Lambda^*\pi$)   & $<$1  \\ &&&&\\ &&&&\\
&&&&\\ &&&&\\ &&&&\\ &&&&\\ &&&&\\
\hline\hline\\[-1.5ex]
\end{tabular}
&
\begin{tabular}{lrr|lrr}
\multicolumn{5}{l}{\boldmath\fbox{\fbox{$\Sigma(2230){3/2^+}$}}
\unboldmath
\hfill * (new)}\\[4ex]
\hline\hline\\[-2.7ex]
\multicolumn{3}{c|}{Pole parameters}&\multicolumn{3}{c}{Breit-Wigner parameters}\\[0.3ex]
\hline\\[-2.7ex]
\multicolumn{3}{l|}{M=2234\er25\qquad\quad\ $\Gamma$=340\er45}    &\multicolumn{3}{l}{M=2240\er27\qquad $\Gamma$=345\er50}\\
\multicolumn{3}{l|}{norm. tran. res. (\%) for $\bar K N\to$ }   & \multicolumn{3}{l}{Branchings (\%)}\\[0.2ex]
$N\bar K$                    & (7\er2)&(25\er15)$^\circ$  & Br($N\bar K$)                &\qquad \quad 6\er2  \\
$\Sigma\pi$              & (3\er2 )&(180\er25) & Br($\Sigma\pi$)          & 2\er1  \\
$\Lambda\pi$             & (11\er5)&-(16\er10)$^\circ$ & Br($\Lambda\pi$)         & 12\er6  \\
$\Xi K$                   & (4\er2)&(155\er20)$^\circ$  & Br($\Xi K$)               & 2\er1  \\
$\Lambda^*\pi_S$     & (12\er5)&-(80\er25)$^\circ$  & Br($\Lambda^*\pi_S$) & 14\er5  \\
$\Lambda^*\pi_D$     & (3\er2)&(160\er30)$^\circ$  & Br($\Lambda^*\pi_D$) & $\sim$1  \\
$\Sigma^*\pi_P$     & (5\er2)& (60\er25)$^\circ$  & Br($\Sigma^*\pi_P$) & 4\er4  \\
$\Sigma^*\pi_F$     & (5\er3)&-(70\er20)$^\circ$  & Br($\Sigma^*\pi_F$) & 3\er2  \\
$\Delta\bar K_P$       & (11\er4)&(60\er15)$^\circ$ & Br($\Delta\bar K_P$)    &14\er5  \\
$\Delta\bar K_F$       & (7\er3)&(90\er25)$^\circ$ & Br($\Delta\bar K_F$)    &8\er2  \\
$N\bar K^*_{P1/2}$        & (8\er4)&(40\er45)$^\circ$ & Br($N\bar K^*_{F1/2}$)     & 8\er 3 \\
$N\bar K^*_{P3/2}$        & (14\er3)&-(40\er45)$^\circ$ &Br($N\bar K^*_{F3/2}$)     &26\er5 \\
$N\bar K^*_{F3/2}$        & (5\er3)& (35\er30)$^\circ$ &Br($N\bar K^*_{F3/2}$)     & 4\er2 \\
\hline\hline\\[-1.5ex]
\end{tabular}\\
\begin{tabular}{lrr|lrr}
\multicolumn{5}{l}{\boldmath\fbox{\fbox{$\Sigma(1915){5/2^+}$}}
\unboldmath
\hfill ****}\\[4ex]
\hline\hline\\[-2.7ex]
\multicolumn{3}{c|}{Pole parameters}&\multicolumn{3}{c}{Breit-Wigner parameters}\\[0.3ex]
\hline\\[-2.7ex]
\multicolumn{3}{l|}{M=1908\er 7\qquad\qquad$\Gamma$=98\er 12} &\multicolumn{3}{l}{M=1918\er 6\qquad\quad $\Gamma$=102\er 12}\\
\multicolumn{3}{l|}{norm. tran. res. (\%) for $\bar K N\to$ } & \multicolumn{3}{l}{Branchings (\%)}\\[0.2ex]
$N\bar K$                    & (8\er2)&-(33\er15)$^\circ$ & Br($N\bar K$)                & \qquad\quad 8\er2 \\
$\Sigma\pi$              & (9\er2)&-(180\er12)$^\circ$ & Br($\Sigma\pi$)          & 10\er2 \\
$\Lambda\pi$             & (7\er2)&-(170\er20)$^\circ$ & Br($\Lambda\pi$)         & 6\er2 \\
$\Xi K$                   & (2\er1)&-(65\er35)$^\circ$ & Br($\Xi K$)               & $<$1  \\
$\Lambda^*\pi_D$     & (8\er2)&-(105\er50)$^\circ$& Br($\Lambda^*\pi_D$) & 8\er 2  \\
$\Lambda^*\pi_G$     & (1\er1)&& Br($\Lambda^*\pi_G$) &  $\sim$0   \\
$\Sigma^*\pi_P$     & (2\er2)&& Br($\Lambda^*\pi_P$) & 2\er2  \\
$\Sigma^*\pi_F$     & (5\er3)&-(30\er50)$^\circ$ & Br($\Lambda^*\pi_F$) &  4\er 2   \\
$\Delta\bar K_P$       & (12\er3)&-(10\er20)$^\circ$ & Br($\Delta\bar K_P$)    & 16\er5 \\
$\Delta\bar K_F$       & (7\er2)&-(35\er25)$^\circ$ & Br($\Delta\bar K_F$)    & 5\er3  \\
$N\bar K^*_{F1/2}$        & (7\er4)&-(60\er45)$^\circ$ & Br($N\bar K^*_{F1/2}$)     & 5\er 3 \\
$N\bar K^*_{F3/2}$        & (7\er3)&-(40\er45)$^\circ$ &Br($N\bar K^*_{F3/2}$)     & 5\er2 \\
\hline\hline\\[-1.5ex]
\end{tabular}
&
\begin{tabular}{lrr|lrr}
\multicolumn{5}{l}{\boldmath\fbox{\fbox{$\Sigma(2030){7/2^+}$}}
\unboldmath
\hfill ****}\\[4ex]
\hline\hline\\[-2.7ex]
\multicolumn{3}{c|}{Pole parameters}&\multicolumn{3}{c}{Breit-Wigner parameters}\\[0.3ex]
\hline\\[-2.7ex]
\multicolumn{3}{l|}{M=2014\er 6\qquad\qquad$\Gamma$=172\er 12} &\multicolumn{3}{l}{M=2032\er 6\qquad $\Gamma$=177\er 12}\\
\multicolumn{3}{l|}{norm. tran. res. (\%) for $\bar K N\to$ } & \multicolumn{3}{l}{Branchings (\%)}\\[0.2ex]
$N\bar K$                    & (20\er4)&-(38\er8)$^\circ$ & Br($N\bar K$)                &\quad  20\er4 \\
$\Sigma\pi$              & (7\er2)&(165\er12)$^\circ$ & Br($\Sigma\pi$)          &  2.5\er0.8 \\
$\Lambda\pi$             & (18\er4)&-(22\er12)$^\circ$ & Br($\Lambda\pi$)         & 17\er 4 \\
$\Xi K$                   & (1\er1)& & Br($\Xi K$)               & $<$1  \\
$\Lambda^*\pi_D$     & (3\er2)&-(100\er40)$^\circ$ & Br($\Lambda^*\pi_D$) & $\sim 1$  \\
$\Lambda^*\pi_G$     & (2\er2 )&& Br($\Lambda^*\pi_G$) & $<1$  \\
$\Sigma^*\pi_F$     & (4\er3)& & Br($\Sigma^*\pi_F$)  & 1\er1  \\
$\Delta\bar K_F$       & (16\er6)&-(130\er20)$^\circ$ & Br($\Delta\bar K_F$)    & 15\er5 \\
$\Delta\bar K_H$       & (4\er2)&-(130\er35)$^\circ$ & Br($\Delta\bar K_H$)    & 1\er1  \\
$N\bar K^*_{F1/2}$        & (2\er2)& & Br($N\bar K^*_{F1/2}$)     & $<1$ \\
$N\bar K^*_{F3/2}$        & (16\er9)&-(160\er40)$^\circ$& Br($N\bar K^*_{F3/2}$)     & 14\er8 \\ \\
\hline\hline\\[-1.5ex]
\end{tabular}
\end{tabular}
\end{scriptsize}
\end{table*}

\paragraph{\boldmath$\Lambda(1800)1/2^-$:} 
The $\Lambda(1800)1/2^-$  Breit-Wigner properties are fully compatible with RPP values, the pole properties are, however,
inconsistent. The real part of the pole position was determined in Ref.~\cite{Zhang:2013sva} to 1729\,MeV 
while we find (1809\er9)\,MeV. The imaginary part is however consistent. 
The product BR for $K^-p\to\Lambda(1800)1/2^-\to\Sigma(1385)\pi$
from Ref.~\cite{Zhang:2013sva} is comparable, 
our $\Sigma\pi$ BR is considerably larger. The values 
from Ref.~\cite{Zhang:2013sva} are consistent with those reported in \cite{Cameron:1978qi} and \cite{Gopal:1976gs}.  
Our BRs add up to (87\er11)\%.\vspace{-6mm}

\paragraph{\boldmath$\Lambda(1810)1/2^+$:} 
The $\Lambda(1810)1/2^+$  pole position 
was determined in Ref.~\cite{Zhang:2013sva} to (1780 - i32)\,MeV, in Ref.~\cite{Kamano:2015hxa} (solution A)
to [(2097$^{+40}_{-1})$ - i(83$^{+32}_{-6})]$  or to $[(1841^{+3}_{-4}) - i(31^{+3}_{-2}) ]$ MeV 
(solution B). 
We find [(1773\er7) - i(19\er7)]\,MeV. Our Breit-Wigner mass and width 
are consistent with RPP. Our $N\bar K$ BR of (2.5\er1.3)\% is much below
 the 20\% to 50\% RPP range. Instead, we find large contributions from $\Sigma(1385)\pi$, 
$\Lambda\sigma$, and $\Sigma\pi$. The $\Sigma(1385)\pi$ branching ratio is also 
found to be large in~\cite{Kamano:2015hxa}.  Our $\Sigma\pi$ BR is compatible with RPP.  
The sum of all BR's is 65-100\%. Larger discrepancies are also seen
in the residues and product branching ratios. 
\vspace{-6mm}

\paragraph{\boldmath$\Lambda(1820)5/2^+$:} 
Our $\Lambda(1820)5/2^+$ mass, with, pole position  
and branching ratios are consistent with RPP values. 
The resonance has a large elasticity: the BR for decays into $N\bar K$ is (58\er12)\%. 
Decays into $\pi\Sigma$ are observed with (19\er4)\%, 
and into $\Sigma(1385)\pi$ with 2\er1\%. These values are not incompatible with 
RPP, the BR sum yields (80\er13)\%. The transition residues from 
Kamano et al. \cite{Kamano:2015hxa} are often in good (sometimes in fair) agreement with our findings.\vspace{-6mm}

\paragraph{\boldmath$\Lambda(1830)5/2^-$:}
Mass, width and most branching ratios of $\Lambda(1830)5/2^-$ are fully compatible with the ranges
given in RPP.  The elasticity is small: our $\Lambda(1830)5/2^-\to N\bar K$ BR is (5.5\er1.0)\%.
But there is a large coupling to $\Sigma\pi$, with a BR of (42\er8)\%. Ref.~\cite{Kamano:2015hxa}
reports 0.6\% and 1.7\% for these two numbers but a very large BR for $\Sigma(1385)\pi$ decays, 
(52\er6)\%. The transition residue for the latter transition is very small (2.37\%). 
The two numbers seem inconsistent. 
In Ref.~\cite{Zhang:2013sva}, a BR of (52\er6)\% is given for the $(\Sigma(1385)\pi)_D$ BR, 
Ref.~\cite{Kamano:2015hxa}
reports 13.4\%, we find (20\er8)\%. The  $\Xi K$ BR of 56.2\% reported in 
Ref.~\cite{Kamano:2015hxa} is not confirmed.
The BR sum of (70\er12)\% indicates some missing intensity.\vspace{-6mm}

\paragraph{\boldmath$\Lambda(1890)3/2^+$:}
The RPP values for mass, width, pole position, and the decay modes of $\Lambda(1890)3/2^+$ into
$N\bar K$ and $\Sigma\pi$ are well confirmed by us. 
We find a strong coupling of $\Lambda(1890)3/2^+\to \Lambda\omega$, the corresponding BR 
vanishes, however, since the sum of $\Lambda$ and $\omega$ masses just exceed the $\Lambda(1890)3/2^+$ mass.
We do not confirm the large $\bar K N\to \Lambda(1890)3/2^+\to\Sigma(1385)\pi$ 
and $\bar K N\to \Lambda(1890)3/2^+\to N\bar K^*$  transition residues from Ref.~\cite{Kamano:2015hxa}. 
There is sizable missing intensity; the BR sum is (48\er7)\% only.\vspace{-6mm}

\paragraph{\boldmath$\Lambda(2070)3/2^+$:}
The $\Lambda(2070)3/2^+$ hyperon is a new resonance with a large coupling to
$N\bar K^*$. The sum of the observed BR's amounts to $\sim$100\%.\vspace{-6mm}

\paragraph{\boldmath$\Lambda(2080)5/2^-$:}
The $\Lambda(2080)5/2^-$ hyperon is a further new hyperon seen with a BR sum of  72-100\%.
\vspace{-6mm}

\paragraph{\boldmath$\Lambda(2100)7/2^-$:} 
This resonance has well defined properties: mass, width, pole position and BR from most analyses 
reported in RPP and our values are consistent. 
A sizable fraction of all decay modes is missing: the sum of measured BR's is (32\er6)\%. The resonance is not
reported in~\cite{Kamano:2015hxa}.\vspace{-6mm}

\paragraph{\boldmath$\Lambda(2110)5/2^+$:}
We find $\Lambda(2110)5/2^+$ with a very large $\Sigma\pi$ BR of (88\er12)\% and little elasticity:
the BR into $N\bar K$ is (2.0\er 0.4)\% only. The sum of all observed BR's is 75-100\%. Note that experimentally, 
the transition $\bar K N\to \Lambda(2110)5/2^+\to \pi\Sigma$
is determined. A factor 2 of the $N\bar K$ BR would change the $\Sigma\pi$ BR by a factor 2.
This would make our observations and those of other groups compatible.   
The Kent group~\cite{Zhang:2013sva} and Cameron {\it et al.} \cite{Cameron:1978qi} find large
contributions from $N\bar K^*$ decays which are not seen by us. The Kent pole mass of 1970\,MeV is low compared to our
finding: (2048\er10)\,MeV.  The pole widths are consistent. \vspace{-6mm}

\paragraph{\boldmath$\Lambda(2090)1/2^-$:}
Finally, we come to a further resonance-like structure which we call
the $\Lambda(2090)1/2^-$. We observe this state with a mass of (2085\er14)\,MeV and a
very broad width of (428\er16)\,MeV. Even though the state is statistically highly significant, we do not
consider this to be a genuine resonance. Rather we believe it to represent a large number of weak
resonances which are expected above 2000\,MeV but which cannot be identified with the presently
available data base. Its properties are not given in Table~\ref{Results-1}.

\subsection{The \boldmath$\Sigma$ hyperons}

The $\Sigma(1620)1/2^-$ is a 1* resonance. It is discussed below jointly with $\Sigma(1750)1/2^-$.\vspace{-6mm}

\paragraph{\boldmath $\Sigma(1660)1/2^+$:}
Our $\Sigma(1660)1/2^+$ has a mass  which is fully compatible with RPP values while our width of \\
(300$^{+140}_{-\ 40})$\,MeV is outside of the RPP range of 40-200\,MeV.  It decays with high probability
to $\Sigma\pi$ -- (37\er10)\%  -- and $\Lambda\pi$ -- (35\er12)\%, and only with (7\er3)\% to $N\bar K$, just reaching
the 10\% to 30\% RPP range. Kamano et al. \cite{Kamano:2015hxa}
find a BR for $\Sigma\pi$ much stronger (86.5\% ) than the one for $\Lambda\pi$ (12.8\%). 
There is no evidence for this resonance from Ref.~\cite{Zhang:2013sva}. 
Our branching ratios add up to 85-100\%. \vspace{-6mm}

\paragraph{\boldmath $\Sigma(1670)3/2^-$:}
Good compatibilty is obtained for all properties of $\Sigma(1670)3/2^-$. However, we do not find 
significant evidence for $\Sigma(1385)\pi$ decays as reported in Ref.~\cite{Kamano:2015hxa} 
while we find some small contribution from $N\bar K^*$ decays. The sum of our BR is 82-100\%. \vspace{-6mm}

\paragraph{\boldmath $\Sigma(1620)1/2^-$} and {\boldmath $\Sigma(1750)1/2^-$:\unboldmath}
The $\Sigma(1620)1/2^-$ to $\Sigma(1750)1/2^-$  region is problematic. If we assume no resonance, 
the fit is unacceptable. A fit with one $1/2^-$ resonance only returns a mass of $M$=(1692\er11)\,MeV and 
$\Gamma$=(208\er18)\,MeV. 
We tentatively identify this resonance with $\Sigma(1750)1/2^-$. 
The real part of our pole position agrees with the ones determined in  
Refs.~\cite{Zhang:2013sva} and~\cite{Kamano:2015hxa}, our imaginary part is larger: we find
$\Gamma_{\rm pole}$=(206\er18)\,MeV instead of 158\,MeV~\cite{Zhang:2013sva} or 
(86$^{+14}_{-\ 4})$\,MeV~\cite{Kamano:2015hxa}. 
Our Breit-Wigner mass does not fall into the range
quoted in the RPP. Also the BRs are inconsistent: our BR for $N\bar K$ is at the upper limit but still compatible
with RPP. For the $\Sigma\pi$ BR, RPP quotes less than 8\%, Kamano et al.~\cite{Kamano:2015hxa} find
37.3\%, we find (16\er4)\%. The BR for $\Sigma(1750)1/2^-\to \Lambda\pi$ decays, the RPP quotes 
{\it seen}, Ref.~\cite{Kamano:2015hxa} finds  43.5\%, we find (14\er5)\%. 
The RPP quotes 15\% to 55\% 
for the $\Sigma(1750)1/2^-\to \Sigma\eta$ BR; there is, however, no measurement listed in the RPP 
supporting this number except for the transition strength 
$(\Gamma_i\Gamma_f)^{1/2} /\Gamma_{\rm tot}$ for the
$N\bar K\to\Sigma(1750)\to\Sigma\eta$~\cite{Jones:1974si} quoting (23\er1)\%. 
By our definition the BR for $\Sigma(1750)1/2^-\to \Sigma\eta$ vanishes. We find a mass of (1692\er11)\,MeV, 
which is below $M_\eta + M_\Sigma$. 
Our BRs  add up to (78\er11)\%.
A fit with two resonances gives a small but significant improvement for a second narrow resonance 
which is found only slightly below $\Sigma(1750)1/2^-$. We list this resonance under $\Sigma(1620)1/2^-$
even though these are likely different objects. We find a sum of branching ratios of (47\er8)\%. 
\vspace{-6mm}

\paragraph{\boldmath $\Sigma(1775)5/2^-$:}
Our $\Sigma(1775)5/2^-$ properties are mostly consistent with those from the RPP. Mass, width
and pole position are close to the RPP central values. However, we
observe a $\Lambda\pi$ BR of (49\er3)\% (instead of the RPP range of 14\% to 20\%). 
We do not observe its decay into $\Sigma(1385)\pi$ which is strongly  
(39.2\%) contributing 
in Ref.~\cite{Kamano:2015hxa}. The sum of our BRs exceeds  84\%. \vspace{-6mm}

\paragraph{\boldmath $\Sigma(1900)1/2^-$:}  
This resonance was first suggested by the Kent group~\cite{Zhang:2013sva} with $M$=(1900\er21)\,MeV,
$\Gamma$=(191\er47) MeV, a large elasticity with a $N\bar K$ BR of (67\er17)\% and to $\Sigma\pi$
of (10\er5)\%. We find $M$=(1938\er12)\,MeV,
$\Gamma$=(155\er30) MeV, a  $N\bar K$ BR of (45\er9)\% and $\Sigma\pi$ BR of of (33\er7)\%. 
In spite of some discrepancies, we consider this result as a confirmation of the Kent result. 
Our BRs add up to more than  92\%. 
\vspace{-6mm}

\paragraph{\boldmath $\Sigma(1915)5/2^+$:}
The results on the $\Sigma(1915)5/2^+$ mass, width and pole position agree mostly well with RPP values.
RPP reports a $N\bar K$ BR in the range from 5\% to 15\%, consistent with our (8\er2)\%. 
$\Lambda\pi$ and $\Sigma\pi$ are {\it seen}. Kamano et al. \cite{Kamano:2015hxa}
find a BR for $\Lambda\pi$ decays almost consitent with our value but a very large 
$\Sigma\pi$ BR (67.8\%) which we do not confirm: we find (10\er2)\%. The normalized
residues for quasi-two-body decay modes of~\cite{Kamano:2015hxa} show some difference but 
have a similar strength. We find a sum of BR's of (69\er 9)\%.\vspace{-6mm}

\paragraph{\boldmath $\Sigma(1940)3/2^-$:} 
Our $\Sigma(1940)3/2^-$  mass of (1878\er12)\,MeV falls ouside of the 1900 -- 1950\,MeV range given in the RPP,
the widths are compatible. In our analysis, it has a very large coupling to $\Sigma\pi$,(86\er21)\%. With our small
branching ratio for $N\bar K$ decays of (3\er2)\%, the 86\% are not incompatible with earlier findings
on the transition element  $\sqrt{\Gamma_i\Gamma_f}/\Gamma_{total}$ for $\bar K N\to \Sigma(1940)3/2^-\to\Sigma\pi$ reported 
in the RPP.  $\Sigma(1940)3/2^-$ was neither seen in~\cite{Zhang:2013sva} nor in~\cite{Kamano:2015hxa}. 
The sum of all observed BR's amounts to 80-100\%. \vspace{-6mm}

\paragraph{\boldmath $\Sigma(2000)3/2^-$:} 
In this partial wave, the RPP lists one resonance above 1670\,MeV which is called $\Sigma(1940)3/2^-$. We find
two states: one at (1878\er12)\,MeV which we identify with $\Sigma(1940)3/2^-$
and a new one at (2005\er14)\,MeV. The latter one shows a very significant $N\bar K*$ decay mode. 
The sum of the BRs of the new resonance amounts to 86-100\%.  \vspace{-6mm}

\paragraph{\boldmath $\Sigma(2030)7/2^+$:}
The results on this resonance on Mass, width and pole position are mostly consistent,  even
though Ref.~\cite{Kamano:2015hxa} gives a somewhat smaller pole width. The BRs for two-body reactions
are mostly consistent as well, only the BRs for quasi-two final-states like $\Sigma(1385)\pi$ or $\Delta \bar K^*$
differ significantly. The BRs add up (72\er11)\%.\vspace{-6mm}

\paragraph{\boldmath $\Sigma(2100)7/2^-$:}
Little was known about $\Sigma(2100)7/2^-$. We observe this resonance at M=(2146\,\er\,17)\,MeV, $\Gamma$\,= (260\er40)\,MeV 
and with a BR sum of (33\er8)\% only. \vspace{-6mm}

\paragraph{\boldmath $\Sigma(2160)1/2^-$:}  
Up to the 2014 edition, the RPP listed under $\Sigma(2000)1/2^-$ all reported resonances
 above $\Sigma(1750)$ $1/2^-$ in this partial wave.
Their masses range from 1755 MeV to 2004\,MeV.
When the analysis of the Kent group was published~\cite{Zhang:2013sva},  a new entry for $\Sigma(1900)1/2^-$
was nevertheless created. We now find weak evidence for a further state at (2165\er23)\,MeV and a width of
(320$^{+300}_{-\ 60}$)\,MeV. It has a large BR to $\Lambda\pi$: (54\er 12)\%. 
With its $N\bar K$ BR of (29\er7)\% and $\Sigma\pi$ BR of
(7\er 2)\%, the sum yields 79-100\%. These properties do not resemble any of
the RPP entries under  $\Sigma(2000)1/2^-$ and we list it as {\it new} resonance. 
It may have a very large width of up to 600\,MeV and 
could play the same role as $\Lambda(2090)1/2^-$: as a
resonance which represents a large number of unidentified resonance above 2100\,MeV.  However, it also might 
have a more natural width of 240\,MeV; hence we keep it as possible new resonance. 
\vspace{-6mm}

\paragraph{\boldmath $\Sigma(2230)3/2^+$:} 
This is a new resonance which we observe at M=(2240\er27)\,MeV and a width of $\Gamma$=(345\er50) MeV. 
It is seen in several decay modes:
$N\bar K$ with (6\er2)\%, $\Lambda\pi$ with (12\er6)\%, $\Xi K$ with (2\er1)\%, and
$\Lambda(1520)\pi$ with (14\er5)\%, and $\Delta(1232)\bar K$ with (22\er5)\%, and $N\bar K^*$ with (38\er6)\%. 
The BRs add to 91-100\%.

\section{\label{sec:class}Classification of hyperon resonances}

\subsection{\label{sec:sym}Symmetries}
\paragraph{The total wave function:} In quark models, baryons are treated as objects
composed of three (constituent) quarks.
The Pauli principle demands that the total wave function should be antisymmetric with respect
to the exchange of any pair of two quarks. The color singlet wave function for three
quarks is antisymmetric, hence the spin-flavor configuration of a baryon
has to be combined with spatial wave functions of the
same symmetry to construct a symmetric spin-space-flavor wave function.
\paragraph{SU(6)}:
The spin-flavor wave function 
\ba
6 \otimes 6 \otimes 6 = 56_S \oplus 70_M \oplus 70_M \oplus 20_A
\ea
can be classified according to their spin SU(2) and SU(3)
representations,
where the symmetric 56 multiplet can be expanded into a  spin-quartet flavor decuplet and
spin-doublet flavor octet
\ba
56\ = \ ^410\,\oplus\,^28,
\ea
the mixed-symmetric 70-plet into a spin-doublet flavor-decuplet, a flavor octet with a spin-quartet and a spin-doublet,
and a spin-doublet flavor-singlet:
\ba
70\ = \ ^210\,\oplus\,^48\,\oplus\,^28\,\oplus\,^21.
\ea
Finally, the antisymmetric 20-plet contains a flavor-octet
spin-doublet and a flavor singlet combined with a spin-quartet:
\ba
20\ = \ ^28\,\oplus\,^41.
\ea
The spin and flavor-content of hyperons is decisive for their properties. A discussion
of the implications of SU(3) symmetry on the masses, widths, and decay fractions can be
found in~\cite{Samios:1974tw,Guzey:2005rx,Guzey:2005vz}.
\paragraph{The spatial wave function:}
The spatial wave function is usually expanded into a series of harmonic oscillator (HO) wave functions.
Often, one of these HO-wave functions provides the leading contribution. From the two oscillators,
wave functions can be constructed which are symmetric ($S$), mixed symmetric ($M_S$),
mixed antisymmetric ($M_A$), or antiysmmetric ($A$). Explicite forms can be found, e.g.,
in~\cite{Loring:2001ky}.

\subsection{\label{sec:Lam}The  $\Lambda^*$ hyperons}

Table~\ref{exp-Lam} presents the $\Lambda^*$ hyperons resonances found in this analysis and
a comparison with the Bonn quark model~\cite{Loring:2001ky}.

There are six negative-partive $\Lambda^*$ resonances found
below 2000\,MeV, three of them with spin-parity $J^P=1/2^-$:
$\Lambda(1405)$, $\Lambda(1670)$, $\Lambda(1800)$;
two with $J^P=3/2^-$: $\Lambda(1520)$ and
$\Lambda(1690)$; and one with $J^P=5/2^-$: $\Lambda(1830)$.

The $\Lambda(1405)1/2^-$ is a highly discussed state; its mass is too low in comparison to quark
models, and the large spin splitting between $\Lambda(1405)1/2^-$ and $\Lambda(1520)3/2^-$ is not
understood. However, this resonance can be constructed dynamically from its decay
products~\cite{Dalitz:1959dn} opening interpretations of $\Lambda(1405)1/2^-$ as molecular state.
Modern approaches based on unitarized chiral perturbation theory exploit a $\bar K\,N$, $\pi\,\Lambda$,
$\pi\,\Sigma$ potential and fit its parameters to data in the low-mass region. Most analyses find a
two-pole structure, with one narrow pole ($\Gamma\approx 20-30$\,MeV) at about 1420\,MeV and one wider
pole ($\Gamma\approx160$\,MeV)~\cite{Oller:2000fj,Jido:2003cb}. These results were confirmed in a number of
publications \cite{Miyahara:2018onh,Hyodo:2011ur,Mai:2012dt,Molina:2015uqp}. 
However, other
analyses interpret the low-mass $N\bar K$ and $\pi\Sigma$ spectra with a single
resonance~\cite{Dong:2016auh,Myint:2018ypc,Hassanvand:2015jia}. The emphasis of the present
analysis is not a study of $\Lambda(1405)1/2^-$ properties: important data on $\pi\Sigma$
interactions below the $K^-p$ threshold \cite{Hemingway:1984pz,Moriya:2013eb,Moriya:2014kpv}, on
the $K^-p$ atom \cite{Bazzi:2011zj,Bazzi:2012eq}, and on $K^-p$ decays at rest \cite{Tovee:1971ga,Nowak:1978au}
are not included in this analysis. For this reason,
 we introduce $\Lambda(1405)1/2^-$ as a single
resonance with fixed parameters from Ref.~\cite{Hyperon-I}.

\begin{table*}[pt]\renewcommand{\arraystretch}{1.5}\setlength{\tabcolsep}{+0.62em}
\caption{\label{exp-Lam}The  $\Lambda^*$ and $\Sigma^*$  hyperons:  The experimental masses (in MeV)
from the {\it RPP}~\cite{Tanabashi:2018oca} give the mass that defines the name of the particle. The mass range
is represented by (asymmetric) uncertainties. The RPP Breit-Wigner mass is compared to our BnGa value
and the quark-model mass $M_{\rm QM}$ from Ref.~\cite{Loring:2001ky}, model A. 
This reference also gives the fractional contributions
(in \%) from different SU(6)$\otimes$O(3) configurations.  Small contributions from opposite parity or higher
configurations are omitted. $^\circ$: not from~\cite{Loring:2001ky}, see text for explanation. $\Sigma(2100)7/2^-$ 
is not included in the table. It must belong to the 3rd excitation band, even though its mass seems to be low. }
\begin{tabular}{ll}
\hspace*{-0.2cm}\begin{minipage}[c]{0.5\textwidth} 
\small
\begin{tabular}{|c|cc|cccc|} \hline
\multicolumn{7}{|c|}{$\Lambda^*$ resonances}\\\hline
$J^\pi$&RPP&BnGa&$M_{\rm QM}$\hspace{-1mm}&\hspace{-1mm}${}^2 1[70]$\hspace{-1mm}&\hspace{-1mm}${}^2 8[70]$\hspace{-1mm}&\hspace{-1mm}${}^4 8[70]$\hspace{-2mm}\\
\hline
$\frac{1}{2}^-$ &\hspace{-1.5mm}\scriptsize1405.1$^{+1.3}_{-1.0}$&1422\,\er\,3&$1524$&{\underline{69.4}}& 26.0 &  0.3 \\
$\frac{3}{2}^-$ &\hspace{-1.5mm}\scriptsize1519.5\,\er1.0&\scriptsize1518.5\er0.5&1508& {\underline{77.7}}& 18.7 &  0.1 \\
\hline
$\frac{1}{2}^-$ &1670\er10&1677\,\er\,2&$1630$ &  29.2& {\underline{61.6}} & 2.1 \\
$\frac{3}{2}^-$ &1690\,\er\,5&1689\,\er\,3&$1662$ &  20.1& {\underline{72.0}} & 2.2 \\
\hline
$\frac{1}{2}^-$ &1800$^{+50}_{-80}$&1811\er10&$1816$ & 0.1 &  3.1 & {\underline{94.9}}\\
$\frac{3}{2}^-$ &-&-&$1775$ &  0.4 &  1.5 & {\underline{96.1}}\\
$\frac{5}{2}^-$ &1830$^{+\ 0}_{-20}$&1821\,\er\,3&$1828$ &  0.0 &  0.0 & {\underline{99.0}}\\
\hline
$\frac{5}{2}^-$ &--&2082\,\er\,13&$2080$  &large$^\circ$& &   \\
$\frac{7}{2}^-$ &2100\er10&2090\er15&$2090$ &large$^\circ$&  &  \\
\hline
$J^\pi$&RPP&BnGa&$M_{\rm QM}$\hspace{-1mm}&\hspace{-1mm}${}^2 1[70]$\hspace{-1mm}&\hspace{-1mm}${}^2 8[56]$\hspace{-1mm}&\hspace{-1mm}${}^2 8[70]$\hspace{-2mm}\\
\hline
$\frac{1}{2}^+$ &1600$^{+100}_{-40}$&1605\,\er\,8&$1677$&  3.7 &{\underline{88.4}}& 6.2 \\
$\frac{1}{2}^+$ &1810$^{+40}_{-60}$&1773\er7 &\multicolumn{4}{c|}{\scriptsize 1747: \underline{91\%} $^21[70]$  
/ 1898: \underline{84\%} $^28[70]$ }\\
\hline
$\frac{3}{2}^+$ &1890$^{+20}_{-40}$&1873\,\er\,5&$1823$& 9.9 &{\underline{60.0}}& 28.2 \\
$\frac{5}{2}^+$ &1820\,\er\,5&1822\,\er\,4&$1834$ & 12.1 & {\underline{57.8}} &  28.3 \\
\hline
$\frac{3}{2}^+$ &-&2070\er24&$1952$&{\underline{84.0}}& 3.8 & 7.6  \\
$\frac{5}{2}^+$ &2110$^{+30}_{-20}$&2086\er12&$1999$ &{\underline{84.1}}& 4.5 & 8.9  \\
\hline
\end{tabular}
\end{minipage} & 
\hspace*{-0.23cm}\begin{minipage}[c]{0.5\textwidth} 
\small
\begin{tabular}{|c|cc|cccc|} \hline
\multicolumn{7}{|c|}{$\Sigma^*$ resonances}\\\hline
$J^\pi$&RPP&BnGa&$M_{\rm QM}$\hspace{-1mm}&\hspace{-1mm}${}^2 8[70]$\hspace{-1mm}&\hspace{-1mm}${}^48[70]$\hspace{-1mm}&\hspace{-1mm}${}^4 10[70]$\hspace{-2mm}\\
\hline
$\frac{1}{2}^-$ &$\sim$1620&1681\er6&$1628$ &{\underline{87.4}} & 2.3 &   3.4 \\
$\frac{3}{2}^-$ &1670$^{+15}_{-\ 5}$&1665\er3&$1669$ &{\underline{89.0}} & 1.2 &   3.4 \\
\hline
$\frac{1}{2}^-$ &1750$^{+50}_{-20}$&1692\er11&$1771$ & 2.9 &{\underline{94.6}} &   1.1 \\
$\frac{3}{2}^-$ &-&-&$1728$ & 0.1 &{\underline{82.7}} &  16.0 \\
$\frac{5}{2}^-$ &1775\,\er\,5&1776\er4&$1770$ & 0.0 &{\underline{99.0}} &   0.0 \\
\hline
$\frac{1}{2}^-$ &$\sim$1900&1938\er12&$1798$ & 2.8 &  1.7 & {\underline{94.4}} \\
$\frac{3}{2}^-$ &1940$^{+10}_{-40}$&1878\er12&$1781$ & 4.4 & 15.0& {\underline{79.3}}  \\
\hline
$J^\pi$&RPP&BnGa&$M_{\rm QM}$\hspace{-1.5mm}&\hspace{-1.5mm}${}^2 8[56]$\hspace{-1.5mm}&\hspace{-1.5mm}${}^2 8[70]$\hspace{-1.5mm}&\hspace{-1.5mm}${}^4 8[70]$\hspace{-2mm}\\
\hline
$\frac{1}{2}^-$ &-&2165\er23&$2111$   & large  & &\\
$\frac{3}{2}^-$ &-&2005\er14&$2139$  & large & &\\
\hline
$\frac{1}{2}^+$ &1660\er30&1665\er20&$1760$ & {\underline{96.1}} &2.3& 0.0\\
\hline
$\frac{3}{2}^+$ &$\sim$1840&-&$1896$ &{\underline{73.9}} & 22.2 &   0.6 \\
$\frac{5}{2}^+$ &1915$^{+20}_{-15}$&1918\er6&$1956$ &{\underline{77.8}} & 18.2 &   0.2 \\
\hline
$J^\pi$&RPP&BnGa&$M_{\rm QM}$\hspace{-1.5mm}&\hspace{-1.5mm}${}^2 8[56]$\hspace{-1.5mm}&\hspace{-1.5mm}${}^4 8[70]$\hspace{-1.5mm}&\hspace{-1.5mm}${}^4 10[56]$\hspace{-2mm}\\
\hline
$\frac{7}{2}^+$ &2030$^{+10}_{-\ 5}$&2032\er6&$2070$ &0.0 & 29.4 &   {\underline{69.6}}\\[-0.1ex]
&&&&&&\\
\hline
\end{tabular}
\end{minipage}\end{tabular}
\end{table*}
$\Lambda(1405)1/2^-$ has a spin partner, $\Lambda(1520)3/2^-$; in quark models, these two states are commonly interpreted
as forming the expected spin doublet, SU(3) singlet.

The four further negative-parity $\Lambda^*$ resonances below 2000\,MeV are $\Lambda(1670)1/2^-$,
$\Lambda(1690)3/2^-$, $\Lambda(1800)1/2^-$, and $\Lambda(1830)5/2^-$. States with identical $J^P$
but different quark spins or in different SU(3) representations can mix. Nevertheless, the lower-mass states
can be assigned to a spin doublet, the higher-mass states could belong to a spin triplet. The comparison with
the quark-model calculation~\cite{Loring:2001ky} suggests that the two states assigned to
a triplet should indeed belong to the $^48[70]$ configuration and that  they have only a small contribution
from spin-doublet configurations. On the other hand, there could be significant singlet-octet mixing as expected for the two lower mass states.

The experimental masses are reasonably consistent with the quark model predictions,
except for the well-known problems with the masses of $\Lambda(1405)1/2^-$ and the Roper-like
resonances $\Lambda(1600)1/2^+$ and $\Sigma(1660)1/2^+$. The two resonances
$\Lambda(1670)1/2^-$ and $\Lambda(1690)3/2^-$ have masses which are about 150\,MeV 
above $N(1535)1/2^-$
and $N(1520)$ $3/2^-$; the mass difference  corresponds to the expected mass difference between
the constituent masses of $u,d$ and $s$-quarks. Correspondingly, we expect a spin triplet of states
150\,MeV above $N(1650)1/2^-$, $N(1700)3/2^-$, $N(1675)$ $5/2^-$, i.e. at about 1825\,MeV. Indeed,
there we observe a $J^P=1/2^-$ state at 1811\,MeV and a $5/2^-$ state at 1821\,MeV. The $3/2^-$ state
is missing; its expected partial width for the $\bar KN$ is 0.2\,MeV only~\cite{Guzey:2005rx,Guzey:2005vz}; given
the limited data base, this partial width is likely too small to be observed in $K^-p$ induced reactions.

There are two further negative-parity $\Lambda^*$ resonances, a new $\Lambda(2080)5/2^-$ and the well known
$\Lambda(2100)7/2^-$. Based on the sign of the $K^-p\to\Lambda(2100)7/2^-\to\pi\Sigma$ amplitude, this
state is assigned to the SU(3) singlet configuration~\cite{sign} with $L=3, S=1/2$ as dominant wave. 
The new $\Lambda(2080)5/2^-$ is likely its spin partner. The mass-square spacing between
$\Lambda(2100)7/2^-$ and $\Lambda(1520)3/2^-$ is (2.1\er0.1)\,GeV$^2$, between $\Lambda(2080)5/2^-$ and 
$\Lambda(1405)1/2^-$ is (2.3\er0.1)\,GeV$^2$; that between $N(2190)7/2^-$ and $N(1535)$ $1/2^-$
is (2.4\er 0.2)\,GeV$^2$. Again, the assignment of $\Lambda(2080)$ $5/2^-$ and
$\Lambda(2100)7/2^-$ to the  $^21[70]$ configuration seems plausible. We identify these two
states with the lowest-mass $\Lambda^*$ resonances with these quantum numbers in the third
excitation shell~\cite{Loring:2001ky}. 

Only one positive-parity $\Lambda$ state with $J^P=1/2^+$ was found to be required in the analysis. 
The $\Lambda(1600)1/2^+$ is likely the first (Roper-like) radial
excitation of the respective ground state. The next state -- called $\Lambda(1810)1/2^+$ --
is not required in this analysis. But it has a 3* rating in the RPP; when included 
in our fits, it is seen with properties (e.g. mass and width) rather similar 
to those found in other analyses. Hence we keep it in the list of resonances.
Its interpretation is ambiguous:  Ref.~\cite{Loring:2001ky} predicts two state in this 
mass region: one state at 1747\,MeV in the $^21[70]$, a second one at 1898\,MeV in the 
$^28[70]$ configuration. The latter state is the analogue 
to $N(1710)1/2^+$. In the $3/2^-$ sector we have seen that singlet and octet states show considerable mixing.
We hence suppose that $\Lambda(1810)1/2^+$ may emerge from the mixing two
quark-model states in the $^21[70]$ and $^28[70]$ configurations. While the mass of (1773\er7)\,MeV would fit better to the dominantly $^21[70]$ quark model state, the strong $\Sigma^*\pi$ decay indicates an 
octet component in the wave function. The state orthogonal
to $\Lambda(1810)1/2^+$ would still need to be discovered. 

The number of expected states with spin-parity $3/2^+$ (seven) and $5/2^+$ (five) in the second excitation shell
is large.  Since we observe no $7/2^+$ state, we assume that $\Lambda(1890)3/2^+$ -- observed here with 
M=(1873\er5) MeV -- and $\Lambda(1820)5/2^+$ form a spin doublet.  Their mean mass
is about 150\,MeV above the mean mass of $N(1720)3/2^+$ and $N(1680)5/2^+$. These latter states 
are usually interpreted as the first orbital
angular momentum excitations with $L=2$ in the $^28[56]$ representation. Thus we assume
that  this interpretation holds for the two $\Lambda$ states as well. 
This assignment is supported by
quark model calculations even though mixing with other states is very significant
(see Table~\ref{exp-Lam} and Ref.~\cite{Loring:2001ky}.

The next two states, $\Lambda(2115)5/2^+$ -- observed here at (2086\er12)\,MeV --
and the new $\Lambda(2070)3/2^+$ again seem to form a doublet; there is no $7/2^+$ 
companion. Their assignment to quark-model states is ambiguous. The next states in mass, 
above $\Lambda(1890)3/2^+$
and $\Lambda(1820)5/2^+$, are predicted at 1952\,MeV and 1999\,MeV in the $^21[70]$
configuration~\cite{Loring:2001ky}, followed by 2045\,MeV and 2078\,MeV in the $^28[70]$.
From the level ordering, the two observed states belong to the SU(3) singlet, from the observed mass to
the octet. Both are predicted to be mixed only modestly. In Table~\ref{exp-Lam} we compare the
experimental findings with the lower-mass singlet states.

\subsection{\label{sec:sig}The  $\Sigma^*$ hyperons}

The assignment of the observed $\Lambda^*$ states to specific configurations provided
a rather consistent picture of the low-mass spectrum. All states predicted~\cite{Loring:2001ky}
below 2000\,MeV are identified
with the exception of a $3/2^-$ state expected at about 1800\,MeV and a further $1/2^+$ state at about 1747/1898\,MeV
which would correspond to the $N(1710)1/2^+$ nucleon mixed with the singlet state 
($N(1710)1/2^+$ was difficult to extract
from $\pi N$ scattering data without the inclusion of photoproduction experiments).

The spectrum of observed $\Sigma^*$ resonances is shown on the right panels of Table~\ref{exp-Lam}.
The two lowest mass states,  $\Sigma(1620)1/2^-$ and $\Sigma(1670)3/2^-$, are easily assigned
to the $^28[70]$ configuration which we have seen already in the $\Lambda$ sector. In the expected triplet of
negative-parity resonances, only  $\Sigma(1750)1/2^-$ and  $\Sigma(1775)5/2^-$ are seen; 
the $3/2^-$ state is missing as in the $\Lambda$ sector. 

However,  the two
$1/2^-$ states at 1692\,MeV ($\Sigma(1750)$) and 1681\,MeV ($\Sigma(1620)$) 
are worrisome. In the first scans, only one $1/2^-$ 
low-mass state was seen at $\sim$1690\,MeV
with high confidence (4*).  When we searched
for the next state, a complicated pattern with two close-by poles at 1681 and 1692\,MeV developed. The
second pole at 1681\,MeV proved to be just statistically significant. When this state was kept 
and the lower mass pole was removed from the fit, $\chi^2$  change was  just 
below the value for which it would be considered as 1* resonance.  If one of these two poles
is fake, one $1/2^-$ state would be missing as well.

The negative-parity resonances which we just discussed have analogue states in the $N^*$ sector. But
there is also a doublet of $\Delta^*$ state: $\Delta(1620)1/2^-$ and $\Delta(1700)3/2^-$. Hence we should
expect a further spin doublet above 1800 MeV. Indeed, there is possibly a further doublet: a $1/2^-$ state
at 1938\,MeV and a $3/2^-$ at 1878\,MeV. Both states are -- compared to~\cite{Loring:2001ky} --
rather high in mass. 
The two states at 2165 and 2005\,MeV could possibly be in the ${}^2 8[56]$ configuration,
analogue to $N(1895)1/2^-$ and $N(1875)$ $3/2^-$,  but this is speculative at the moment.

Four positive-parity $\Sigma$ state were found to be required in the analysis.
The $\Sigma(1660)1/2^+$ is likely the first (Roper-like) radial
excitation of the ground state. States corresponding to $\Delta(1600)3/2^+$ and $N(1710)1/2^+$ are missing.
The first orbital angular momentum excitations with $L=2$ are
$N(1720)3/2^+$ and $N(1680)5/2^+$.
 $\Sigma(1915)5/2^+$ is likely one of the analogue states with its spin partner with $J^P=3/2^+$ missing. We
assign $\Sigma(2030)7/2^+$ to be partner of $\Delta(1950)7/2^+$.
The interpretation of the $3/2^+$ state at 2230 MeV is open.

\subsection{Discussion}
The agreement between the spectrum of hyperon resonances and quark-model predictions
is remarkable. It should be noted that in each partial wave, all quark-model resonances 
are listed in Table~\ref{exp-Lam} up to the largest observed mass. Decisive for this interpretation
is the removal of ``spurious'' signals stemming from a variety of different analyses. Particularly  
interesting is the identification of three spin-doublets which can be assigned to the spectrum
of SU(3) singlet baryons. Possibly, also the negative-parity spin-doublet of $\Sigma$ decuplet
has been identified.


\section{\label{Summary}Summary}
We have performed a coupled-channel analysis of available data on $K^-p$ induced reactions.
Data on two-body reactions were reported in the preceding paper where also the analysis method
is described.  The emphasis of this paper
is laid upon the inclusion of three-body data -- which were analyzed event-by-event in a likelihood
fit -- and on quasi-two-body final states. For these, the differential cross sections and the $\rho$ density
matrix elements are available in the form of associated Legendre polynomes.

In this paper we present Tables of the properties of hyperon resonances as observed in the
BnGa analysis. The branching ratios of most lower-mass resonances add up to unity. We report
pole position and normalized transition residues as well as Breit-Wigner properties such as
mass, width and branching ratios.

The comparison with the results from other analyses often show larger discrepancies than allowed by
statistics. In particular there is little agreement for the quasi-two-body decay modes. 
These are obviously not
sufficiently constrained and the results seem to depend on the particular choice of the model.   

The most important result of this analysis is the systematic check of the significance of resonances. 
It turns out that a large number of resonances reported in the Review of Particle Physics is not required
to achieve a reasonable fit. In total, 20 resonances or ``bumps'' are found to make no significant
improvement of the fit. 

The spectrum is compared to the Bonn quark model which uses a linear confinement potential
and instanton interactions between constituent quarks in a relativistic kinematic. Generally, the comparison
gives  good agreement. It is remarkable that six $\Lambda$ states have to be assigned to the SU(3)
singlet system, and that two of them are observed here for the first time. \vspace{2mm}

This work was supported by the \textit{Deutsche Forschungsgemeinschaft} (SFB/TR110)
and the \textit{Russian Science Foundation} (RSF 16-12-10267).

  \end{document}